\documentclass{aa}  

\usepackage{graphicx}
\usepackage{longtable}
\usepackage{xcolor}
\usepackage{fontawesome}
\usepackage{mathtools}
\definecolor{dgreen}{rgb}{0.0, 0.5, 0.0}
\usepackage{txfonts}
\usepackage{hyperref}
\hypersetup{
    colorlinks=true,
    linkcolor=blue,
    filecolor=blue,
    citecolor=blue,
    urlcolor=blue,
}
\begin{document}

   \title{Simulation of CH$_3$OH ice UV photolysis under laboratory conditions}

   \author{W. R. M. Rocha\inst{1,2},
          P. Woitke\inst{3},
          S. Pilling\inst{4},
          W. -F. Thi\inst{5}
          J. K. J{\o}rgensen\inst{2},
          L. E. Kristensen\inst{2},
          G. Perotti\inst{2,6},
          \and
          I. Kamp\inst{7}
          }

   \institute{Laboratory for Astrophysics, Leiden Observatory, Leiden University, P.O. Box 9513, NL 2300 RA Leiden, The Netherlands.\\
              \email{rocha@strw.leidenuniv.nl}
         \and
             Niels Bohr Institute \& Centre for Star and Planet Formation, University of Copenhagen, {\O}ster Voldgade 5$-$7, 1350 Copenhagen K., Denmark
         \and
             Space Research Institute, Austrian Academy of Sciences, Schmiedlstrasse 6, A-8042 Graz, Austria
        \and
             Instituto de Pesquisa e Desenvolvimento (IP\&D), Universidade do Vale do Paraíba, Av. Shishima Hifumi 2911, CEP 12244-000, São José dos Campos, SP, Brazil
        \and
             Max-Planck-Institut für extraterrestrische Physik, Giessenbachstrasse 1, 85748 Garching, Germany
        \and
             Max Planck Institute for Astronomy, K{\"o}nigstuhl 17, D-69117 Heidelberg, Germany
         \and
             Kapteyn Astronomical Institute, University of Groningen, The Netherlands
             }

   \date{Received xxxx; accepted yyyy}

 
  \abstract
   {Methanol is the most complex molecule securely identified in interstellar ices and is a key chemical species for understanding chemical complexity in astrophysical environments. Important aspects of the methanol ice photochemistry are still unclear such as the branching ratios and photo-dissociation cross-sections at different temperatures and irradiation fluxes.}
   {This work aims at a quantitative agreement between laboratory experiments and astrochemical modelling of the CH$_3$OH ice UV photolysis. Ultimately, this work allows us to better understand which processes govern the methanol ice photochemistry present in laboratory experiments.}
   {We use the \texttt{ProDiMo} code to simulate the radiation fields, pressures and pumping efficiencies characteristic of laboratory measurements. The simulations start with simple chemistry consisting only of methanol ice and helium to mimic the residual gas in the experimental chamber. A surface chemical network enlarged by photo-dissociation reactions is used to study the chemical reactions within the ice. Additionally, different surface chemistry parameters such as surface competition, tunnelling, thermal diffusion and reactive desorption are adopted to check those that reproduce the experimental results.}
   {The chemical models with the \texttt{ProDiMo} code including surface chemistry parameters can reproduce the methanol ice destruction via UV photodissociation at temperatures of 20, 30, 50 and 70~K as observed in the experiments. We also note that the results are sensitive to different branching ratios after photolysis and to the mechanisms of reactive desorption. In the simulations of a molecular cloud at 20~K, we observed an increase in the methanol gas abundance of one order of magnitude, with a similar decrease in the solid-phase abundance.}
   {Comprehensive astrochemical models provide new insights into laboratory experiments as the quantitative understanding of the processes that govern the reactions within the ice. Ultimately, those insights can help to better interpret astronomical observations.}

   \authorrunning{Rocha et al.}

   \maketitle
%

\section{Introduction}
Methanol (CH$_3$OH) ice has been securely detected in the interstellar medium (ISM) towards different lines of sight \citep[e.g.,][]{Schutte1991, Chiar2000, Pontoppidan2004, Boogert2011, Perotti2020, Perotti2021, Chu2020, Goto2021, Kim2022ApJ}. We highlight recent detection made with {\it James Webb} Space Telescope (JWST) observations, towards the Class 0 protostar IRAS 15398-3359 \citep[program 2151, PI: Y.-L. Yang;][]{Yang2022} and in the densest regions of the molecular clouds Chameleon I \citep{Mcclure2023} as part of the Early Release Science (ERS) Ice Age program (PIs: M. Mcclure, A. Boogert, H. Linnartz). This simple alcohol is mostly formed via successive CO hydrogenation \citep[][]{Watanabe2002, Fuchs2009} through ``dark ice chemistry'' \citep[][]{Ioppolo2021}, a term used to characterize molecular synthesis without the influence of ionizing radiation. Additionally, other surface reactions at low temperature have been proposed in the literature. \citet{Qasim2018} found experimentally that CH$_3$OH is formed via a sequential surface reaction chain at low temperature, i.e., CH$_4$ + OH $\rightarrow$ CH$_3$ + H$_2$O and CH$_3$ + OH $\rightarrow$ CH$_3$OH. More recently, \citet{Santos2022} confirmed via experiments that CH$_3$OH also forms in the ice via CH$_3$O + H$_2$CO $\rightarrow$ CH$_3$OH + HCO at temperatures similar to the interior of molecular clouds. Additionally, methanol ice is formed by energetic processing of simple molecules in the ice mantle, such as H$_2$O, CO and CO$_2$ \citep[e.g.,][]{Jimenez2016}. Despite its low abundance toward low-mass star-forming regions \citep[5$-$12\% with respect to H$_2$O ice;][]{Oberg2011}, methanol is crucial for understanding the chemical complexity in stellar nurseries because it is the precursor of large and Complex Organic Molecules - COMs \citep[][]{Oberg2009, Oberg2016}. In the Solar System, methanol has also been detected in comets \citep[e.g., 67P/Churyumov-Gerasimenko][]{LeRoy2015} and Kuiper Belt Objects \citep[e.g.,][]{Grundy2020}. In the former, mono- and di-deuterated methanol detection in a cometary coma is also reported by \citet{Drozdovskaya2021}. In the latter case, spectral analysis of Arrokoth suggests a water-poor surface and high abundance of methanol \citep[][]{Grundy2020}.

In order to provide qualitative information on how frozen methanol promotes the synthesis of COMs, systematic laboratory experiments simulating the effects of UV radiation on the CH$_3$OH ice were performed by \citet{Oberg2009}. From that study, the formation of several COMs such as CH$_3$CH$_2$OH, CH$_3$OCH$_3$, HCOOCH$_3$ and HOCH$_2$CHO was observed. As suggested by the authors, this is the result of CH$_3$OH photolysis and radical recombination after diffusion. Nevertheless, a few aspects remained unclear, such as the reaction channel that dominates at different temperatures and fluences (flux integrated over time), as well as the branching ratios (BRs), i.e., the fraction of the radicals formed after UV photolysis. Additionally, the photodissociation cross-sections derived from this experimental study are underestimated because of the immediate methanol ice reformation. Complementary chemical kinetic modelling is, therefore, necessary to completely quantify the photochemistry of methanol ices, as pointed out by the authors. 

Combining careful modelling of ices under laboratory conditions aided by the experiments might provide a comprehensive understanding of the processes that govern reactions within the ices. With this goal, \citet{Shingledecker2019} studied the proton-irradiation of O$_2$ and H$_2$O ices using astrochemical-type models implemented with the \texttt{MONACO} code \citep[][]{Vasyunin2017}, that uses a multiphase Monte Carlo approach to calculate chemical abundances in the context of gas-grain interaction. The models demonstrated that cosmic-ray-driven chemical reactions can reproduce the abundances of O$_2$, O$_3$ and H$_2$O$_2$ observed in the experiments if the radicals produced by proton-irradiation react quickly and locally in the ice. Similarly, the formation of S$_8$ and other sulfur-bearing species are addressed with the same method by \citet{Shingledecker2020}, showing again that these kinetic studies of chemical reactions in the ices are in agreement with the abundances of sulfur-bearing species measured toward Comet 67P/C-G by the ROSINA \citep[{\it Rosetta} Orbiter Spectrometer for
Ion and Neutral Analysis; ][]{Balsiger2007} instrument on ESA's Rosetta mission \citep[][]{Calmonte2016}. In a lower energy regime, \citet{Mullikin2021} successfully simulated the ozone (O$_3$) formation due to UV irradiation of O$_2$ ice using a Monte Carlo-based model. In their simulations, chemical processes such as photoionization, photoexcitation, and the formation of electronically excited species are considered in the models. Despite the good agreements between experiments and simulations, the caveats in these approaches are the uncertainties in the chemical networks, the reaction rates and BRs for surface chemistry processes included in the models. Further theoretical works exploring these uncertainties and using a reduced network have been carried out by \citet{Pilling2022}, \citet{Carvalho2022} and \citet{Pilling2023} (accepted) aiming at the determination of effective rate constants of reactions in the solid phase.

Inside dense molecular clouds, energetic cosmic rays excite the gas-phase H$_2$, which results in Lyman and Werner emission lines \citep[][]{Prasad1983} $-$ the secondary UV photons. The typical flux in dense starless cores ranges from 10$^3$ to 10$^4$ cm$^{-2}$ s$^{-1}$ \citep[e.g.,][]{Shen2004}. Although it represents a mild flux, over the timescale of molecular clouds \citep[$<$ 10 Myr - e.g.,][]{Clark2012}, this results in fluence of around 10$^{18}$ cm$^{-2}$. This is comparable with the fluence adopted by \citet{Oberg2009} to photolyse the methanol ice and to form a variety of COMs. Additionally, other laboratory experiments with UV as an irradiation source have reported the formation of prebiotic molecules, such as ribose \citep[][]{Meinert2016} and deoxyribose \citep[][]{Nuevo2018}. In spite of the importance of chemical processes within ices triggered by UV radiation, little is known about those processes. In addition, JWST is dedicating many hours to observe ices, which makes it crucial to improve our understanding of the mechanisms involved in ice photolysis. To shed light on this problem, and to shrink the gap between laboratory and chemical models, we introduce in this paper systematic modelling of methanol ice photochemistry triggered by UV radiation using the \texttt{ProDiMo} code \citep[][]{Woitke2009, Kamp2010, Thi2011}, that is currently equipped with surface chemistry \citep{Thi2020_feb, Thi2020_mar}. In particular, we have modelled the methanol ice destruction under laboratory conditions shown in \citet{Oberg2009}. Once the best results are found, they are used to simulate the chemical evolution in a molecular cloud environment. The differences and similarities in the models including or excluding methanol ice photochemistry are discussed.

This paper is outlined as follows. Section~2 introduces the chemical model used to reproduce the methanol UV photolysis. In Section~3, the set-up of the models considering different surface chemistry parameters is presented as well as the selection method of the best model. The results of this study are shown in Section~4 and Section~5 is focused on the discussion and astrophysical implications.

\section{Chemical model}
Our two-phase chemical model was constructed with the \texttt{ProDiMo} code, which is originally a thermochemical radiative transfer code for the modelling of the physics and chemistry of protoplanetary disks and molecular clouds. In particular, we have used the surface chemistry module, recently developed by \citet{Thi2020_feb, Thi2020_mar}. We changed the setup of these models to simulate the radiation fields, pressures, and the composition and thickness of ice layers as they are prepared in laboratory experiments. Regarding the surface chemistry processes, the models include photolysis (photodissociation) with different BRs, thermal and non-thermal desorption, and diffusion mechanisms via tunnelling and thermal hopping. Further details about these parameters and processes are given in the next subsections.

\subsection{UV flux}
The experiment performed by \citet{Oberg2009} used UV irradiation to destroy the methanol ice in their experiments. In \texttt{ProDiMo}, the chemical models adopt a UV flux as a function of the values in the ISM. The standard UV radiation field in the ISM in units of erg~cm$^{-3}$ is formally defined by \citet{ Draine1978, Draine_Bertoldi1996}:
 \begin{equation}
     \lambda u_{\lambda}^{\rm{Draine}} = 6.84 \times 10^{-14} \frac{31.016 \lambda_3^2 - 49.913\lambda_3 + 19.897}{\lambda_3^5}
     \label{Draine_energy}
 \end{equation}
where $\lambda$ is the wavelength, $u_{\lambda}^{\rm{Draine}}$ is the spectral photon energy ($\frac{4\pi}{c}J_{\lambda}$), $c$ the speed of light, $J_{\lambda}$ the mean intensity and $\lambda_3 = \lambda/100$ nm. The standard FUV flux of the ISM is found by integrating Equation~\ref{Draine_energy} between 91.2~nm and 205~nm:
\begin{equation} \label{Draine_flux}
\begin{split}
 F_{\mathrm{Draine}} & = \frac{1}{h} \int_{91.2~nm}^{205~nm} \lambda u_{\lambda}^{\rm{Draine}} d\lambda \\
 & = 1.9921 \times 10^{8} \; \; \mathrm{[cm^{-2} s^{-1}]}.
\end{split}
\end{equation}
where $h$ is Planck's constant. The integration interval spans over crucial physical processes such as (i) the photoexcitation and photodissociation of many molecules including H$_2$, (ii) photoionization of heavy elements, and (iii) the ejection of electrons from dust grains. Often, the FUV radiation strength ($\chi$) of local regions in the ISM is normalized by the radiation field from Draine, and it is given by:
\begin{equation}
    \chi = \frac{\int_{91.2~nm}^{205~nm} F_{\lambda} d\lambda}{ F_{\rm{Draine}}}
    \label{Drfield1}
\end{equation}
where $F_{\lambda}$ is the local FUV flux. By definition, the strength of the the standard FUV flux in the ISM is $\chi_{\rm{ISM}} = 1$.

In laboratory experiments, the UV radiation is produced by H$_2$ or D$_2$ microwave discharge lamps. The spectrum usually covers 110$-$180~nm and is characterized by a strong Ly~$\alpha$ peak at 121~nm and fluorescence emission around 160~nm. As found by \citet{Ligterink2015}, the Ly~$\alpha$ peak intensity can be increased, and the fluorescence decreased if helium (He) is added with H$_2$ or D$_2$, respectively.   
 
\begin{figure}
   \centering
   \includegraphics[width=\hsize]{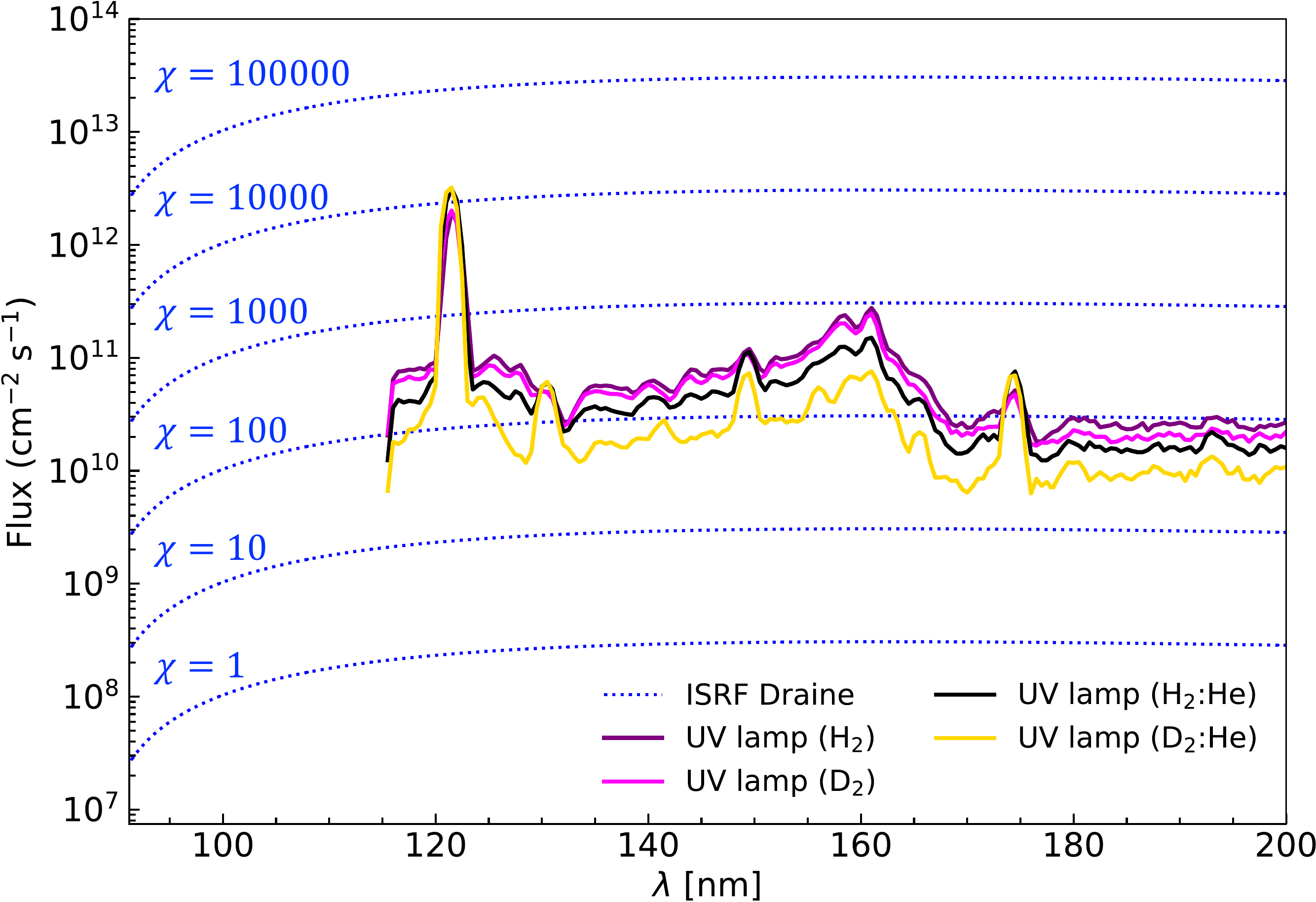}
      \caption{Comparison between the UV flux distribution of the interstellar radiation field ($\chi$) from \citet{Draine1978} and from the lamps used in the laboratory for ice photo-chemistry.}
         \label{lamp_ism}
   \end{figure}

Figure~\ref{lamp_ism} compares the UV spectra in the laboratory and ISM. The typical laboratory UV flux integrated from 110~nm to 200~nm is around 10$^{12-13}$~cm$^{-2}$ s$^{-1}$, i.e. four or five orders of magnitude higher than the typical ISM value (see Equation~\ref{Draine_flux}). Given such difference, it is postulated in the literature that a high flux in the laboratory in a short period of time produces the same chemistry as in the ISM, which has a low flux over a longer period. This can be translated to the term ``fluence'', which means the flux integrated over time. As a test case, \citet{Oberg2009} compared two experiments with different fluxes (factor of four difference), but with the same fluences. As the authors observed, the IR spectra are identical, within the experimental uncertainties. In this paper, we used $\chi = 5 \cdot 10^4$ to simulate the UV flux of $1.1 \cdot 10^{13}$ cm$^{-2}$ s$^{-1}$ used by \citet{Oberg2009}.
 
\subsection{Pump simulator}
In laboratory experiments with astrophysical ices, high or ultra-high vacuum conditions are required to avoid (i) atmospheric contamination, (ii) re-condensation of the photo-products released to the gas phase and keep the evolution of the ice composition under control. The typical vacuum inside the experimental chamber is around 10$^{-10}$~mbar, which is usually achieved by combining diffusion and turbo-molecular pumps.

In our model, the gas pressure is calculated by the following equation:
\begin{equation}
    P_{\mathrm{g}} = n_{\mathrm{tot}}k_{\mathrm{B}}T_{\mathrm{g}}
    \label{pressure}
\end{equation}
where $n_{\mathrm{tot}}$ is the total gas density, $k_{\mathrm{B}}$ the Boltzmann constant, and $T_{\mathrm{g}}$ the gas temperature. To simulate the same laboratory conditions, an initial pressure of 10$^{-10}$~mbar due to the residual gas (e.g. He) is assumed.

During the simulations, the gas pressure increases because of the chemical species that migrate from the ice to the gas phase. To keep the pressure constant, a correction factor $f$ of around 0.01 is often applied to Equation~\ref{pressure}.

\subsection{Surface chemistry}
\label{surfchem}
Simulating the chemical reactions within the ices under laboratory conditions with astrochemical-type models are challenging, and inevitably introduces caveats. For example, the grain surface chemistry in \texttt{ProDiMo} assumes that ices mantles are formed on a dust grain, whereas in the laboratory ices are condensed on a plain substrate (e.g., CsI, ZnSe, KBr), which leads to differences in the binding energies in the interface between substrate/molecule and grain/molecule. However, this effect is minimized for thick ice layers, where the chemical species are mostly bound by van der Waals forces. In the models, the majority of the ice chemistry occurs in the statically well-mixed
mantle, where all adsorptions/desorptions are on physisorption sites (see Section~\ref{sec_diffusion}). There is no distinction between ice surface and bulk. However the ice chemistry itself in \texttt{ProDiMo} is considered multi-phase, because of the chemistry that can occur in the ice mantle (phase one), interface mantle and core (phase two) or in the core itself (phase three), where only chemisorption sites are considered.

In this subsection, we describe the details of the chemical network adopted in our models, as well as the surface chemistry parameters we have adopted to simulate the methanol ice photolysis. Those parameters range from photodissociation, radical migration, and thermal and non-thermal desorption mechanisms. The different models including or excluding those processes are described in Section~\ref{models_setup}.

\subsubsection{Network and branching ratios}
Both gas-phase and solid-phase reactions are included in the network of chemical reactions considered in this paper to model the methanol ice photodissociation. If not explicitly discussed, the chemical reaction pathways and their rates are taken from the KIDA database, namely, \texttt{kida.uva.2014}. Appendix~\ref{all_elements} lists the gas and ice chemical species used in this work. Appendix~\ref{ice_photo} contain the list of newly added surface photolysis and Appendix~\ref{bimol} shows the bimolecular surface reactions and binding energies used in this paper cross-checked against \citet{Wakelam2017}. We highlight that we assume the binding energy as a single value, and not as a distribution as discussed by \citet{Shimonishi2018}, \citet{Grassi2020}, \citet{Bovolenta2022}, \citet{Villadsen2022}, and \citet{Minissale2022}. In addition to the methanol formation and destruction pathways included in these lists, we have added the methanol ice photodissociation routes (see Table~\ref{table_brs}). 

Currently, there is no consensus about the methanol ice BRs. The ``standard'' BRs used in astrochemical models were proposed by \citet{Garrod2008} and assume a destruction channel favouring the methyl channel (CH$_3$) for both gas- and solid-phase. Although the overall rate coefficient for methanol photodissociation used in \citet{Garrod2008} models was derived from gas-phase experimental work, the BRs are not constrained by laboratory work. Such a measurement in the ice phase was carried out by \citet{Oberg2009} using infrared and mass spectrometry analysis. Contrary to the previous BRs proposed by \citet{Garrod2008}, the experiments demonstrated that solid-phase methanol is mainly destroyed via the hydroxymethyl (CH$_2$OH) channel.

By using rate-equation-based chemical models, \citet{Laas2011} addressed the impact of the BRs in the production of complex molecules observed toward Sgr B2(N). In addition to the ``standard'' BRs previously proposed, other two values were also adopted to test extreme cases where the CH$_3$, CH$_2$OH or the OCH$_3$ is the dominant channel of destruction. It was observed that for some COMs, CH$_3$ or OCH$_3$ results in better agreement with the gas-phase observations. Conversely, the CH$_2$OH channel results in worse matching, which is contrary to the experimental results found by \citet{Oberg2009}. 

More sensitive measurements in the ice phase were made by \citet{Paardekooper2016} using time-of-flight mass spectrometry analysis. Initially, the authors traced the abundance of the photoproducts CH$_2$OH, CH$_3$ and OCH$_3$ to follow the chemical network proposed by \citet{Oberg2009}. Next, the abundance of new molecules formed by the recombination of the photoproducts was investigated. Only OH-dependent species were excluded from this tracking because of water formation during the experiments which is an important sink for the hydroxyl (OH) radical. The BR deduced from this analysis indicates that the CH$_2$OH branch is the dominant photo-destruction route of methanol ice. While this result qualitatively agrees with the BRs found by \citet{Oberg2009}, the experiments showed 35\% more icy methanol destruction via the CH$_3$ channel and $\sim$5\% less destruction via the CH$_2$OH channel.

A different technique is used by \citet{Weaver2018} to estimate the methanol photolysis BRs in the gas phase. In a vacuum chamber, methanol is photodissociated by a VUV laser in the throat of a supersonic expansion. After that, the products are analysed with millimeter/submillimeter spectroscopy to derive abundances in the gas phase. This study showed the same photoproducts as observed in the experiments by \citet{Oberg2009} and \citet{Paardekooper2016}, with the addition of formaldehyde (H$_2$CO), which was not observed in the experiments in the solid phase as a direct photo-product of the methanol ice. However, this agrees with \citet{Hagege1968} that identified H$_2$CO as one of the photoproducts of gaseous methanol.

Finally, an experiment performed by \citet{Yocum2021} suggests that the OCH$_3$ radical is the dominant branch of the methanol ice photolysis. The study is carried out by combining solid-phase analysis using standard infrared spectroscopy and submillimeter/far-infrared spectroscopy when the ice is warmed-up to 300~K. However, no quantification of the BRs is estimated.

The percentage values of these different BRs of methanol ice photodissociation are summarized in Table~\ref{table_brs}. They are explored in this paper along with other parameters detailed in this section to find the best solution that matches the experimental data.

\begin{table*}
\caption{\label{table_brs}Branching ratios of methanol ice photo-dissociation. Symbol ``\#'' indicates chemical species in the solid phase.}
\renewcommand{\arraystretch}{1.5}
\centering
\small\addtolength{\tabcolsep}{-0.5pt}
\begin{tabular}{ccccccccccccc}
\hline\hline
Reactions & & & & & \multicolumn{8}{c}{Branching ratios (\%)} \\
\hline
 & & & & & GA08$^a$ & OB09$^b$ & LA11-M$^c$ & LA11-MX$^d$ & LA11-HM$^e$ & PA16$^f$ & WW18-1$^g$ & WW18-2$^h$\\
\cline{6-13}
CH$_3$OH\# + h$\nu$ & $\rightarrow$ & $\rm{CH_3\#}$ &+& $\rm{OH\#}$     & 60 & 15 & 90 & 5 & 5 & 20 & 5 & 12\\
                    & $\rightarrow$ & $\rm{OCH_3\#}$ &+& $\rm{H\#}$     & 20 & 14 & 5 & 90 & 5 & 13 & 39 & 36\\
                    & $\rightarrow$ & $\rm{CH_2OH\#}$ &+& $\rm{H\#}$    & 20 & 71 & 5 & 5 & 90 & 67 & 53 & 49\\
                    & $\rightarrow$ & $\rm{H_2CO\#}$ &+& $\rm{H_2\#}$    & -/- & -/- & -/- & -/- & -/- & -/- & 2 & 2\\

\hline
\end{tabular}
\tablefoot{$^a$\citet{Garrod2008}, $^b$\citet{Oberg2009}, $^c$\citet{Laas2011} - methyl BR, $^d$\citet{Laas2011} - methoxy BR, $^e$\citet{Laas2011} - hydroxymethyl BR, $^f$\citet{Paardekooper2016}, $^g$\citet{Weaver2018} - assuming CH$_3$ = 5\%, $^h$\citet{Weaver2018} - assuming CH$_3$ = 12\%.}
\end{table*}

\subsubsection{Photodissociation}
\label{sec_photodiss}
The absorption of UV photons by a molecule in the ice changes its energy state to an excited level, which might lead to molecular fragmentation. However, the dissociation probability decreases as the number of atoms in the molecule increase because of the high number of vibrational modes \citep[][]{Bixon1968, TRAMER1979281, Leger1989, vanDishoeck2011}. When a photon is absorbed by a large molecule, there is a rapid internal vibrational energy redistribution, and the molecule ends up at a vibrational level highly excited. Subsequently, the molecule relaxes by emission of infrared photons or fluorescence before arriving in the electronic ground state. The probability that the molecule finds a path to dissociation is rather small. For example, experiments performed by \citet{Jochims1994} show that the photostability of polycyclic aromatic hydrocarbons is significantly reduced if the number of atoms is lower than 30-40. For small and intermediate-sized molecules containing less than 10 atoms (including CH$_3$OH), \citet{Ashfold2010} concluded from theoretical work that their dissociative excited states are comparable to those of H$_2$O and NH$_3$, regardless of the level of structural complexity.

In laboratory conditions, the FUV flux is not dependent on the visual extinction ($A_{\rm{V}}$), and the photodissociation rate is equal to the photo-rate $\alpha$ calculated in Appendix~\ref{ap_alpha}, i.e.,
\begin{equation}
\label{eq_phd_rate}
R_i^{\rm{phd}} = \overline{\sigma} \cdot F_{\rm{lamp}} \;\;\; [\rm{s^{-1}}]
\end{equation}
where $\overline{\sigma}$ is the average photodissociation cross-section (unit cm$^{-2}$) measured in laboratory adopting an H$_2$ (or D$_2$) lamp as UV source with flux $F_{\rm{lamp}}$. The flux lamp can be scaled to the strength of the FUV field from Draine by $F_{\rm{lamp}} = \chi F_{\rm{Draine}}$. Equation~\ref{eq_phd_rate} becomes $R_i^{\rm{phd}} [\rm{s^{-1}}] = \overline{\sigma} \cdot 1.9921 \times 10^8$ when $\chi = 1$. 

\subsubsection{Chemisorption, physisorption and diffusion}
\label{sec_diffusion}
There are two potential energy regimes that keep chemical species stuck on the catalytic surface, namely, chemisorption ({\bf c}) and physisorption ({\bf p}). In the former case, short-range forces associated with the overlap of the wave functions promote strong bonds between the grain surface and the first ice layers above. In the latter case, large distance forces take place, and the attraction due to the van der Walls force promotes weak binding at upper ice layers. It is worth noting that the binding energies in the two cases are vastly different, as well. In computational modelling, the physisorption sites only become available after the chemisorption sites are filled.

In this model, atoms and molecules can diffuse by changing states (physisorption or chemisorption), as well as it can migrate from physisorption to chemisorption sites, and vice-versa. The probability to overcome an adsorption activation barrier for chemisorption is described by the tunnelling-corrected Arrhenius formula called
Bell's formula \citep{Bell1980,Bell1982}, namely, $Q_{Bell}$, which is given by:
\begin{equation}
    Q_{Bell}\left(a_i, E_i, m_i, T_g \right) = \frac{\beta \mathrm{exp}(-\alpha) - \alpha \mathrm{exp}(-\beta)}{\beta - \alpha}
\end{equation}
where $a_i$ is the width of the rectangular activation barrier of height $E_i$, $m_i$ the mass of the atoms or molecule, and $T_g$ the gas temperature, assumed here equal to the dust temperature ($T_d$). The parameters $\alpha$ and $\beta$ are defined as:
\begin{equation}
    \alpha = \frac{E_i}{k T_g},
\end{equation}
and
\begin{equation}
    \beta = \frac{4\pi a_i}{h} \sqrt{2m_iE_i}.
\end{equation}

Finally, the diffusion rate is given by:
\begin{equation}
R_i^{\rm{diff}} = \begin{cases}
 \nu_i^{\rm{osc}} Q_{\rm{Bell}}, &\text{if {\bf p}$\rightarrow${\bf p} or {\bf c}$\rightarrow${\bf c}: $\Delta E_{if} = 0$}\\
\nu_i^{\rm{osc}} Q_{Bell} \;\mathrm{exp} \left(\frac{-\Delta E_{ij}}{k_{\rm{B}} T_s}  \right) \bigg/ nb_{site}, &\text{if {\bf p/c}$\rightarrow${\bf c/p}: $\Delta E_{if} \ne 0$}
\label{eq_diff}
\end{cases}
\end{equation}
where $nb_{site}$ is the number of adsorption sites per ice monolayer equal to $4\pi N_{\mathrm{surf}} a^2$ and $\Delta E_{if}$ is the energy difference between the sites $i$ and $f$. $N_{\rm{surf}}$ is the surface density of adsorption sites, i.e. $1.5 \times 10^{15}$~cm$^{-2}$, also known as the Langmuir number, and $a^2$ is the grain radius squared.

As the chemical species are migrating either over the catalytic surface, they can react via the Langmuir-Hinshelwood mechanism, following the rate:
\begin{equation}
    k_{ij} = \kappa_{ij} \left( \frac{R_i^{\rm{diff}} + R_j^{\rm{diff}}}{n_d} \right)
\end{equation}
where $\kappa_{ij}$ is the reaction probability to occur after the species $i$ and $j$ has diffused and it is formally defined as:
\begin{equation}
    \kappa_{ij} = \frac{Q_{\mathrm{Bell}}}{Q_{\mathrm{Bell}} + w_i^{\mathrm{diff}} + w_j^{\mathrm{diff}}},
\end{equation}
in which $w_{i,j}^{\mathrm{diff}} = R_{i,j}^{\mathrm{diff}}/\nu_{0,i,j}$. The frequency $\nu$ is defined by the equation of a rectangular barrier given by: 
\begin{equation}
    \nu_{0,i,j} = \sqrt{\frac{2N_{\mathrm{surf}}E_{i,j}}{\pi^2 m_{i,j}}},
\end{equation}
where $E_i$ is the height of the barrier, and $N_{\mathrm{surf}}$ is the surface site density, which is equal to 1.5$\times$10$^{15}$ sites cm$^{-2}$. In the cases of barrierless reactions, and {\it diffusion limited} conditions, i.e. when only thermal hopping or diffusion tunnelling dominate, $\kappa_{ij} \rightarrow 1$. The energy barrier of H to migrate from one physisorbed site is not negligible, but instead, it ranges from 256~K \citep{Kuwahata2015} to 341~K \citep{Congiu2014}, which allows H recombine locally when radicals are abundant because of barrieless reactions ($\kappa_{ij} = 1$) or after diffusion since the diffusion energy is around 30$-$50\% of the desorption energy \citep[e.g.,][]{Ruffle2000, Garrod2008, Thi2020_feb}. The amount of reactions per second (rate) is directly proportional to the number species available of the reactants ($n_i$, $n_j$):
\begin{equation}
    \frac{dn(ij)}{dt} = k_{ij}n_{ij}
    \label{rate_eq}
\end{equation}

On the other hand, if there is a {\it surface competition} between the two diffusion mechanisms, $\kappa_{ij} \ne 1$. For instance, in a surface competition scenario of H$_2$, $\kappa_{ij} \simeq 1/3$. Surface competition occurs when the surface reaction process competes with diffusion and desorption as defined by \citet{Smith2008}, \citet{Garrod2011} and \citet{Cuppen2017}.

The gas-phase atom or radical recombination with an adsorbed species from the grain surface $-$ Eley-Rideal reactions $-$ are not considered in this paper. Additionally, the sticking coefficient of all molecules in the gas phase is deliberately set to zero to certify that molecules desorbing from the ice do not adsorb again. In this approach, is reasonable to ignore those processes to a better agreement with the experimental conditions.

\subsubsection{Desorption}
\label{sec_desorption}
The desorption mechanism is characterized by the release to the gas phase of a chemical species from the catalytic surface when its internal energy is greater than the binding energy with the sticking surface. In this model, three desorption pathways are considered:

{\it Thermal desorption}.
The thermal desorption rates of chemical species in physisorption or chemisorption regimes are given by:
\begin{equation}
R_i^{\rm{des,th}} = \begin{cases}
 \nu_i^{\rm{osc}} \mathrm{exp}\left(\frac{-E_i^{\rm{des}}}{k_{\rm{B}} T_s}  \right), &\text{for physisorption regime}\\
\nu_i^{\rm{osc}} Q_{Bell} \;\mathrm{exp}\left(\frac{-E_i^{\rm{des}}}{k_{\rm{B}} T_s}  \right), &\text{for chemisorption regime}
\end{cases}
\end{equation}
where $\nu_i^{\rm{osc}}$ is the vibrational frequency of the species in the surface potential well of ice species {\it i}, $T_s$ is the surface temperature, and $-E_i^{\rm{des}}$ is the desorption energy, defined as the sum of the binding energy and the activation energy, i.e. $E_i^{\rm{des}} = E_i^{\rm{b}} + E_i^{\rm{des,act}}$. In the physisorption regime, $E_i^{\rm{des}} = E_i^{\rm{b}}$, whereas the chemisorption bonds require activation energy $E_i^{\rm{des,act}}$ to be broken. The number of active surface places in the ice per volume subject to the thermal desorption is given by:
\begin{equation}
    n_{\rm{act}} = 4\pi \langle a^2 \rangle n_d N_{\rm{surf}}N_{\rm{Lay}}
    \label{nlay}
\end{equation}
where $\langle a^2 \rangle$ is the squared mean grain size value, $n_d$ is the dust grain numerical density, $N_{\mathrm{surf}}$ is the surface density of adsorption sites in one monolayer of ice, also known as Langmuir number, and $N_{\rm{Lay}}$ is the number of surface layers considered as active. In the literature, the values attributed to $N_{\rm{Lay}}$ are 2 \citep[][]{Aikawa1996}, 4 \citep[][]{Vasyunin2013}, and 6 \citep[][]{Oberg2009}. In this paper we assume $N_{\rm{Lay}} = 6$ to be consistent with the experiments performed by \citet{Oberg2009}.

{\it Photodesorption}.
The internal energy of a chemical species on the grain surface can be increased by the absorption of a UV photon, and the photodesorption of species {\it i} is given by:
\begin{equation}
    R_i^{\rm{des,ph}} = \pi \langle a^2 \rangle \frac{n_d}{n_{\rm{act}}} Y_i \; \chi F_{\rm{Draine}}
\end{equation}
where $Y_i$ is the photodesorption yield, $\chi$ is the Draine field parameter. Generally, $Y_i$ depends on the ionizing source and the composition and thickness of the ice \citep[e.g.,][]{Brown1984, Westley1995, Oberg2009yield, Fayolle2011, Dartois2018}. For pure CH$_3$OH ice, \citet{Oberg2009} found $Y_i = 10^{-3}$, whereas \citet{Bertin2016} has found $Y_i = 10^{-5}$. Both values were tested in this paper.

{\it Reactive desorption}. Due to the exothermicity of some surface reactions, the bond between the chemical species and the surface is broken, thus resulting in non-thermal desorption, experimentally confirmed by \citet{Dulieu2013}. This mechanism has been introduced into astrochemical models by \citet{Garrod2007} and the desorption probability ($P$) is quantified by the Rice-Ramsperger-Kessel (RRK) theory \citep[e.g.,][]{Holbrook1996}, and defined as:
\begin{equation}
    P = \left[1 - \frac{E_{\rm{b}}}{E_{\rm{reac}}} \right]^{s-1}
\end{equation}
where $E_{\rm{b}}$ is the binding energy of the product molecule, $E_{\rm{reac}}$ is the energy released in the formation and $s$ is the number of vibrational modes of the molecule/surface-bound system. Once $P$ is known, the fraction of reactive desorbed products is calculated by:
\begin{equation}
    f_G = \frac{e P}{1-eP}
\end{equation}
where $e$ is an empirical parameter that relates the frequencies of molecule-surface bonds and of the energy transferred to the grain surface. The ``classical'' value adopted in the literature is 1\% \citep[e.g.,][]{Aikawa2008,Vasyunin2013, Shingledecker2020} because it provides better agreement with chemical abundances derived from astronomical observations. It is worth mentioning that molecular dynamic calculations indicate that a value larger than 1\% is unlikely \citep{Kroes2006}, and a lower percentage cannot be rejected \citep{Kristensen2010}. 

Another formalism has been proposed by \citet{Minissale2016}, in which the fraction of reactive desorption depends on the substrate and the mass of the product. In this case, the fraction is given by:
\begin{equation}
    f_M = \rm{exp}\left(-\frac{E_{\rm{D}}}{\epsilon E_{\rm{reac}}/N} \right)
\end{equation}
where $N$ is the number of degrees of freedom of the product species and $\epsilon = [(M-m)/(M+m)]^2$ is the energy kept by the product of mass $m$ on a surface of mass $M$.

\subsection{Models set-up}
\label{models_setup}
To replicate the CH$_3$OH ice photodissociation as measured by \citet{Oberg2009}, seven models considering different surface chemistry parameters described in Section~\ref{surfchem} were adopted. Additionally, since the silicate core is not simulated in this work, the surface reactions are limited to the physisorption regime. A summary of each model is shown in Table~\ref{models}. Thermal diffusion is considered in all models. Models~1 and 2 represent the scenarios with surface competition and diffusion/reaction tunnelling while adopting the approaches from \citet{Garrod2007} and \citet{Minissale2016}, respectively. In Models~3$-$4, the surface competition is switched off, but diffusion/reaction tunnelling to overcome the activation energy barrier in the two-body reaction is kept. This allows assessing the impact of surface competition in Models 1 and 2. Similarly, in Models~5$-$6 we enable surface competition and switch off tunnelling for diffusion and reaction. In this way, the effect of diffusion/reaction tunnelling in Models~1 and 2 are evaluated. Finally, Model~7 represents a scenario considering only thermal diffusion and reaction desorption according to the approach of \citet{Garrod2007}.

\begin{table}
\caption{\label{t7} Surface chemistry parameters used in the set-up of seven models. The present and absent mechanisms in each model are given by the green check markers and red crosses, respectively.}
\renewcommand{\arraystretch}{1.1}
\scalebox{0.95}{
\centering
\begin{tabular}{ccccc}
\hline\hline
Model & Surface & Diff./react. & Thermal& Reactive\\
& competition & tunnelling  & diffusion& desorption$^a$\\
\hline
\#1 & \textcolor{dgreen}\faCheckCircle & \textcolor{dgreen}\faCheckCircle & \textcolor{dgreen}\faCheckCircle & G07\\
\#2 & \textcolor{dgreen}\faCheckCircle & \textcolor{dgreen}\faCheckCircle & \textcolor{dgreen}\faCheckCircle & M16\\
\#3 & \textcolor{red}\faTimesCircleO & \textcolor{dgreen}\faCheckCircle & \textcolor{dgreen}\faCheckCircle & G07\\
\#4 & \textcolor{red}\faTimesCircleO & \textcolor{dgreen}\faCheckCircle & \textcolor{dgreen}\faCheckCircle & M16\\
\#5 & \textcolor{dgreen}\faCheckCircle & \textcolor{red}\faTimesCircleO & \textcolor{dgreen}\faCheckCircle & G07\\
\#6 & \textcolor{dgreen}\faCheckCircle & \textcolor{red}\faTimesCircleO & \textcolor{dgreen}\faCheckCircle & M16\\
\#7 & \textcolor{red}\faTimesCircleO & \textcolor{red}\faTimesCircleO & \textcolor{dgreen}\faCheckCircle & G07\\
\hline
\end{tabular}}
\tablefoot{$^a$G07: reactive desorption expression from \citet{Garrod2007} is adopted. M16: reactive desorption expression from \citet{Minissale2016} is used.}
\label{models}

\end{table}

Each model listed in Table~\ref{models} was performed assuming temperatures of 20, 30, 50 and 70~K to mimic the experiments carried out by \citet{Oberg2009}. The experimental photo-dissociation cross-sections of methanol ice calculated at these four temperatures are $2.6\pm0.9 \times 10^{-18} \; \rm{cm^{-2}}$ at 20~K, $2.4\pm0.8 \times 10^{-18} \; \rm{cm^{-2}}$ at 30~K, $3.3\pm1.1 \times 10^{-18} \; \rm{cm^{-2}}$ at 50~K and $3.9\pm1.3 \times 10^{-18} \; \rm{cm^{-2}}$ at 70~K. However, because of the efficient methanol ice reformation, these values are underestimated \citep[][]{Oberg2009}. \citet{Roncero2018} also found a similar trend in the cross-section of methanol reaction in the gas phase with OH. They observed that reactive cross-section has a minimum value when increasing the temperature, and then it starts increasing again. In their case, they interpret this trend as the formation of a CH$_3$OH-OH complex that only disappears via tunnelling toward the products.

To check the $\sigma_{\rm{phd}}$ that provides the best agreement with the experiments, the interval between the lower and upper bounds of photodissociation cross-sections calculated by \citet{Oberg2009} are used to create a grid of unshielded photorates ($\alpha$) for different BRs listed in Table~\ref{table_brs}. Specifically, the lower and upper bounds interval of $\sigma_{\rm{phd}}$ are evenly spaced, and subsequently, the rates associated with each BRs (Table~\ref{table_brs}) are used in the simulations. A summary of the physical parameters adopted in the model set-up is given in Table~\ref{sc_params}. It also includes the size of the dust grain ($a$) that acts as the substrate in the experiments and the dust-to-gas ratio ($\delta$), that is used in the simulations.
\begin{table}
\caption{\label{sc_params}Physical parameters used in the chemical simulation of CH$_3$OH ice photolysis}
\renewcommand{\arraystretch}{1.3}
\scalebox{0.95}{
\centering
\begin{tabular}{llll}
\hline\hline
Parameter & Symbol & Values & Units\\
\hline
Gas density (He) & $n$ & 10$^8$ & cm$^{-3}$\\
Temperature & $T_{\mathrm{dust}}$ = $T_{\mathrm{gas}}$ & 20, 30, 50, 70 & K\\
Pressure & $P_g$ & $10^{-10}$ & mbar\\
Thickness & $d$ & 20 & ML\\
Extinction & $A_{\rm{V}}$ & 0 & mag\\
Strength of ISM UV & $\chi$ & 5.4 $\times$ 10$^4$ & \\
Grain radius & $a$ & 0.1 & $\mu$m\\
Dust-to-gas mass ratio & $\delta$ & 5.55 & \\
\hline
\end{tabular}
}
\end{table}

\section{Results}
\label{results}

\subsection{Grid of photodissociation curves}
\begin{figure*}
   \centering
   \includegraphics[width=\hsize]{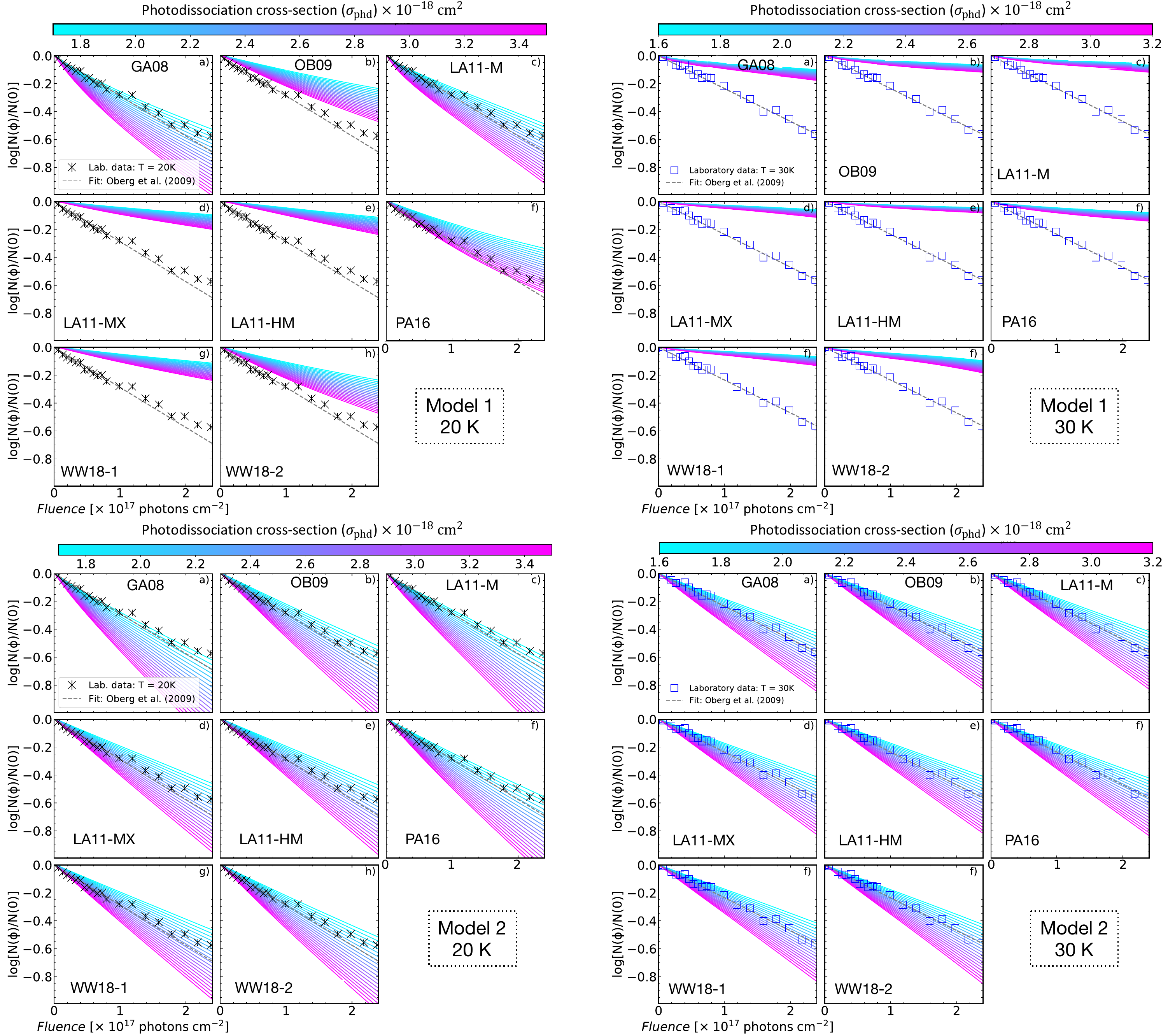}
      \caption{Grid of methanol ice photodissociation in Models 1 and 2 at 20~K (left panel), and at 30~K (right panel). The models assume different photodissociation cross-sections and BRs (see Table~\ref{table_brs}). These models assume surface competition, diffusion and reaction tunnelling, and thermal diffusion. In addition, Models 1 and 2 adopt reactive desorption formalism from \citet{Garrod2007} and \citet{Minissale2016}, respectively.}
         \label{M1}
   \end{figure*}

A grid of 1,120 photodissociation curves was simulated in order to find a good match with the experimental data. A series of photodissociation curves simulated with \texttt{ProDiMo} by adopting different BRs in Models~1 and 2 at 20~K are shown in the left panels of Figure~\ref{M1}, whereas the right panels shows the results when simulations are performed with temperature equal to 30~K. These curves show that the variation in the BRs and in the surface chemistry parameters leads to different results that may or may not fit the experimental data.

Despite the intrinsic differences among the models, we note that the reactive desorption from \citet{Minissale2016} promotes more destruction of the CH$_3$OH ice than the mechanism from \citet{Garrod2007} at all four temperatures. Nevertheless, it is the only scenario in which the fits match the experiments at 30~K. The top right panel in Figure~\ref{M1} shows the methanol ice photo-dissociation is rather underestimated at 30~K when the reactive desorption from \citet{Garrod2007} is adopted. On the other hand, the bottom right panels in Figure~\ref{M1} show that the models match the experimental data if reactive desorption from \citet{Minissale2016} is considered. We observed the repetition of this trend for the four temperatures addressed in this paper. An analysis of the chemical reactions that lead to these results are shown in Section~\ref{reactions_network}.

Appendix~\ref{Ap_all_models} shows the gallery of fits for Models~3$-$7 at 20~K and 30~K, as well as for Models~1$-$7 at 50~K and 70~K, which are omitted in this subsection. To decide which case best fit the experiments from all simulations, we used the approach described in Section~\ref{BR_selec}.

\subsection{BR selection}
\label{BR_selec}
As most of the models can fit the experimental data in some scenarios, the best fit was selected based on two criteria: first, the Maximum Likelihood Estimator (MLE), calculated by 
\begin{equation}
    \mathrm{MLE} \equiv \mathrm{ln} p\left(y^{\mathrm{lab}}|y^{\mathrm{mod}} \right) = -\frac{1}{2} \sum_{i} \left[\left(\frac{y_i^{\mathrm{mod}} - y_i^{\mathrm{lab}}}{\delta_i}\right)^2 + \mathrm{ln}(2\pi\delta_i^2)\right],
\end{equation}
where $y_{\rm{lab}}$ and $y_{\rm{mod}}$ are the normalized methanol ice column densities from the laboratory and the model, respectively, and $\delta_i$ is the absolute error of the measurements. The second criteria is that $\sigma_{\rm{phd}}^{\rm{model}} \geq \sigma_{\rm{phd}}^{\rm{lab}}$ as discussed by \citet{Oberg2009}.

\begin{figure}
   \centering
   \includegraphics[width=9cm]{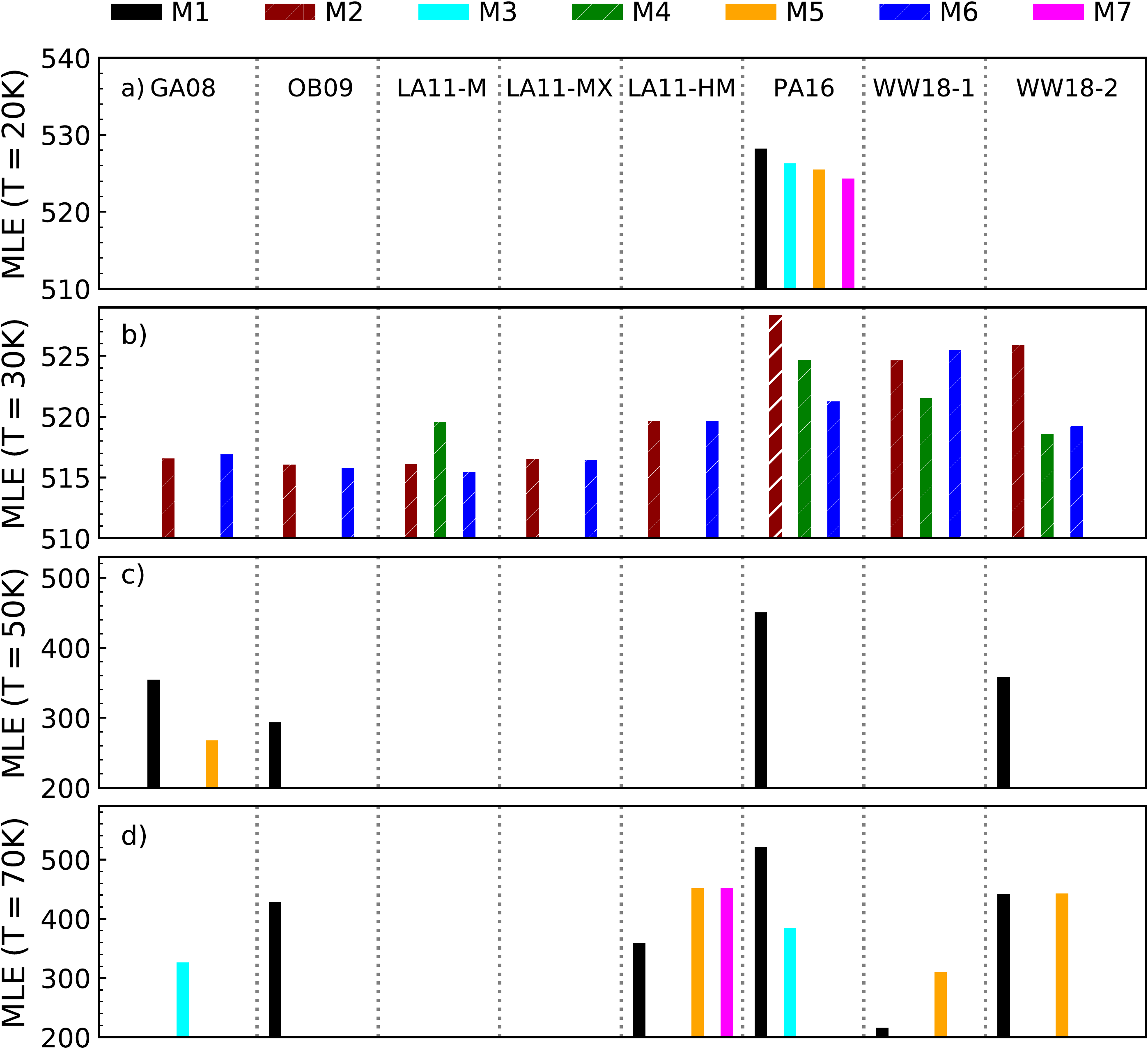}
      \caption{Accuracy of the CH$_3$OH ice photo-dissociation fit for Models 1$-$7. Higher bars indicate the best models, respectively. Non-visible bars indicate {\it MLE} values lower than the {\it y}-axis threshold. The hatched 
      and non-hatched bars indicate the models assuming the reactive desorption mechanism by \citet{Minissale2016} and \citet{Garrod2007}, respectively.}
         \label{models_fitness}
   \end{figure}

Figure~\ref{models_fitness} shows the $MLE$ values in a bar diagram for Models~1$-$7 at four temperatures and different BRs. The higher values of $MLE$ indicate the best model. The absent bars refer to models where the best fit is obtained with $\sigma_{\rm{phd}}^{\rm{model}} < \sigma_{\rm{phd}}^{\rm{lab}}$, which does not satisfy the second criteria. In all scenarios, the BR from \citet{Paardekooper2016} provides the best agreement with the experimental data. This suggests that methanol ice is mostly destroyed by the hydroxymethyl branch (CH$_2$OH\# + H\#), than by the methyl and methoxy pathways. Concerning the reactive desorption mechanism, models adopting the reactive desorption formalism from \citet{Garrod2007} provide better fits at $T =$ 20, 50, and 70~K, whereas at $T =$ 30~K, the formalism from \citet{Minissale2016} agrees better with the data. As seen in the bottom left panel in Figure~\ref{M1} and Figures~\ref{M3_30}, \ref{M5_30} and \ref{M7_30} (Appendix~\ref{Ap_all_models}), none of the fits adopting reactive desorption from \citet{Garrod2007} can match the experimental data at 30~K.

Among the models adopting the BR from \citet{Paardekooper2016}, Model~1 always provide the best fit at 20~K, 50~K and 70~K, whereas at 30~K, Model~2 results in better agreement with the experiments. In common, these two models account for surface competition and diffusion/reaction tunnelling for migration and reaction on the grain surface. However, at 20~K, the difference among the $MLE$ values in Models 1, 3, 5, and 7 is less than 5\%, thus suggesting that surface competition and diffusion/reaction tunnelling are not critical at this temperature. The same trend is observed at 30~K, where the difference between Models 2, 4 and 6 is still less than 5\%. Conversely, at 50~K only Model~1 provides a good match with the experiments, whereas at 70~K, Model~1 is better than Model~3 in 30\%. This indicates that surface competition and diffusion/reaction tunnelling are essential for models at high temperatures.

Although the photodissociation curves adopting the Paardekooper's BRs lead to better fits at the four temperatures, a few other cases also result in high fitness scores. For example, at 30~K, some models adopting the BRs from \citet{Weaver2018} can also fit well the experiments. At 50~K and 70~K, models considering other BRs also provide a relatively good match with the experiments. However, removing this degeneracy is not straightforward, and we decided to adopt the models providing the highest $MLE$ value for the subsequent analysis in this paper. 
   
\subsection{Photodissociation rates}
The methanol ice photodissociation 
rates were calculated using  Equation~\ref{eq_phd_rate}, which relates the integrated UV radiation field with the average photo-dissociation cross-section. Based on the results shown in Figure~\ref{models_fitness}, the photodissociation cross-sections based on the models are $\overline{\sigma}_{20~K}=3.0 \times 10^{-18} \; \rm{cm^{-2}}$, $\overline{\sigma}_{30~K}=2.5 \times 10^{-18} \; \rm{cm^{-2}}$, $\overline{\sigma}_{50~K}=3.8 \times 10^{-18} \; \rm{cm^{-2}}$, and $\overline{\sigma}_{70~K}=4.4 \times 10^{-18} \; \rm{cm^{-2}}$. Despite these values are about 10$-$15\% higher than the experimental $\sigma_{\rm{pd}}$, they indicate reasonable agreement and the small $\sigma_{\rm{pd}}$ at 30~K is also obtained with the simulations. Table~\ref{table_rates} shows the photo-rates of CH$_3$OH ice with \citet{Paardekooper2016} BR and at four temperatures. Again, at 30~K the photo-rate of the path CH$_3$OH\# + h$\nu$ $\rightarrow$ CH$_3$\# + OH\# is one order of magnitude lower than the cases at 20, 50 and 70K.

As pointed out by \citet{Oberg2009}, a likely reason for the 10$-$15\% higher photodissociation cross-section in the simulations is that  methanol can still be reformed locally because of the so-called ``cage effect'' \citep{Franck_Rabinowitsch_1934}. It states that the diffusion of the radicals formed during photolysis is inhibited because of the surrounding molecules. Consequently, fast recombination of the radicals occurs at the site of the photodissociation. In the case of pure CH$_3$OH, CO and H$_2$O ices, \citet{Laffon2010} show that back reactions limit the destruction effect.

\begin{table*}
\caption{\label{table_rates}Methanol ice photo-dissociation pathways and unshielded photo-rates ($\alpha = R_i^{\rm{phd}}$) taken from the best models.}
\renewcommand{\arraystretch}{1.5}
\centering
\begin{tabular}{cccccccccc}
\hline\hline
Reaction & & & & & BR (\%) & $\alpha_{\rm{20K}}$ ($\rm{s^{-1}}$) & $\alpha_{\rm{30K}}$ ($\rm{s^{-1}}$) & $\alpha_{\rm{50K}}$ ($\rm{s^{-1}}$) & $\alpha_{\rm{70K}}$ ($\rm{s^{-1}}$)\\
\hline
CH$_3$OH\# + h$\nu$ & $\rightarrow$ & $\rm{CH_3\#}$ &+& $\rm{OH\#}$     & 20 & 1.2 $\times$ 10$^{-10}$ & 1.0 $\times$ 10$^{-11}$ & 1.5 $\times$ 10$^{-10}$ & 1.7 $\times$ 10$^{-10}$\\
                    & $\rightarrow$ & $\rm{OCH_3\#}$ &+& $\rm{H\#}$     & 13 & 7.8 $\times$ 10$^{-11}$ & 6.6 $\times$ 10$^{-11}$ & 9.8 $\times$ 10$^{-11}$ & 1.1 $\times$ 10$^{-10}$\\
                    & $\rightarrow$ & $\rm{CH_2OH\#}$ &+& $\rm{H\#}$    & 67 & 4.0 $\times$ 10$^{-10}$ & 3.4 $\times$ 10$^{-10}$ & 5.1 $\times$ 10$^{-10}$ & 5.9 $\times$ 10$^{-10}$\\

\hline
\end{tabular}
\end{table*}

\subsection{Methanol ice destruction}
\begin{figure*}
   \centering
   \includegraphics[width=\hsize]{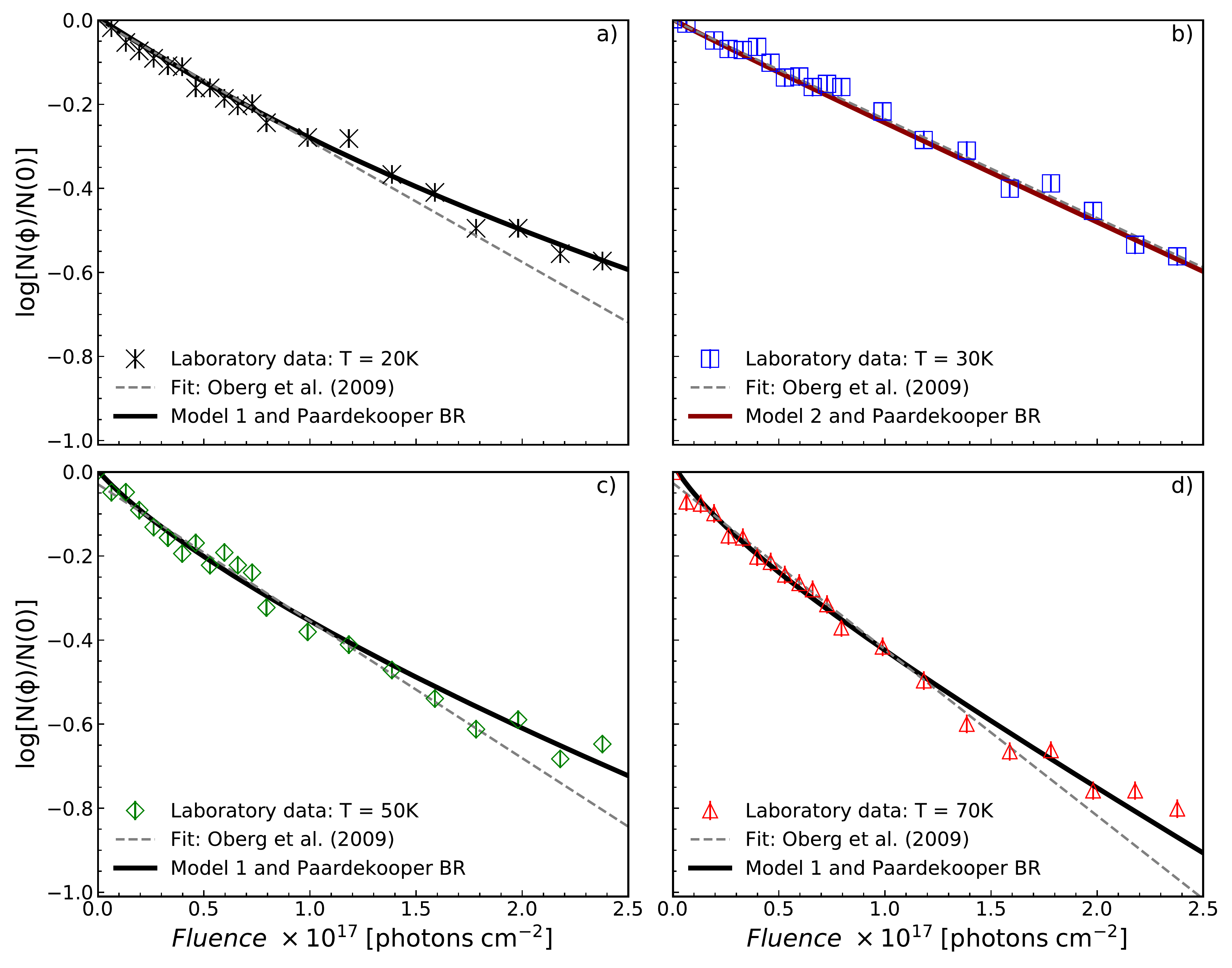}
      \caption{Natural logarithm of the normalized CH$_3$OH ice column density as a function of the UV fluence at 20, 30, 50, 70~K, in panels {\it a}-{\it d}, respectively. The symbols show the laboratory data from \citet{Oberg2009} and the lines indicate the models providing the best matching with the experiments.}
         \label{fits}
   \end{figure*}

Figure~\ref{fits} shows the methanol ice destruction simulated with the ProDiMo code compared with the laboratory measurements at four temperatures during the first half-hour of the experiment. The photodissociation rates used in the simulations are shown in Table~\ref{table_rates}. In this figure, the grey-dashed lines are the fits performed by \citet{Oberg2009} to calculate the $\sigma_{\rm{phd}}$ assuming an exponential decrease of the initial ice column density, given by:
\begin{equation}
    N(\phi) = N(0)\rm{exp}\left(-\phi \sigma_{\rm{phd}}\right)
    \label{column_density}
\end{equation}
where $N(\phi)$ and $N(0)$ are the ice column densities at fluence $\phi$ ($flux \; \times \; time$) and at the beginning of the experiment, respectively.

Equation~\ref{column_density} provides good match with the experiments at 20, 50 and 70~K for fluences below $1\times10^{17}$~photons~cm$^{-2}$. However, at fluences higher than $2\times10^{17}$~photons~cm$^{-2}$, such an exponential function overestimates the ice destruction, which can be indicating methanol ice reformation. A combination of exponential and linear analytical functions was used by \citet{Oberg2009} to fit the ice destruction. They are respectively associated with bulk photolysis and photodesorption. In contrast, the ice loss at 30~K is fully fitted by Equation~\ref{column_density}, thus arising the question of the methanol photo-dissociation at this temperature following the same process at other temperatures.

In order to compare the ProDiMo results with the experiments, the normalized ice column density was calculated by the following equation:
\begin{equation}
    \frac{N(\phi)}{N(0)} = \frac{n(\phi)}{n(0)} \cdot \frac{d(\phi)}{d(0)}
\end{equation} 
where $n(\phi)$ and $n(0)$ are the respective numeric densities of methanol ice at fluence $\phi$ and at the beginning of the experiment. The same for the ice thickness $d(\phi)$ and $d(0)$, given by Equation~\ref{nlay}. 

The models simulated with ProDiMo can reproduce the experimental measurements at the entire fluence range and four temperatures, as shown in Figure~\ref{fits}. In all cases, model M1 resulted in better fits at 20, 50~K and 70~K, whereas model M2 fits the data at 30~K. Models M1 and M2 include surface competition, diffusion/reaction tunnelling and thermal diffusion, which indicates that such surface chemistry processes may be occurring within the ice during the photolysis. The difference between them is the reactive desorption formalism as shown in Table~\ref{t7}. Additionally, these models assume BRs from \citet{Paardekooper2016}. Figure~\ref{fits} also shows a good agreement between the simulations and the trend of the normalized column density values at all four temperatures, which was not possible by using Equation~\ref{column_density}. This highlights that methanol ice is indeed reformed more efficiently at 20, 50 and 70~K for fluences above 1$\times$10$^{17}$ photons cm$^{-2}$. On the other hand, an additional methanol destruction route occurs in the experiment at 30~K. Below, we comment on the reason that Models M2$-$M7 do not fit the data.

\underline{20K:} At this temperature, the models assuming reactive desorption via the \citet{Minissale2016} formalism, i.e., M2, M4 and M6 always intensify the photo-destruction of the methanol ice and cannot fit the experimental data. In this case, we comment on models M3, M5 and M7 that adopt the same reactive desorption efficiency as model M1. When surface competition is switched off (M3), the $MLE$ is only 2\% lower than in model M1, which indicates that diffusion may still occur in the ice. The lower agreement in model M5, where no tunnelling is considered, shows that at low temperatures, this effect is important. Finally, model M7, which does not account for surface competition and tunnelling results in the least agreement with the experiments.

\underline{30K:} Differently from the cases at 20, 50 and 70~K, the data at 30~K is only fitted by models M2, M4 and M6, where M2 produces the best fit. The surface competition is switched off in model M4, and methanol ice is slightly less destroyed than in model M2. We found the same result in model M6, where no diffusion and reaction tunnelling is occurring.

\underline{50K:} Similarly to the cases of 20~K, models M2, M4 and M6 enhance the photo-destruction of methanol ice. On the other hand, models M3, M5 and M7, cannot induce enough photolysis in the ice. Model M3 does not consider surface competition, and methanol ice is more efficiently reformed than expected from the experiments. The effect of tunnelling diffusion and reaction is checked in model M5. We note that despite methanol ice being more destroyed than in model M3, the photodissociation cross-section to fit the data is higher than the upper limit estimated by \citet{Oberg2009}, which does not satisfy our goodness criteria defined in Section~\ref{BR_selec}. Finally, model M7 also shows that methanol is more reformed than expected.

\underline{70K:} Again, models M2, M4 and M6 promote efficient photo-destruction of the methanol ice beyond what is expected at 70~K. When surface competition is disabled in model M3, methanol ice is efficiently reformed, and cannot fit the experiments. When tunnelling is switched off in model M5, there is more destruction of the methanol ice. A similar result is found with model M7.

\subsection{Growth curves of the photoproducts}
\label{g_curves}
\begin{figure*}
   \centering
   \includegraphics[width=\hsize]{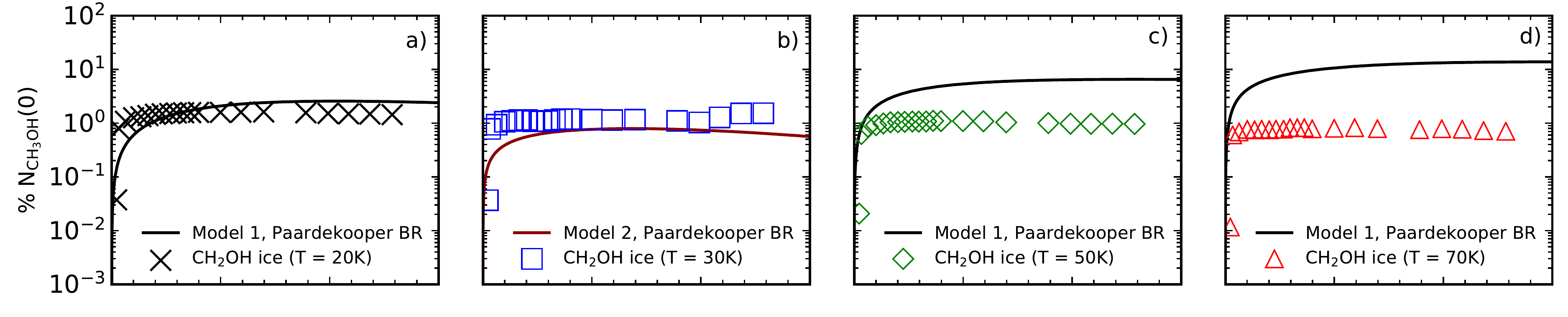}
   \includegraphics[width=\hsize]{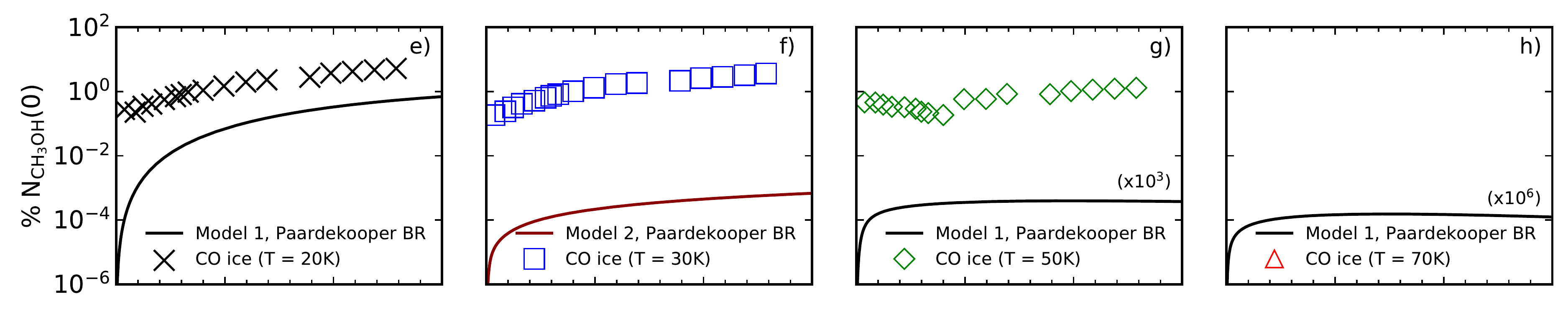}
   \includegraphics[width=\hsize]{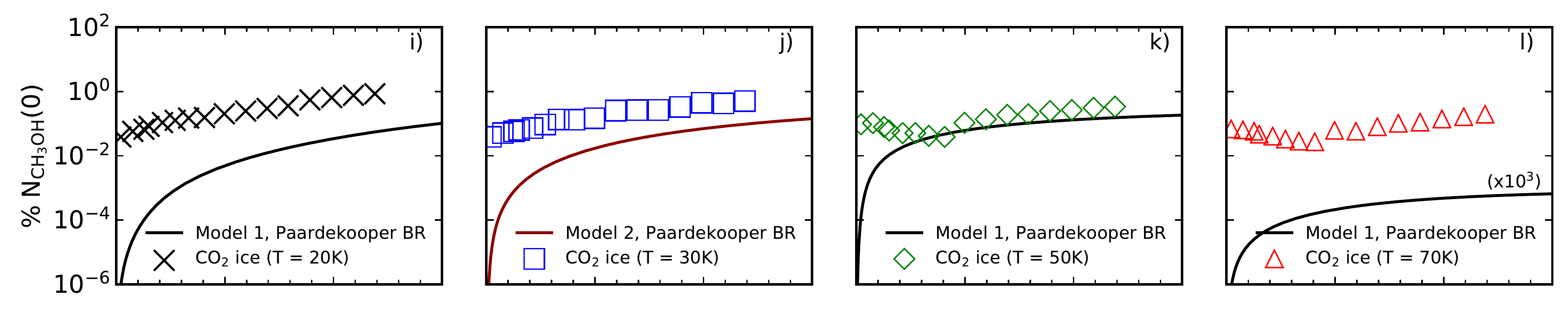}
   \includegraphics[width=\hsize]{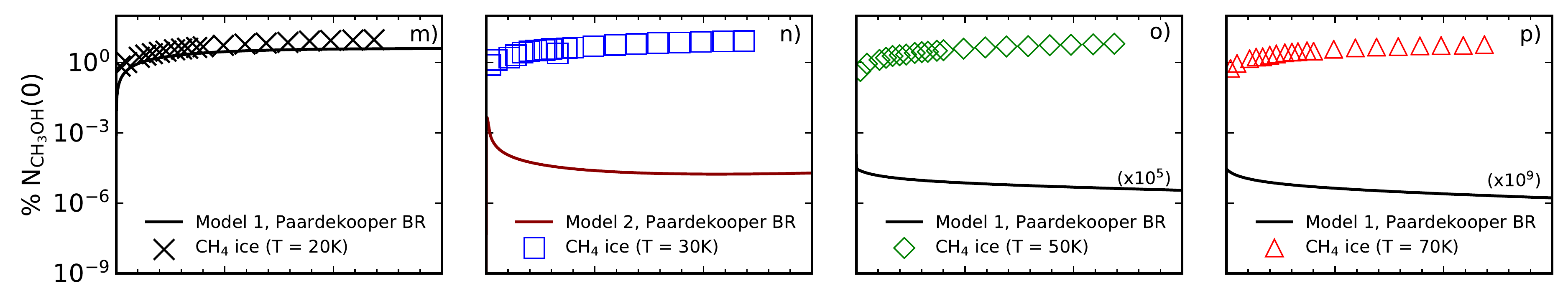}
   \includegraphics[width=\hsize]{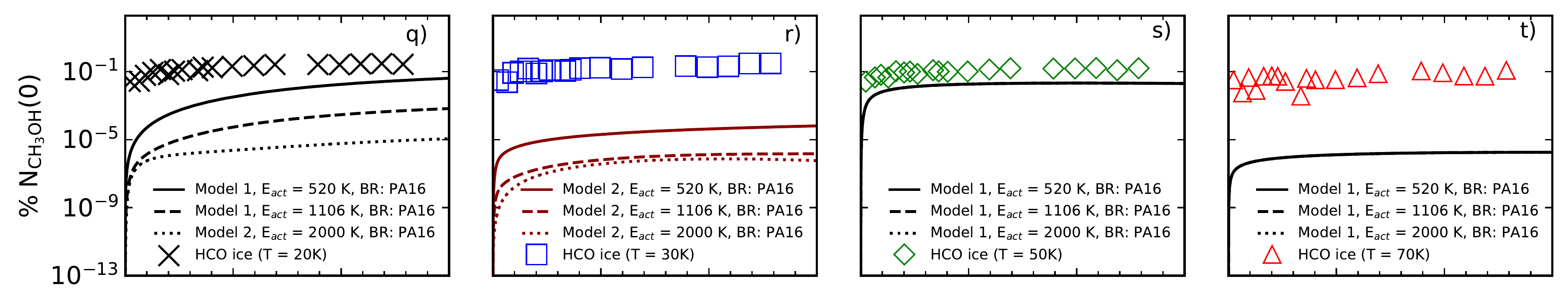}
   \includegraphics[width=\hsize]{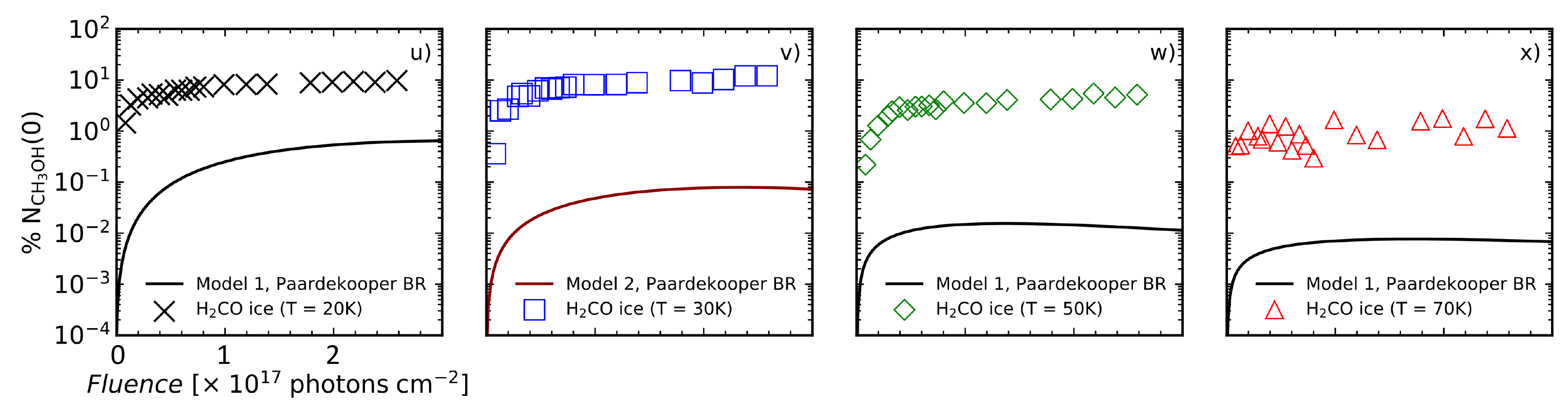}
      \caption{Formation curves of selected daughter species (CH$_2$OH, CO, CO$_2$, CH$_4$, HCO and H$_2$CO) due to CH$_3$OH photolysis over fluence with respect to the initial methanol ice column density at four temperatures. The symbols are the experimental measurements and the solid lines are the results of ProDiMo simulations of the best photodissociation models (see Figure~\ref{fits}).}
         \label{growth}
   \end{figure*}

The photolysis of ices leads to the formation of several daughter species as evidenced by {\it in situ} analysis of laboratory experiments \citep[see review by][]{Oberg2016}. The growth curves of some chemical species calculated with ProDiMo are compared with the experimental measurements in Figure~\ref{growth} at four temperatures. The growing curves of products formed just after irradiation (first generation) are different from products formed after radical migration and reaction (second or more generations). In addition, molecules formed from dissociation of first or second-generation photo-products show a yet different formation curve. Temperature is another physical parameter able to modify the growth curve because of the diffusion of the chemical species synthesized during the ice photolysis. 

Figure~\ref{growth}{\it a}-{\it d} shows the cases of CH$_2$OH ice. This is a first-generation radical formed efficiently by the methanol ice photolysis. The experimental data indicate that the hydroxymethyl ice abundance with respect to the initial methanol column density is around 1\% at 20, 30, 50~K and 70~K. A good match is seen with the ProDiMo models at 20 and 30~K, whereas a difference of around one order of magnitude is observed at 50 and 70~K. As shown in reaction R1, hydroxymethyl is also formed via two-body reactions, and the mismatch between measurement and modelling can be associated with inaccuracies in the rates involved in this reaction at high temperatures.

Figure~\ref{growth}{\it e}-{\it h} shows the cases of CO ice. The experimental abundance with respect to initial methanol column density is at around 1\% at 20, 30, and 50~K, whereas it is not detected at 70~K. Because the desorption temperature of pure CO ice has been calculated in around 20-25~K \citep{Collings2004, Acharyya2007}, its presence at 30 and 50~K is likely due to molecular migration into the porous ice matrix and thus, being desorbed at a higher temperature. The models were not able to reproduce the experiments at any temperature. However, the difference between the simulated growing curve and the experiment at 20~K is smaller than in the cases at 30 and 50~K. Such a higher difference at 30 and 50~K $-$ 3 and 7 order of magnitude, respectively $-$ can be explained by the lack of physical modelling with ProDiMo of the ice structure. On the other hand, the small difference at 20~K is likely due to inaccuracies in the reaction rates in the network, as in the case of hydroxymethyl. For example, during the methanol ice photo-dissociation, the CO ice is formed from the destruction of second-generation molecules:
\begin{subequations}
\makeatletter
\def\@currentlabel{R}
\makeatother
\renewcommand{\theequation}{R\arabic{equation}}
\setcounter{equation}{0}
\begin{align}
&\rm{CH_3OH\#} + h\nu \rightarrow \rm{CH_3O\#} + H\#\\
&\rm{CH_3O\#} + h\nu \rightarrow \rm{H_2CO}\# + H\#\\
&\rm{H_2CO\#} + h\nu \rightarrow \rm{CO}\# + H_2\#
\end{align}
\end{subequations}

The CO$_2$ ice formation is shown in Figure~\ref{growth}{\it i}-{\it l}. It is mostly formed via OH\# + CO\# $\longrightarrow$ CO$_2$\# + H\#, and O\# + HCO\# $\longrightarrow$ CO$_2$\# + H\# at 20 and 30~K. At 50 and 70~K, reaction OH\# + CO\# $\longrightarrow$ CO$_2$\# + H\# is the main formation pathway of carbon dioxide. These reactions depend on the CO and HCO ice formation (Reactions 1$-$3) and the OH photo-product. A relatively good fit is observed at 30 and 50~K. At 20~K, the abundance from the model is one order of magnitude lower than the experiments, whereas at 70~K, the abundance of CO$_2$ ice is seven orders of magnitude lower. Methane is also formed late because it depends on the photodissociation of two molecules as already described. In the models shown in Figure~\ref{growth}{\it m}-{\it p}, only the CH$_4$ formation at 20~K is well reproduced. At 30, 50 and 70~K the methane abundance is more than three orders of magnitude lower than measured in the experiments. Its formation indicates efficient diffusion of the radicals to the ice matrix porous, where they can react to form methane.

The formation pathways of HCO ice shown in Figures~\ref{growth}{\it q}-{\it t} change with the temperature. At 20 and 30~K, they are formed via two-body reactions of H\# + CO\# $\longrightarrow$ HCO\#. There are different activation energies in the literature for this reaction. For this reason, we checked the differences assuming $\Delta E$ = 520~K \citep{Fuchs2009}, 1106~K \citep{Rimola2014} and 2000~K \citep{Awad2005}. We note that Model M1 with a lower activation energy provides HCO\# abundances relatively close to the experimental results at 20~K, but the increase in the abundance is not significant at 30~K. Additionally, higher activation energies make the mismatch larger. At 50 and 70~K, the reaction H$_2$CO\# + OH\# $\longrightarrow$ HCO\# + H$_2$O\# dominates over any other reaction. A good match with the experiments is found only at 50~K. Since the reaction H\# + CO\# does not play any role, the growth curves are the same regardless of the activation energy adopted. H$_2$CO formation is shown in Figures~\ref{growth}{\it u}-{\it x}. In the models presented in this paper, it is formed via photolysis of the methoxy radical, CH$_3$O\# + $h\nu$ $\longrightarrow$ H$_2$CO\# + H\#. The photodissociation rate of this reaction is assumed $1 \times 10^{-9}$ s$^{-1}$ and can be overestimating the methoxy photodissociation.

The mismatch between the model and experiments indicates not only that further laboratory studies are necessary to accurately reproduce surface reactions, but also that improvements to the models are needed. For example, chemical models including more realistic energy redistribution when absorbing UV photons could provide useful insights into this framework. In the models shown in this paper, some differences are rather notable. At 20~K, the chemical abundances of some molecules and radicals in the models are lower measured from the experiments by one order of magnitude, except for the cases of CH$_2$OH, HCO and CH$_4$. At 30~K, except in the cases of CH$_2$OH and CO$_2$, the abundances can be 2$-$5 orders of magnitude lower than measured by \citet{Oberg2009}. At 50~K, the abundances of CO, CH$_4$ and H$_2$CO are underestimated by 2 to several orders of magnitude. At 70~K, the models do not reproduce the measurements in any case.

\section{Discussion}
\label{discussion}
\subsection{Chemical pathways at different fluences and temperatures}
\label{reactions_network}
The most efficient chemical pathways to form and destroy methanol ice can be addressed by analysing the curvature in the molecule formation plots 20, 50 and 70~K (see Figure~\ref{fits}) and the absence of this trend at 30~K. In this regard, the chemical reactions at fluences of 0.5, 1.5 and 2.4 $\times$ 10$^{17}$ photons cm$^{-2}$ are analysed.

For the four temperatures and selected fluences, methanol ice photodissociates following the pathways and branch ratios shown in Table~\ref{table_rates}. This indicates that bulk ice photolysis dominates over other destruction processes either at the beginning of the experiment or after a half-hour of constant irradiation in the case of thin ices (20 monolayers). We noted that the photodesorption mechanism is not significant at this range of fluence regardless the yield is $Y= 10^{-3}$ or $Y= 10^{-5}$. At 50~K and 70~K, methanol ice is further destroyed via a two-body surface reaction \citep{Esplugues2016}:

\begin{subequations}
\makeatletter
\def\@currentlabel{R}
\makeatother
\renewcommand{\theequation}{R\arabic{equation}}
\setcounter{equation}{3}
\begin{align}
&\rm{CH_3OH\#} + \rm{OH\#} \longrightarrow \rm{CH_2OH\#} + \rm{H_2O\#}
\end{align}
\end{subequations}
where the activation barrier of this reaction is equal to 1000~K and makes reaction R4 inefficient at low temperatures. This reaction leads to the formation of H$_2$O ice, which is one of the products observed from the experiments \citep{Oberg2009}. At 20~K and 30~K, water ice is formed via the two-body reaction between H\# + OH\#, and CH$_4$\# + OH\# $\longrightarrow$ CH$_3$\# + H$_2$O\#. CH$_4$\# readily forms via two-body reaction in the ice (CH$_3$\# + H\# $\longrightarrow$ CH$_4$\#). Those pathways could be an explanation for the H$_2$O ice formation in the experiments with methanol ice.

The models with ProDiMo show that a fraction of methanol ice is immediately reformed after photodissociation in each of the selected fluences. The three reformation reaction pathways are:
\begin{subequations}
\makeatletter
\def\@currentlabel{R}
\makeatother
\renewcommand{\theequation}{R\arabic{equation}}
\setcounter{equation}{4}
\begin{align}
&\rm{H\#} + \rm{CH_2OH\#} \longrightarrow \rm{CH_3OH\#}\\
&\rm{H\#} + \rm{CH_3O\#} \longrightarrow \rm{CH_3OH\#}\\
&\rm{OH\#} + \rm{CH_3\#} \longrightarrow \rm{CH_3OH\#}
\end{align}
\end{subequations}

\begin{figure}
   \centering
   \includegraphics[width=\hsize]{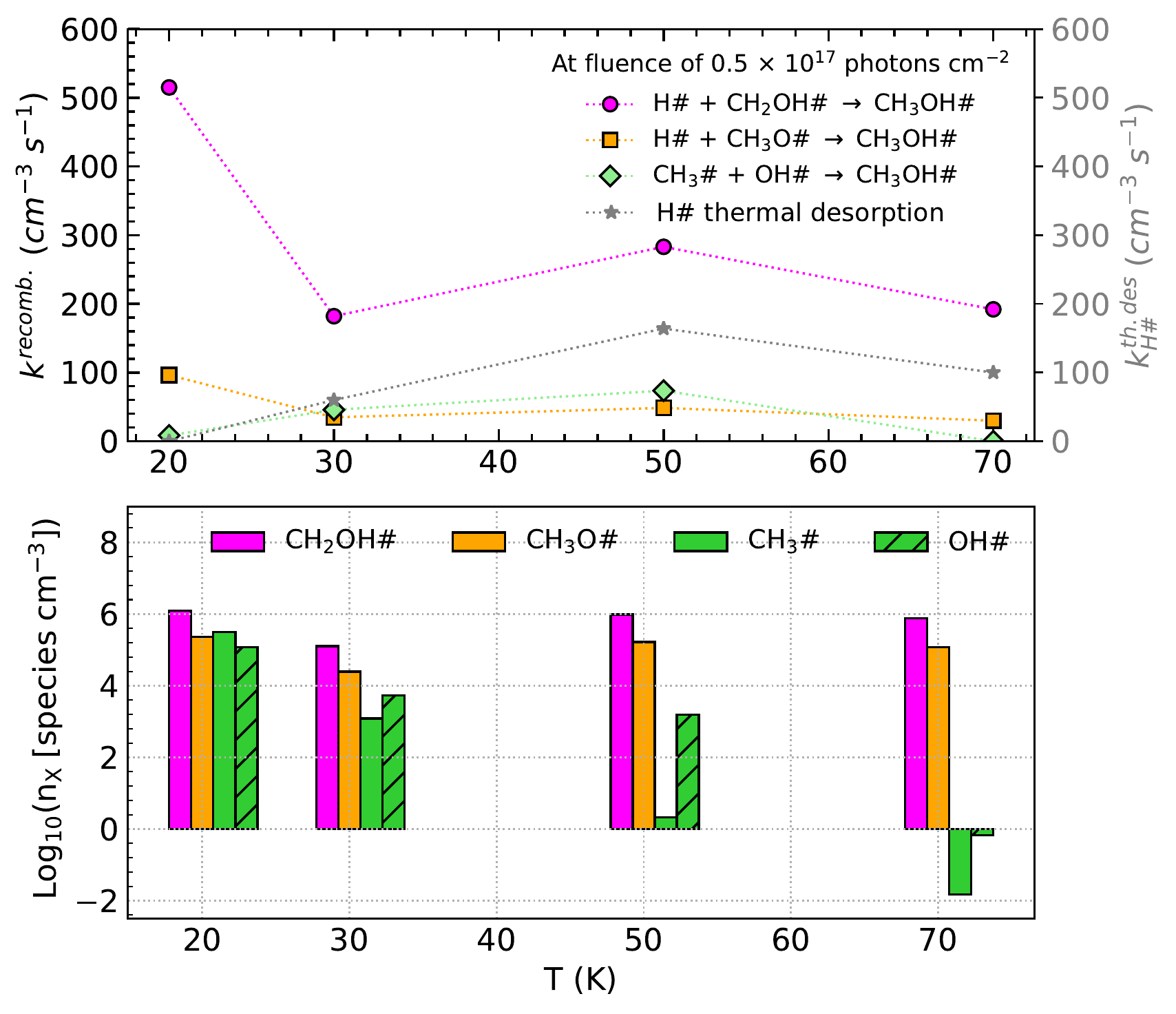}
      \caption{Top: Methanol ice recombination rate as a function of temperature occurring at a selected fluence. At all temperatures, reaction R5 (violet circles) occurs faster than Reactions R6 (orange squares) and R7 (green diamonds). Hydrogen thermal desorption is indicated by the grey star symbols and dotted lines. Bottom: Logarithm number density of the radicals ($X$) formed after CH$_3$OH ice photolysis.}
         \label{Hvstemp}
   \end{figure}

The recombination rates of Reactions R5$-$R7 as a function of temperature are shown in the top panel of Figure~\ref{Hvstemp}. The rate is defined as the product of the rate coefficient by the particle density of the reactants (see Equation~\ref{rate_eq}). The bottom panel of Figure~\ref{Hvstemp} shows the number density of radicals available to recombine. At all four temperatures, methanol ice recombines faster via Reaction R5 than via Reactions R6 and R7 due to the large amount of CH$_2$OH\#. The trend observed in Reaction R5 agrees with the number densities of CH$_2$OH\# formed after photolysis. Additionally, there is less H\# available for recombination due to hydrogen desorption as indicated by the grey symbols in the right y-axis in the top panel of Figure~\ref{Hvstemp}. The lower recombination rate at 30~K is because of the small methanol ice photolysis cross-section, and, consequently, the photo-rates at this temperature. As a result, fewer photoproducts are available in the ice to recombine. The rate decreasing trend as a function of temperature is also observed for reaction R6 (H\# + CH$_3$O\#). The lower rates are associated with the lower BRs for this reaction (13\%; see Table~\ref{table_rates}). The BRs also explain the lower rate values for reaction R7 (OH\# + CH$_3$\#), but a different behaviour is observed for the recombination reaction trend. At 20~K, the low methanol ice reformation via this reaction is due to the efficient reaction between H\# and CH$_3$\# to form CH$_4$\#. At 30 and 50~K methane is less abundant in our models, which increases the number of CH$_3$ and OH in the ice to reform CH$_3$OH\#. Despite the low number density of CH$_3$\#, OH\# is high enough to increase the recombination rate at 50~K. At 70~K, the CH$_3$\# desorbs from the ice which significantly decreases the methanol ice reformation via reaction R7. Additionally, at this temperature, the reaction between CH$_3$OH\# and OH\# takes place.

Despite the decreasing recombination rate with the increasing of temperature for reactions R5 and R6, Figure~\ref{Hvstemp} indicates that a fraction of H remains in the ice to recombine mostly with CH$_2$OH\# to reform CH$_3$OH\# at all four temperatures. The methane ice formation in the experiments shown in Section~\ref{g_curves} also gives a hint that H is present in the ice, and reacts with other species at all four temperatures (e.g., CH$_4$\# and HCO\#).

\begin{figure}
   \centering
   \includegraphics[width=\hsize]{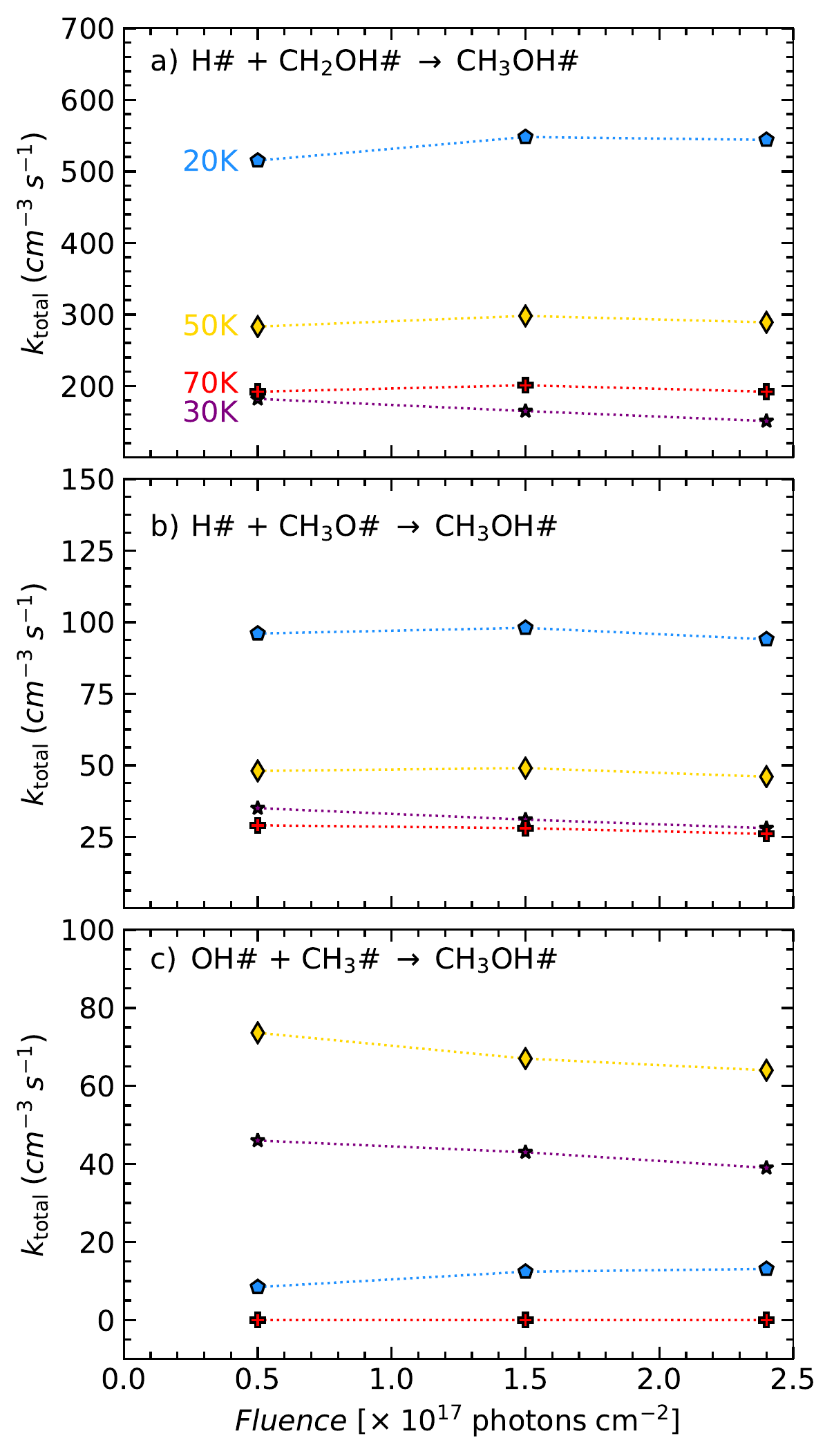}
      \caption{Methanol ice recombination rate as a function of temperature and fluence. Blue pentagons, purple stars, gold diamond, and red cross indicate the rates at temperatures of 20, 30, 50 and 70~K, respectively. Panels a, b and c show the recombination reactions R5-R7, respectively.}
         \label{Hreac}
   \end{figure}

The effect of the fluence in the recombination reactions R5$-$R7 is shown in Figure~\ref{Hreac}. One can note that the reformation reaction rates have little variation over the selected fluences. On the other hand, a significant variation is noted with the temperature. At 20~K, reactions R5 and R7 have a small increase with the fluence, whereas reaction R6 remains almost constant. At 30~K, all recombination rates decrease with the fluence, which consequently intensifies the methanol ice destruction. At 50~K, frozen methanol is mostly reformed via reaction R5 over the fluences, whereas the rates of reactions R6 and R7 are constant and decrease, respectively. Finally, at 70~K, the rates of reactions R5 and R6 are constant during the experiments, and virtually null for reaction R7.

Reactions R4$-$R7 are crucial for understanding the methanol ice destruction via photolysis, but they cannot explain the trend in the destruction curve in the experiment at 30~K for fluences above 1.5$\times$10$^{17}$ photons cm$^{-2}$. To further investigate this trend at 30~K and why the methanol ice recombination rate decreases over fluence (Figure~\ref{Hreac}), we analyse other chemical reactions involving the photo-products. This analysis shows that only at 30~K, the photo-products are further desorbed from the ice to the gas phase via reactive desorption by a factor of two compared to the experiments at 20~K, 50~K and 70~K. The reaction scheme at 30~K is given by:
\begin{subequations}
\makeatletter
\def\@currentlabel{R}
\makeatother
\renewcommand{\theequation}{R\arabic{equation}}
\setcounter{equation}{7}
\begin{align}
&\rm{H\#} + \rm{CH_2OH\#} \longrightarrow \rm{CH_3OH}\\
&\rm{H\#} + \rm{CH_3O\#} \longrightarrow \rm{CH_3OH}\\
&\rm{CH_3\#} + \rm{OH\#} \longrightarrow \rm{CH_3OH}
\end{align}
\end{subequations}
where the reactive desorption formalism from \citet{Minissale2016} is assumed in this case. Conversely, at 20~K, 50~K and 70~K, methanol ice recombination via two-body reactions (R5$-$R7) dominate over the reactive desorption by a factor of three, where reactive desorption from \citet{Garrod2007} is adopted. At 30~K, other species also leave the ice via reactive desorption at a lower rate compared to reactions R8$-$R10 and reduce the number of reactants to reform methanol in the ice, that are the cases of: 
\begin{subequations}
\makeatletter
\def\@currentlabel{R}
\makeatother
\renewcommand{\theequation}{R\arabic{equation}}
\setcounter{equation}{10}
\begin{align}
&\rm{H\#} + \rm{CH_3\#} \longrightarrow \rm{CH_4}\\
&\rm{H\#} + \rm{HCO\#} \longrightarrow \rm{H_2CO}\\
&\rm{H\#} + \rm{H_2CO\#} \longrightarrow \rm{CH_3O}
\end{align}
\end{subequations}

Nevertheless, it is unclear why reactive desorption from \citet{Minissale2016} is only required at 30~K. An inaccurate determination of the methanol column densities at this temperature could be an explanation for the lack of the curved profile at longer fluences. However, the right panels of Figure~\ref{M1}, and Figures~\ref{M3_30}-\ref{M7_30} in Appendix~\ref{Ap_all_models} show that only models adopting reactive desorption from \citet{Minissale2016} can fit the experimental data at 30~K, where no curvature in the destruction profile is observed.

\begin{figure}
   \centering
   \includegraphics[width=\hsize]{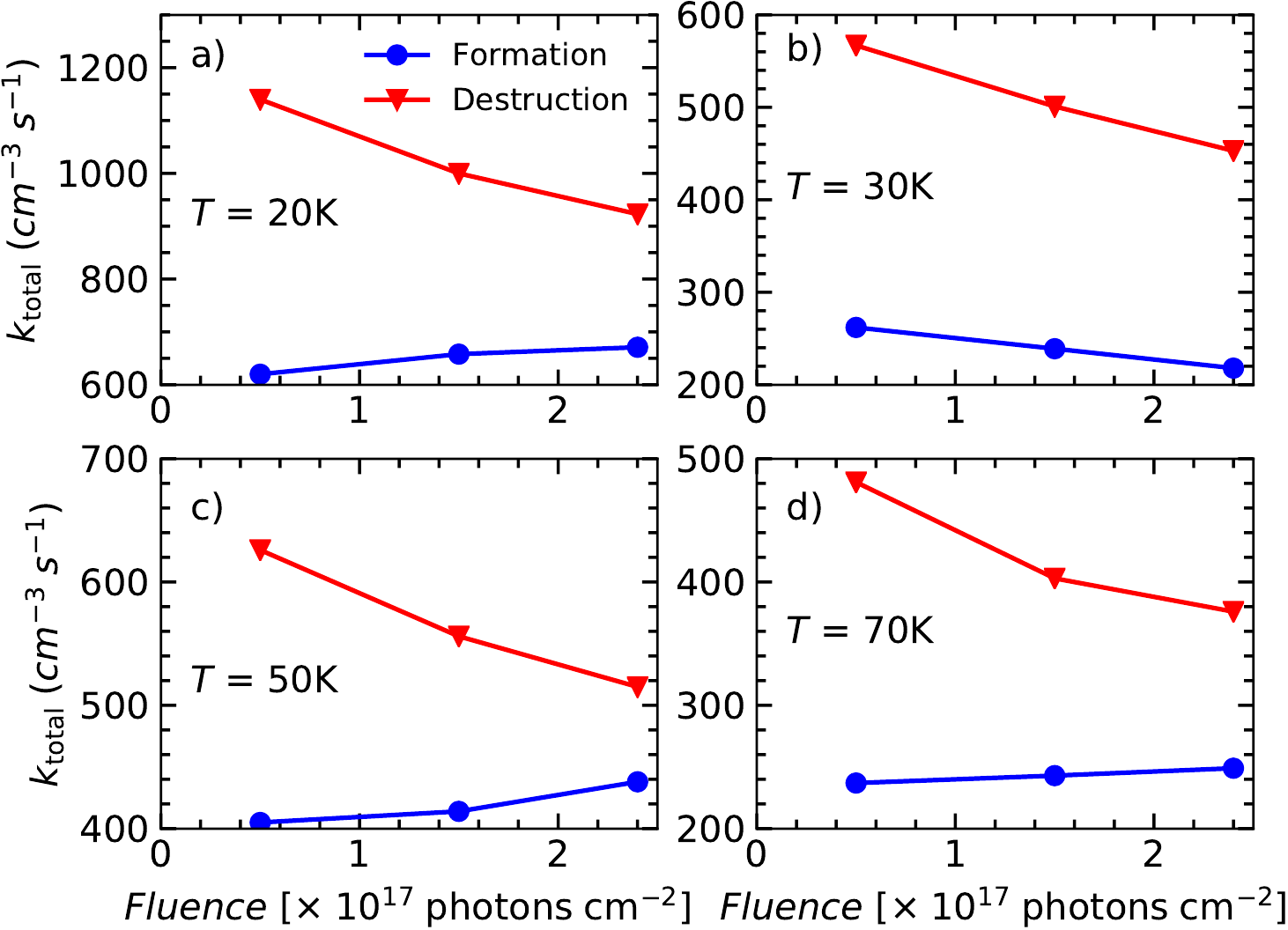}
      \caption{Total formation and destruction rates of CH$_3$OH ice at selected fluences. Except at 30~K, the formation rate (blue line/symbols) increases with fluence while the destruction rate (red line/symbols) decreases. The lack of this trend at 30~K is attributed to the efficient reactive desorption at all fluences (see text).}
         \label{rates}
   \end{figure}

Figure~\ref{rates} shows the total methanol ice destruction and formation rates, and complements Figures~\ref{Hvstemp} and \ref{Hreac}. At 20, 50 and 70~K the trends are anti-correlated, whereas, at 30~K, both rates decrease with the fluence. This indicates that, except at 30~K, the chemical reactions in the methanol ice start converging towards the chemical equilibrium at the first half-hour of the experiment. The efficient reactive desorption observed at 30~K prevents the same trend before fluence of 2.5 $\times$ 10$^{17}$ photons cm$^{-2}$.

\subsection{Astrophysical implications}
Based on the findings obtained in the previous sections, now we assess the impact of the methanol ice photolysis in the chemical abundances of a molecular cloud subjected to UV radiation field with different strengths. This is worth analysing because, in most published astrochemical models, the photolysis of methanol ice is not included.

With ProDiMo, the molecular cloud model is benchmarked against ALCHEMIC and NAUTILUS \citep[][]{Semenov2010}, and performs 0D time-dependent chemical simulations where $A_{\rm{V}} = 10$ mag and $T_{\rm{dust}} = T_{\rm{gas}}$ are adopted. In particular, we use the same temperatures in laboratory experiments. The chemical variations are checked for different physical environments, namely, numerical densities of $n = 10^{4}$ cm$^{-3}$ and $n = 10^{5}$ cm$^{-3}$, cosmic ray ionization rates of $\zeta = 10^{-16}$ s$^{-1}$ and $\zeta = 10^{-17}$ s$^{-1}$, and UV radiation strength from $\chi = 1$, that represents inner regions of molecular clouds to $\chi = 10^5$ which illustrates the high UV field in photon-dominated regions. For completeness, all physical parameters used for the molecular cloud simulations, including the dust grain size ($a$) and the dust-to-gas mass standard ratio ($\delta$) are given in Table~\ref{mc_params}. The elemental abundances assumed in these models are shown in Table~\ref{elements}, and the chemical elements, gas-phase species and the additional photolysis reactions are shown in Appendix~\ref{all_elements} and \ref{ap_photdiss}, respectively.

\begin{table}
\caption{\label{mc_params}Physical parameters used in the chemical simulation of the molecular cloud.}
\scalebox{0.95}{
\centering
\begin{tabular}{llll}
\hline\hline
Parameter & Symbol & Values & Units\\
\hline
Gas density & $n$ & 10$^4$, 10$^5$ & cm$^{-3}$\\
Temperature & $T_{\mathrm{dust}}$ = $T_{\mathrm{gas}}$ & 20, 30, 50, 70 & K\\
Extinction & $A_{\rm{V}}$ & 10 & mag\\
Strength of ISM UV & $\chi$ & 1 & \\
CR H$_2$ ionisation rate & $\zeta_{\rm{CR}}$ & 10$^{-17}$, 10$^{-16}$ & s$^{-1}$\\
Grain radius & $r$ & 0.1 & $\mu$m\\
Dust-to-gas mass ratio & $\delta$ & 0.01 & \\
\hline
\end{tabular}
}
\end{table}

\begin{table}
\caption{\label{elements}Elements and abundances adopted in the molecular cloud models.}
\renewcommand{\arraystretch}{1.}
\centering
\begin{tabular}{ccc}
\hline\hline
Elements & Initial species & Abundances\\
\hline
H & H$^+$ & 12.00\\
He & He & 10.98\\
C & C$^+$ & 8.14\\
N & N & 7.90\\
O & O & 8.48\\
Ne & Ne & 7.95\\
Na & Na$^+$ & 3.36\\
Mg & Mg$^+$ & 4.03\\
Ar & Ar & 6.08\\
Fe & Fe$^+$ & 3.24\\
S & S$^+$ & 5.27\\
Si & Si$^+$ & 4.24\\
\hline
\end{tabular}
\tablefoot{The abundances are calculated as 12 + log $\epsilon$, where $\epsilon$ is the mass fraction of each element.}
\end{table}

\begin{figure}
   \centering
   \includegraphics[width=\hsize]{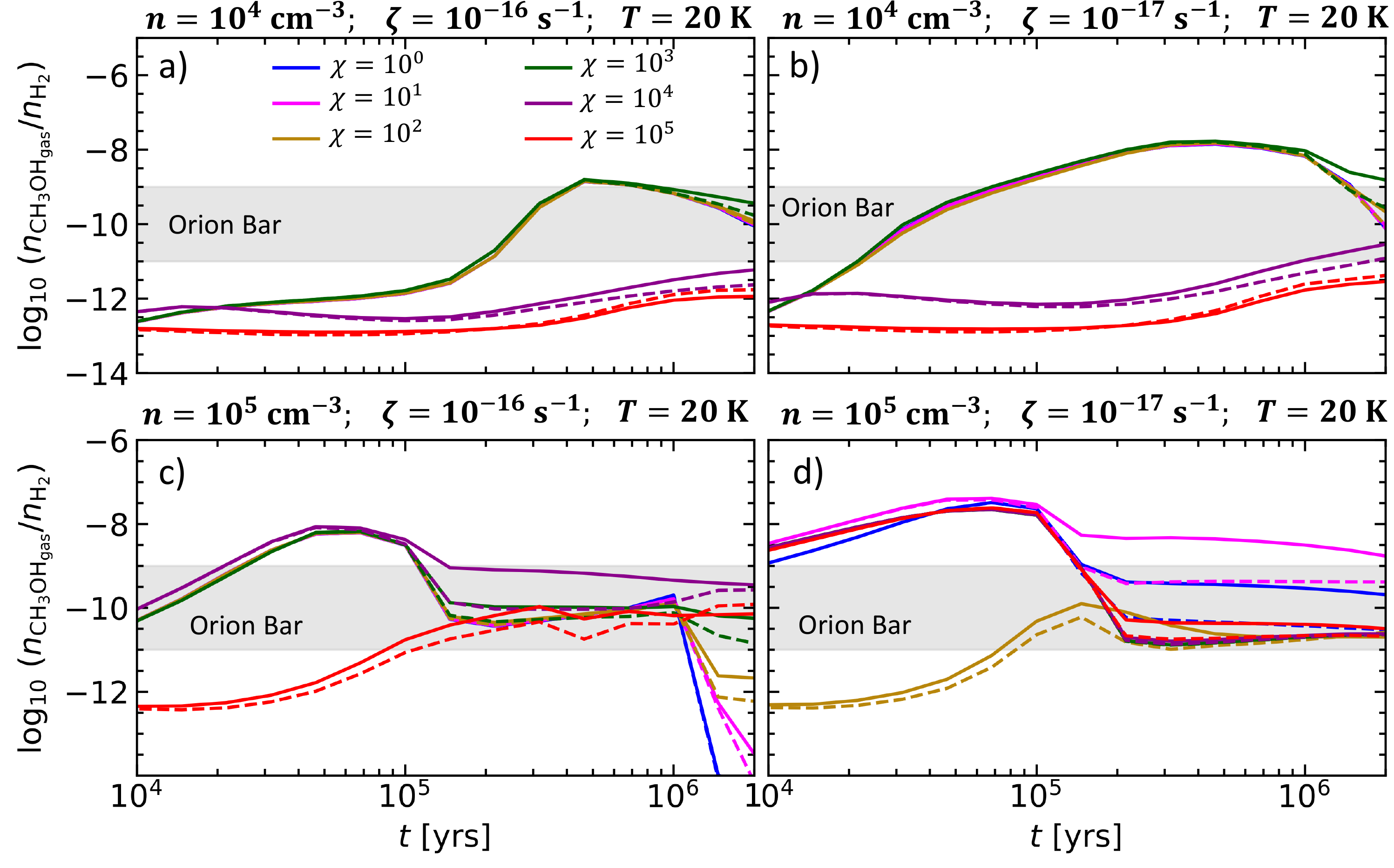}
      \caption{Abundances of gas-phase methanol at different physical conditions and $T =$ 20~K. The solid and dashed lines show the abundances in the models with and without methanol ice photolysis, respectively. The line colours indicate the strength of the UV radiation field. The grey shaded area indicate methanol abundance in the Orion Bar PDR region \citep[][]{Cuadrado2017} only for comparison purposes with an astrophysical environment.}
         \label{abund}
   \end{figure}

Figure~\ref{abund} shows the gas-phase methanol abundance when CH$_3$OH ice photodissociation is included (solid lines) or excluded (dashed lines) in the chemical simulations of a molecular cloud irradiated by external sources. The abundance observed in Orion Bar is only for comparison purposes since the molecular cloud model adopted in this paper does not include the physics involved in PDRs. Yet, the models adopting a temperature of 20~K produce abundances compatible with that astrophysical environment. Qualitatively, the models with and without methanol ice photolysis produce similar abundances for $n = 10^{4}$ cm$^{-3}$ and, $\zeta = 10^{-16}$ s$^{-1}$ or $\zeta = 10^{-17}$ s$^{-1}$ (panels {\it a} and {\it b}, respectively). On the other hand, when the same H$_2$ ionization rates are adopted for $n = 10^{5}$ cm$^{-3}$, major differences are observed (panels {\it c} and {\it d}).

\begin{figure}
   \centering
   \includegraphics[width=\hsize]{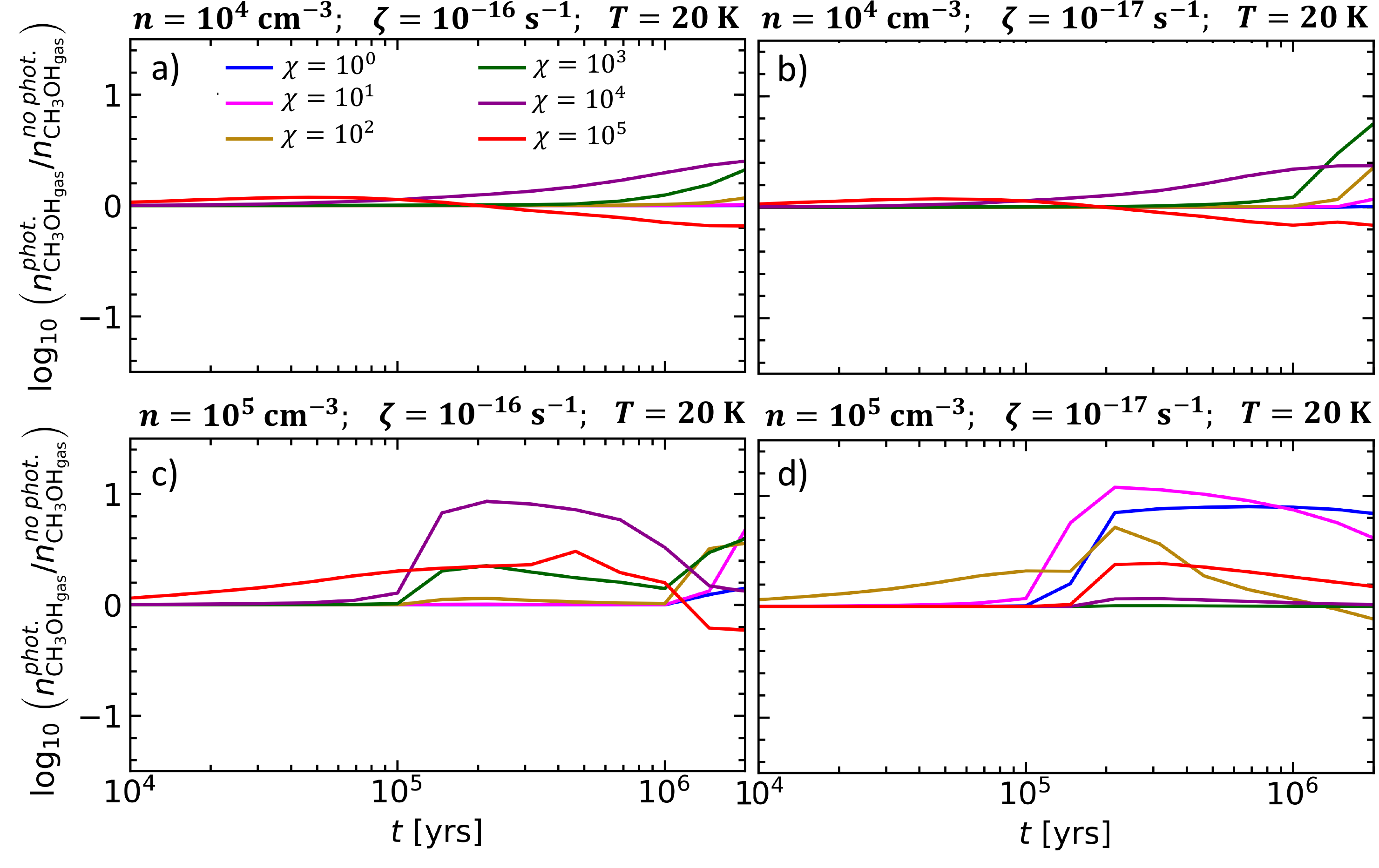}
      \caption{Methanol gas-phase abundance difference in logarithm scale between the models including and excluding CH$_3$OH ice photolysis at different physical conditions and $T =$ 20~K. The line colours indicate the strength of the UV radiation field.}
         \label{diff_gas}
   \end{figure}

The quantitative differences of the models with and without methanol ice photolysis are given in Figure~\ref{diff_gas} for the gas phase. Panels {\it a} and {\it b} show that the chemical differences are lower than a factor of five in regions where $n = 10^{4}$ cm$^{-3}$ and, $\zeta = 10^{-16,-17}$ s$^{-1}$. On the other hand, in denser regions where $n = 10^{5}$ cm$^{-3}$ (panels {\it c} and {\it d}), models including methanol ice photolysis result in up to one order of magnitude more gas-phase methanol after 1 $\times$ 10$^5$ years. However, the strength of the UV radiation field strength of $\chi = 10^4$ is required to enhance the methanol gas-phase abundance if the ionization field is $\zeta = 10^{-16}$ s$^{-1}$. Conversely, assuming a lower H$_2$ ionization rate ($\zeta = 10^{-17}$ s$^{-1}$), less UV radiation field ($\chi \leq 10$) is needed to produce such an increment in the gas-phase methanol abundance. The main mechanism to release methanol to the gas phase in these models is the reactive desorption following the reactions R5 and R6 and photodesorption.  

This analysis was repeated for temperatures of 30~K, 50~K and 70~K, where reactive desorption from \citet{Minissale2016} is adopted at 30~K for consistency reasons with the results in Section~\ref{results}. However, only in the case assuming $T$ = 30~K, $n = 10^{5}$ cm$^{-3}$, $\zeta = 10^{-16}$ s$^{-1}$, and $\chi = 10^4$, a difference of one order of magnitude in the gas-phase methanol abundance is observed for the models including methanol ice photolysis. Overall, this simple analysis indicates that higher densities at 20 and 30~K are needed to observe the effect of the methanol ice photodissociation, whereas at 50 and 70~K, no difference occurs when methanol ice photolysis is taken into account (See Appendix~\ref{Ap_mc_highT}). A trivial explanation is that at low temperatures, the recombination of dissociated CH$_3$OH is more efficient than at higher temperatures because of the amount of radical available. Consequently, the high number of recombination leads to more methanol desorbed to the gas phase via reactive desorption. Conversely, in the molecular cloud model at high temperature, reactive desorption does not play a role because of the low amount of radicals to recombine and the lower recombination rates.

\section{Conclusions}
The ProDiMo code has been used to model the methanol ice photolysis by UV radiation under laboratory conditions. The experiments performed by \citet{Oberg2009} are used to set up the models introduced in this paper. Different surface chemistry mechanisms and branch ratios were used to run a grid of models and quantify the dependence of the temperature in the ice photolysis. The main results of this work are:

\begin{enumerate}
    \item The photodissociation cross-sections $\sigma_{\rm{phd}}$ of the methanol ice derived in this paper are in reasonable agreement with the experiments. The difference of 10-15\% between modelling and laboratory can be associated with multiple causes, including the ``cage effect'', i.e., that the radical diffusion is inhibited by the environment, which leads to a fast CH$_3$OH ice reformation. Based on the $\sigma_{\rm{phd}}$ that provides the maximum likelihood between experiment and models, we have calculated the temperature-dependent photo-rates for methanol ice photolysis.
    
    \item Surface chemistry mechanisms such as diffusion/reaction tunnelling, surface competition and thermal diffusion are required to reproduce the methanol ice photodissociation observed in the experiments. In addition, the models including reactive desorption from \citet{Minissale2016} fit well the experiment at 30~K, whereas the cases at 20, 50 and 70~K are better fitted when reactive desorption from \citet{Garrod2007} is considered. The reason for this difference is not clear from the analysis introduced in this paper.
    
    \item The methanol ice BRs calculated by \citet{Paardekooper2016} provide the best agreement with the experimental data when compared with the results adopting other BRs estimation from the literature. This indicates that the methanol ice is mostly destroyed by the hydroxymethyl (CH$_2$OH) channel than by the methyl branch, as often assumed in the chemical modelling of the ISM.
    
    \item The abundance of the chemical species formed from the methanol ice photolysis is well replicated in the cases of CH$_2$OH at 20~K and 30~K, CO$_2$ at 50~K and CH$_4$ at 20~K. A small difference between experiment and model is obtained for the formation of HCO in the ice when adotping a low activation energy (520~K) for the reaction H\# + CO\# $\longrightarrow$ HCO\#. The model fails to reproduce the other cases addressed in this paper, especially at 70~K. Although this indicates that more accurate surface chemistry approaches are needed, it is also a warning about the accuracy of physicochemical parameters in the reaction network, such as binding energy, radical diffusion rates and photolysis of other chemical species.
    
    \item In typical conditions of molecular clouds (high density, low temperature, and standard strength of UV radiation field), CH$_3$OH ice photolysis leads to an enhancement of gas-phase CH$_3$OH abundance by up to one order of magnitude via chemical desorption.
    
\end{enumerate}

This study shows how the computational modelling of chemical reactions in the ice under laboratory conditions can be used to shrink the gap between experiments and models to interpret astronomical observations. In particular, the analysis of the current observations with JWST is improving our knowledge of the composition and chemical environment of interstellar ice mantles. Therefore, a quantitative understanding of the processes that govern ice chemistry is crucial. Moreover, this work provides a template for future studies simulating the UV radiation of ice samples.

\begin{acknowledgements}
We thanks the anonymous referee for providing a throughout constructive report that improved the quality of this paper. This work benefited from support from the European Research Council (ERC) under the European Union's Horizon 2020 research and innovation program through ERC Consolidator Grant ``S4F'' (grant agreement No~646908), and from the ERC ``MOLDISK'' grant (grant agreement No~101019751) . WRMR acknowledges the São Paulo Research Foundation (grant No~16/23054-7) and the financial support from Leiden Observatory. WRMR also thanks Carlos M. R. Rocha and Ko-Ju Chuang for insightful discussions. P. W. acknowledges funding from the European Union H2020-MSCA-ITN-2019 under Grant Agreement no. 860470 (CHAMELEON). The research of LEK is supported by a research grant (19127) from VILLUM FONDEN. 
\end{acknowledgements}

\bibliographystyle{aa}
\bibliography{References}

\begin{thebibliography}{118}
\expandafter\ifx\csname natexlab\endcsname\relax\def\natexlab#1{#1}\fi

\bibitem[{{Acharyya} {et~al.}(2007){Acharyya}, {Fuchs}, {Fraser}, {van
  Dishoeck}, \& {Linnartz}}]{Acharyya2007}
{Acharyya}, K., {Fuchs}, G.~W., {Fraser}, H.~J., {van Dishoeck}, E.~F., \&
  {Linnartz}, H. 2007, \aap, 466, 1005

\bibitem[{{Aikawa} {et~al.}(1996){Aikawa}, {Miyama}, {Nakano}, \&
  {Umebayashi}}]{Aikawa1996}
{Aikawa}, Y., {Miyama}, S.~M., {Nakano}, T., \& {Umebayashi}, T. 1996, \apj,
  467, 684

\bibitem[{{Aikawa} {et~al.}(2008){Aikawa}, {Wakelam}, {Garrod}, \&
  {Herbst}}]{Aikawa2008}
{Aikawa}, Y., {Wakelam}, V., {Garrod}, R.~T., \& {Herbst}, E. 2008, \apj, 674,
  984

\bibitem[{{Ashfold} {et~al.}(2010){Ashfold}, {King}, {Murdock}, {Nix},
  {Oliver}, \& {Sage}}]{Ashfold2010}
{Ashfold}, M. N.~R., {King}, G.~A., {Murdock}, D., {et~al.} 2010, Physical
  Chemistry Chemical Physics (Incorporating Faraday Transactions), 12, 1218

\bibitem[{{Awad} {et~al.}(2005){Awad}, {Chigai}, {Kimura}, {Shalabiea}, \&
  {Yamamoto}}]{Awad2005}
{Awad}, Z., {Chigai}, T., {Kimura}, Y., {Shalabiea}, O.~M., \& {Yamamoto}, T.
  2005, \apj, 626, 262

\bibitem[{{Balsiger} {et~al.}(2007){Balsiger}, {Altwegg}, {Bochsler},
  {Eberhardt}, {Fischer}, {Graf}, {J{\"a}ckel}, {Kopp}, {Langer}, {Mildner},
  {M{\"u}ller}, {Riesen}, {Rubin}, {Scherer}, {Wurz}, {W{\"u}thrich}, {Arijs},
  {Delanoye}, {de Keyser}, {Neefs}, {Nevejans}, {R{\`e}me}, {Aoustin},
  {Mazelle}, {M{\'e}dale}, {Sauvaud}, {Berthelier}, {Bertaux}, {Duvet},
  {Illiano}, {Fuselier}, {Ghielmetti}, {Magoncelli}, {Shelley}, {Korth},
  {Heerlein}, {Lauche}, {Livi}, {Loose}, {Mall}, {Wilken}, {Gliem}, {Fiethe},
  {Gombosi}, {Block}, {Carignan}, {Fisk}, {Waite}, {Young}, \&
  {Wollnik}}]{Balsiger2007}
{Balsiger}, H., {Altwegg}, K., {Bochsler}, P., {et~al.} 2007, \ssr, 128, 745

\bibitem[{Bell(1980)}]{Bell1980}
Bell, R.~P. 1980, The Tunnel Effect in Chemistry (New York: Springer, 2013)

\bibitem[{{Bell} \& {Le Roy}(1982)}]{Bell1982}
{Bell}, R.~P. \& {Le Roy}, R.~J. 1982, Physics Today, 35, 85

\bibitem[{{Bertin} {et~al.}(2016){Bertin}, {Romanzin}, {Doronin}, {Philippe},
  {Jeseck}, {Ligterink}, {Linnartz}, {Michaut}, \& {Fillion}}]{Bertin2016}
{Bertin}, M., {Romanzin}, C., {Doronin}, M., {et~al.} 2016, \apjl, 817, L12

\bibitem[{{Bixon} \& {Jortner}(1968)}]{Bixon1968}
{Bixon}, M. \& {Jortner}, J. 1968, \jcp, 48, 715

\bibitem[{{Boogert} {et~al.}(2011){Boogert}, {Huard}, {Cook}, {Chiar}, {Knez},
  {Decin}, {Blake}, {Tielens}, \& {van Dishoeck}}]{Boogert2011}
{Boogert}, A.~C.~A., {Huard}, T.~L., {Cook}, A.~M., {et~al.} 2011, \apj, 729,
  92

\bibitem[{{Bovolenta} {et~al.}(2022){Bovolenta}, {Vogt-Geisse}, {Bovino}, \&
  {Grassi}}]{Bovolenta2022}
{Bovolenta}, G.~M., {Vogt-Geisse}, S., {Bovino}, S., \& {Grassi}, T. 2022,
  \apjs, 262, 17

\bibitem[{{Brown} \& {Bolina}(2007)}]{Brown2007}
{Brown}, W.~A. \& {Bolina}, A.~S. 2007, \mnras, 374, 1006

\bibitem[{{Brown} {et~al.}(1984){Brown}, {Augustyniak}, {Marcantonio},
  {Simmons}, {Boring}, {Johnson}, \& {Reimann}}]{Brown1984}
{Brown}, W.~L., {Augustyniak}, W.~M., {Marcantonio}, K.~J., {et~al.} 1984,
  Nuclear Instruments and Methods in Physics Research B, 1, 307

\bibitem[{{Calmonte} {et~al.}(2016){Calmonte}, {Altwegg}, {Balsiger},
  {Berthelier}, {Bieler}, {Cessateur}, {Dhooghe}, {van Dishoeck}, {Fiethe},
  {Fuselier}, {Gasc}, {Gombosi}, {H{\"a}ssig}, {Le Roy}, {Rubin}, {S{\'e}mon},
  {Tzou}, \& {Wampfler}}]{Calmonte2016}
{Calmonte}, U., {Altwegg}, K., {Balsiger}, H., {et~al.} 2016, \mnras, 462, S253

\bibitem[{{Carvalho} {et~al.}(2022){Carvalho}, {Pilling}, \&
  {Galv{\~a}o}}]{Carvalho2022}
{Carvalho}, G.~A., {Pilling}, S., \& {Galv{\~a}o}, B. R.~L. 2022, \mnras, 515,
  3760

\bibitem[{{Cazaux} \& {Tielens}(2002)}]{Cazaux2002}
{Cazaux}, S. \& {Tielens}, A.~G.~G.~M. 2002, \apjl, 575, L29

\bibitem[{{Chaabouni} {et~al.}(2012){Chaabouni}, {Minissale}, {Manic{\`o}},
  {Congiu}, {Noble}, {Baouche}, {Accolla}, {Lemaire}, {Pirronello}, \&
  {Dulieu}}]{Chaabouni2012}
{Chaabouni}, H., {Minissale}, M., {Manic{\`o}}, G., {et~al.} 2012, \jcp, 137,
  234706

\bibitem[{{Chiar} {et~al.}(2000){Chiar}, {Tielens}, {Whittet}, {Schutte},
  {Boogert}, {Lutz}, {van Dishoeck}, \& {Bernstein}}]{Chiar2000}
{Chiar}, J.~E., {Tielens}, A.~G.~G.~M., {Whittet}, D.~C.~B., {et~al.} 2000,
  \apj, 537, 749

\bibitem[{{Chu} {et~al.}(2020){Chu}, {Hodapp}, \& {Boogert}}]{Chu2020}
{Chu}, L. E.~U., {Hodapp}, K., \& {Boogert}, A. 2020, \apj, 904, 86

\bibitem[{{Clark} {et~al.}(2012){Clark}, {Glover}, {Klessen}, \&
  {Bonnell}}]{Clark2012}
{Clark}, P.~C., {Glover}, S. C.~O., {Klessen}, R.~S., \& {Bonnell}, I.~A. 2012,
  \mnras, 424, 2599

\bibitem[{{Collings} {et~al.}(2004){Collings}, {Anderson}, {Chen}, {Dever},
  {Viti}, {Williams}, \& {McCoustra}}]{Collings2004}
{Collings}, M.~P., {Anderson}, M.~A., {Chen}, R., {et~al.} 2004, \mnras, 354,
  1133

\bibitem[{{Congiu} {et~al.}(2014){Congiu}, {Minissale}, {Baouche}, {Chaabouni},
  {Moudens}, {Cazaux}, {Manic{\`o}}, {Pirronello}, \& {Dulieu}}]{Congiu2014}
{Congiu}, E., {Minissale}, M., {Baouche}, S., {et~al.} 2014, Faraday
  Discussions, 168, 151

\bibitem[{{Cuadrado} {et~al.}(2017){Cuadrado}, {Goicoechea}, {Cernicharo},
  {Fuente}, {Pety}, \& {Tercero}}]{Cuadrado2017}
{Cuadrado}, S., {Goicoechea}, J.~R., {Cernicharo}, J., {et~al.} 2017, \aap,
  603, A124

\bibitem[{{Cuppen} {et~al.}(2017){Cuppen}, {Walsh}, {Lamberts}, {Semenov},
  {Garrod}, {Penteado}, \& {Ioppolo}}]{Cuppen2017}
{Cuppen}, H.~M., {Walsh}, C., {Lamberts}, T., {et~al.} 2017, \ssr, 212, 1

\bibitem[{{Dartois} {et~al.}(2018){Dartois}, {Chabot}, {Id Barkach}, {Rothard},
  {Aug{\'e}}, {Agnihotri}, {Domaracka}, \& {Boduch}}]{Dartois2018}
{Dartois}, E., {Chabot}, M., {Id Barkach}, T., {et~al.} 2018, \aap, 618, A173

\bibitem[{{Draine}(1978)}]{Draine1978}
{Draine}, B.~T. 1978, \apjs, 36, 595

\bibitem[{{Draine} \& {Bertoldi}(1996)}]{Draine_Bertoldi1996}
{Draine}, B.~T. \& {Bertoldi}, F. 1996, \apj, 468, 269

\bibitem[{{Drozdovskaya} {et~al.}(2021){Drozdovskaya}, {Schroeder I}, {Rubin},
  {Altwegg}, {van Dishoeck}, {Kulterer}, {De Keyser}, {Fuselier}, \&
  {Combi}}]{Drozdovskaya2021}
{Drozdovskaya}, M.~N., {Schroeder I}, I. R.~H.~G., {Rubin}, M., {et~al.} 2021,
  \mnras, 500, 4901

\bibitem[{{Dulieu} {et~al.}(2013){Dulieu}, {Congiu}, {Noble}, {Baouche},
  {Chaabouni}, {Moudens}, {Minissale}, \& {Cazaux}}]{Dulieu2013}
{Dulieu}, F., {Congiu}, E., {Noble}, J., {et~al.} 2013, Scientific Reports, 3,
  1338

\bibitem[{{Edridge}(2010)}]{Edridge2010}
{Edridge}, J.~L. 2010, PhD thesis, University of London, University College
  London (United Kingdom)

\bibitem[{{Esplugues} {et~al.}(2016){Esplugues}, {Cazaux}, {Meijerink},
  {Spaans}, \& {Caselli}}]{Esplugues2016}
{Esplugues}, G.~B., {Cazaux}, S., {Meijerink}, R., {Spaans}, M., \& {Caselli},
  P. 2016, \aap, 591, A52

\bibitem[{Fahr \& Laufer(1995)}]{Fahr1995}
Fahr, A. \& Laufer, A.~H. 1995, The Journal of Physical Chemistry, 99, 262

\bibitem[{{Fayolle} {et~al.}(2011){Fayolle}, {Bertin}, {Romanzin}, {Michaut},
  {{\"O}berg}, {Linnartz}, \& {Fillion}}]{Fayolle2011}
{Fayolle}, E.~C., {Bertin}, M., {Romanzin}, C., {et~al.} 2011, \apjl, 739, L36

\bibitem[{Franck \& Rabinowitsch(1934)}]{Franck_Rabinowitsch_1934}
Franck, J. \& Rabinowitsch, E. 1934, Trans. Faraday Soc., 30, 120

\bibitem[{{Fuchs} {et~al.}(2009){Fuchs}, {Cuppen}, {Ioppolo}, {Romanzin},
  {Bisschop}, {Andersson}, {van Dishoeck}, \& {Linnartz}}]{Fuchs2009}
{Fuchs}, G.~W., {Cuppen}, H.~M., {Ioppolo}, S., {et~al.} 2009, \aap, 505, 629

\bibitem[{{Garrod} \& {Herbst}(2006)}]{Garrod2006}
{Garrod}, R.~T. \& {Herbst}, E. 2006, \aap, 457, 927

\bibitem[{{Garrod} \& {Pauly}(2011)}]{Garrod2011}
{Garrod}, R.~T. \& {Pauly}, T. 2011, \apj, 735, 15

\bibitem[{{Garrod} {et~al.}(2007){Garrod}, {Wakelam}, \& {Herbst}}]{Garrod2007}
{Garrod}, R.~T., {Wakelam}, V., \& {Herbst}, E. 2007, \aap, 467, 1103

\bibitem[{{Garrod} {et~al.}(2008){Garrod}, {Widicus Weaver}, \&
  {Herbst}}]{Garrod2008}
{Garrod}, R.~T., {Widicus Weaver}, S.~L., \& {Herbst}, E. 2008, \apj, 682, 283

\bibitem[{{Goto} {et~al.}(2021){Goto}, {Vasyunin}, {Giuliano},
  {Jim{\'e}nez-Serra}, {Caselli}, {Rom{\'a}n-Z{\'u}{\~n}iga}, \&
  {Alves}}]{Goto2021}
{Goto}, M., {Vasyunin}, A.~I., {Giuliano}, B.~M., {et~al.} 2021, \aap, 651, A53

\bibitem[{{Goumans} {et~al.}(2008){Goumans}, {Uppal}, \& {Brown}}]{Goumans2008}
{Goumans}, T.~P.~M., {Uppal}, M.~A., \& {Brown}, W.~A. 2008, \mnras, 384, 1158

\bibitem[{{Grassi} {et~al.}(2020){Grassi}, {Bovino}, {Caselli}, {Bovolenta},
  {Vogt-Geisse}, \& {Ercolano}}]{Grassi2020}
{Grassi}, T., {Bovino}, S., {Caselli}, P., {et~al.} 2020, \aap, 643, A155

\bibitem[{{Grundy} {et~al.}(2020){Grundy}, {Bird}, {Britt}, {Cook},
  {Cruikshank}, {Howett}, {Krijt}, {Linscott}, {Olkin}, {Parker}, {Protopapa},
  {Ruaud}, {Umurhan}, {Young}, {Dalle Ore}, {Kavelaars}, {Keane}, {Pendleton},
  {Porter}, {Scipioni}, {Spencer}, {Stern}, {Verbiscer}, {Weaver}, {Binzel},
  {Buie}, {Buratti}, {Cheng}, {Earle}, {Elliott}, {Gabasova}, {Gladstone},
  {Hill}, {Horanyi}, {Jennings}, {Lunsford}, {McComas}, {McKinnon}, {McNutt},
  {Moore}, {Parker}, {Quirico}, {Reuter}, {Schenk}, {Schmitt}, {Showalter},
  {Singer}, {Weigle}, \& {Zangari}}]{Grundy2020}
{Grundy}, W.~M., {Bird}, M.~K., {Britt}, D.~T., {et~al.} 2020, Science, 367,
  aay3705

\bibitem[{Hagege {et~al.}(1968)Hagege, Roberge, \& Vermeil}]{Hagege1968}
Hagege, J., Roberge, P.~C., \& Vermeil, C. 1968, Trans. Faraday Soc., 64, 3288

\bibitem[{{Hasegawa} \& {Herbst}(1993)}]{Hasegawa1993}
{Hasegawa}, T.~I. \& {Herbst}, E. 1993, \mnras, 263, 589

\bibitem[{{Herbst} \& {Millar}(2008)}]{Smith2008}
{Herbst}, E. \& {Millar}, T.~J. 2008, {Low Temperatures and Cold Molecules},
  ed. I.~W.~M. Smith (London: Imperial College Press), 1

\bibitem[{{Herrero} {et~al.}(2010){Herrero}, {G{\'a}lvez}, {Mat{\'e}}, \&
  {Escribano}}]{Herrero2010}
{Herrero}, V.~J., {G{\'a}lvez}, {\'O}., {Mat{\'e}}, B., \& {Escribano}, R.
  2010, Physical Chemistry Chemical Physics (Incorporating Faraday
  Transactions), 12, 3164

\bibitem[{Holbrook {et~al.}(1996)Holbrook, Pilling, \&
  Robertson}]{Holbrook1996}
Holbrook, K. A. K.~A., Pilling, M.~J., \& Robertson, S.~H. 1996, Unimolecular
  reactions / Kenneth A. Holbrook, Michael J. Pilling, Struan H. Robertson.,
  2nd edn. (Chichester: John Wiley \& Sons)

\bibitem[{{Ioppolo} {et~al.}(2008){Ioppolo}, {Cuppen}, {Romanzin}, {van
  Dishoeck}, \& {Linnartz}}]{Ioppolo2008}
{Ioppolo}, S., {Cuppen}, H.~M., {Romanzin}, C., {van Dishoeck}, E.~F., \&
  {Linnartz}, H. 2008, \apj, 686, 1474

\bibitem[{{Ioppolo} {et~al.}(2021){Ioppolo}, {Fedoseev}, {Chuang}, {Cuppen},
  {Clements}, {Jin}, {Garrod}, {Qasim}, {Kofman}, {van Dishoeck}, \&
  {Linnartz}}]{Ioppolo2021}
{Ioppolo}, S., {Fedoseev}, G., {Chuang}, K.~J., {et~al.} 2021, Nature
  Astronomy, 5, 197

\bibitem[{{Jim{\'e}nez-Escobar} {et~al.}(2016){Jim{\'e}nez-Escobar}, {Chen},
  {Ciaravella}, {Huang}, {Micela}, \& {Cecchi-Pestellini}}]{Jimenez2016}
{Jim{\'e}nez-Escobar}, A., {Chen}, Y.~J., {Ciaravella}, A., {et~al.} 2016,
  \apj, 820, 25

\bibitem[{{Jochims} {et~al.}(1994){Jochims}, {Ruhl}, {Baumgartel}, {Tobita}, \&
  {Leach}}]{Jochims1994}
{Jochims}, H.~W., {Ruhl}, E., {Baumgartel}, H., {Tobita}, S., \& {Leach}, S.
  1994, \apj, 420, 307

\bibitem[{{Kamp} {et~al.}(2010){Kamp}, {Tilling}, {Woitke}, {Thi}, \&
  {Hogerheijde}}]{Kamp2010}
{Kamp}, I., {Tilling}, I., {Woitke}, P., {Thi}, W.~F., \& {Hogerheijde}, M.
  2010, \aap, 510, A18

\bibitem[{{Kim} {et~al.}(2022){Kim}, {Lee}, {Jeong}, {Kim}, {Aikawa}, {Noble},
  {Choi}, {Lee}, {Dunham}, {Kim}, \& {Koo}}]{Kim2022ApJ}
{Kim}, J., {Lee}, J.-E., {Jeong}, W.-S., {et~al.} 2022, \apj, 935, 137

\bibitem[{{Kristensen} {et~al.}(2010){Kristensen}, {van Dishoeck}, {van
  Kempen}, {Cuppen}, {Brinch}, {J{\o}rgensen}, \&
  {Hogerheijde}}]{Kristensen2010}
{Kristensen}, L.~E., {van Dishoeck}, E.~F., {van Kempen}, T.~A., {et~al.} 2010,
  \aap, 516, A57

\bibitem[{{Kroes} \& {Andersson}(2005)}]{Kroes2006}
{Kroes}, G.~J. \& {Andersson}, S. 2005, in IAU Symposium, Vol. 231,
  Astrochemistry: Recent Successes and Current Challenges, ed. D.~C. {Lis},
  G.~A. {Blake}, \& E.~{Herbst}, 427--442

\bibitem[{{Kuwahata} {et~al.}(2015){Kuwahata}, {Hama}, {Kouchi}, \&
  {Watanabe}}]{Kuwahata2015}
{Kuwahata}, K., {Hama}, T., {Kouchi}, A., \& {Watanabe}, N. 2015, \prl, 115,
  133201

\bibitem[{{Laas} {et~al.}(2011){Laas}, {Garrod}, {Herbst}, \& {Widicus
  Weaver}}]{Laas2011}
{Laas}, J.~C., {Garrod}, R.~T., {Herbst}, E., \& {Widicus Weaver}, S.~L. 2011,
  \apj, 728, 71

\bibitem[{{Laffon} {et~al.}(2010){Laffon}, {Lasne}, {Bournel}, {Schulte},
  {Lacombe}, \& {Parent}}]{Laffon2010}
{Laffon}, C., {Lasne}, J., {Bournel}, F., {et~al.} 2010, Physical Chemistry
  Chemical Physics (Incorporating Faraday Transactions), 12, 10865

\bibitem[{{Lamberts} {et~al.}(2014){Lamberts}, {Cuppen}, {Fedoseev}, {Ioppolo},
  {Chuang}, \& {Linnartz}}]{Lamberts2014}
{Lamberts}, T., {Cuppen}, H.~M., {Fedoseev}, G., {et~al.} 2014, \aap, 570, A57

\bibitem[{{Lamberts} {et~al.}(2016){Lamberts}, {Samanta}, {K{\"o}hn}, \&
  {K{\"a}stner}}]{Lamberts2016}
{Lamberts}, T., {Samanta}, P.~K., {K{\"o}hn}, A., \& {K{\"a}stner}, J. 2016,
  Physical Chemistry Chemical Physics (Incorporating Faraday Transactions), 18,
  33021

\bibitem[{{Lara} {et~al.}(2014){Lara}, {Lellouch}, {Gonz{\'a}lez}, {Moreno}, \&
  {Rengel}}]{Lara2014}
{Lara}, L.~M., {Lellouch}, E., {Gonz{\'a}lez}, M., {Moreno}, R., \& {Rengel},
  M. 2014, \aap, 566, A143

\bibitem[{{Lavvas} {et~al.}(2008){Lavvas}, {Coustenis}, \&
  {Vardavas}}]{Lavvas2008}
{Lavvas}, P.~P., {Coustenis}, A., \& {Vardavas}, I.~M. 2008, \planss, 56, 27

\bibitem[{{Le Roy} {et~al.}(2015){Le Roy}, {Altwegg}, {Balsiger}, {Berthelier},
  {Bieler}, {Briois}, {Calmonte}, {Combi}, {De Keyser}, {Dhooghe}, {Fiethe},
  {Fuselier}, {Gasc}, {Gombosi}, {H{\"a}ssig}, {J{\"a}ckel}, {Rubin}, \&
  {Tzou}}]{LeRoy2015}
{Le Roy}, L., {Altwegg}, K., {Balsiger}, H., {et~al.} 2015, \aap, 583, A1

\bibitem[{{Leger} {et~al.}(1989){Leger}, {D'Hendecourt}, {Boissel}, \&
  {Desert}}]{Leger1989}
{Leger}, A., {D'Hendecourt}, L., {Boissel}, P., \& {Desert}, F.~X. 1989, \aap,
  213, 351

\bibitem[{{Li} {et~al.}(2004){Li}, {Zhang}, \& {Zhang}}]{Li2004}
{Li}, Q.~S., {Zhang}, Y., \& {Zhang}, S. 2004, \jcp, 121, 9474

\bibitem[{{Ligterink} {et~al.}(2015){Ligterink}, {Paardekooper}, {Chuang},
  {Both}, {Cruz-Diaz}, {van Helden}, \& {Linnartz}}]{Ligterink2015}
{Ligterink}, N.~F.~W., {Paardekooper}, D.~M., {Chuang}, K.~J., {et~al.} 2015,
  \aap, 584, A56

\bibitem[{{McClure} {et~al.}(2023){McClure}, {Rocha}, {Pontoppidan}, {Crouzet},
  {Chu}, {Dartois}, {Lamberts}, {Noble}, {Pendleton}, {Perotti}, {Qasim},
  {Rachid}, {Smith}, {Sun}, {Beck}, {Boogert}, {Brown}, {Caselli}, {Charnley},
  {Cuppen}, {Dickinson}, {Drozdovskaya}, {Egami}, {Erkal}, {Fraser}, {Garrod},
  {Harsono}, {Ioppolo}, {Jim{\'e}nez-Serra}, {Jin}, {J{\o}rgensen},
  {Kristensen}, {Lis}, {McCoustra}, {McGuire}, {Melnick}, {{\"O}berg},
  {Palumbo}, {Shimonishi}, {Sturm}, {van Dishoeck}, \&
  {Linnartz}}]{Mcclure2023}
{McClure}, M.~K., {Rocha}, W.~R.~M., {Pontoppidan}, K.~M., {et~al.} 2023,
  Nature Astronomy [\eprint[arXiv]{2301.09140}]

\bibitem[{{Meinert} {et~al.}(2016){Meinert}, {Myrgorodska}, {de Marcellus},
  {Buhse}, {Nahon}, {Hoffmann}, {d'Hendecourt}, \&
  {Meierhenrich}}]{Meinert2016}
{Meinert}, C., {Myrgorodska}, I., {de Marcellus}, P., {et~al.} 2016, Science,
  352, 208

\bibitem[{{Minissale} {et~al.}(2022){Minissale}, {Aikawa}, {Bergin}, {Bertin},
  {Brown}, {Cazaux}, {Charnley}, {Coutens}, {Cuppen}, {Guzman}, {Linnartz},
  {McCoustra}, {Rimola}, {Schrauwen}, {Toubin}, {Ugliengo}, {Watanabe},
  {Wakelam}, \& {Dulieu}}]{Minissale2022}
{Minissale}, M., {Aikawa}, Y., {Bergin}, E., {et~al.} 2022, ACS Earth and Space
  Chemistry, 6, 597

\bibitem[{{Minissale} {et~al.}(2016){Minissale}, {Dulieu}, {Cazaux}, \&
  {Hocuk}}]{Minissale2016}
{Minissale}, M., {Dulieu}, F., {Cazaux}, S., \& {Hocuk}, S. 2016, \aap, 585,
  A24

\bibitem[{{Minissale} {et~al.}(2015){Minissale}, {Loison}, {Baouche},
  {Chaabouni}, {Congiu}, \& {Dulieu}}]{Minissale2015}
{Minissale}, M., {Loison}, J.~C., {Baouche}, S., {et~al.} 2015, \aap, 577, A2

\bibitem[{{Mullikin} {et~al.}(2021){Mullikin}, {Anderson}, {O'Hern}, {Farrah},
  {Arumainayagam}, {van Dishoeck}, {Gerakines}, {Vasyunin}, {Majumdar},
  {Caselli}, \& {Shingledecker}}]{Mullikin2021}
{Mullikin}, E., {Anderson}, H., {O'Hern}, N., {et~al.} 2021, \apj, 910, 72

\bibitem[{{Nuevo} {et~al.}(2018){Nuevo}, {Cooper}, \& {Sandford}}]{Nuevo2018}
{Nuevo}, M., {Cooper}, G., \& {Sandford}, S.~A. 2018, Nature Communications, 9,
  5276

\bibitem[{{Oba} {et~al.}(2012){Oba}, {Watanabe}, {Hama}, {Kuwahata}, {Hidaka},
  \& {Kouchi}}]{Oba2012}
{Oba}, Y., {Watanabe}, N., {Hama}, T., {et~al.} 2012, \apj, 749, 67

\bibitem[{{\"O}berg(2016)}]{Oberg2016}
{\"O}berg, K.~I. 2016, Chemical reviews, 116, 9631

\bibitem[{{{\"O}berg} {et~al.}(2011){{\"O}berg}, {Boogert}, {Pontoppidan}, {van
  den Broek}, {van Dishoeck}, {Bottinelli}, {Blake}, \& {Evans}}]{Oberg2011}
{{\"O}berg}, K.~I., {Boogert}, A.~C.~A., {Pontoppidan}, K.~M., {et~al.} 2011,
  \apj, 740, 109

\bibitem[{{{\"O}berg} {et~al.}(2009{\natexlab{a}}){{\"O}berg}, {Garrod}, {van
  Dishoeck}, \& {Linnartz}}]{Oberg2009}
{{\"O}berg}, K.~I., {Garrod}, R.~T., {van Dishoeck}, E.~F., \& {Linnartz}, H.
  2009{\natexlab{a}}, \aap, 504, 891

\bibitem[{{{\"O}berg} {et~al.}(2009{\natexlab{b}}){{\"O}berg}, {van Dishoeck},
  \& {Linnartz}}]{Oberg2009yield}
{{\"O}berg}, K.~I., {van Dishoeck}, E.~F., \& {Linnartz}, H.
  2009{\natexlab{b}}, \aap, 496, 281

\bibitem[{Opansky \& Leone(1996)}]{Opansky1996}
Opansky, B.~J. \& Leone, S.~R. 1996, The Journal of Physical Chemistry, 100,
  19904

\bibitem[{{Paardekooper} {et~al.}(2016){Paardekooper}, {Bossa}, \&
  {Linnartz}}]{Paardekooper2016}
{Paardekooper}, D.~M., {Bossa}, J.~B., \& {Linnartz}, H. 2016, \aap, 592, A67

\bibitem[{{Perotti} {et~al.}(2021){Perotti}, {J{\o}rgensen}, {Fraser},
  {Suutarinen}, {Kristensen}, {Rocha}, {Bjerkeli}, \&
  {Pontoppidan}}]{Perotti2021}
{Perotti}, G., {J{\o}rgensen}, J.~K., {Fraser}, H.~J., {et~al.} 2021, \aap,
  650, A168

\bibitem[{{Perotti} {et~al.}(2020){Perotti}, {Rocha}, {J{\o}rgensen},
  {Kristensen}, {Fraser}, \& {Pontoppidan}}]{Perotti2020}
{Perotti}, G., {Rocha}, W.~R.~M., {J{\o}rgensen}, J.~K., {et~al.} 2020, \aap,
  643, A48

\bibitem[{{Pilling} {et~al.}(2022){Pilling}, {Carvalho}, \&
  {Rocha}}]{Pilling2022}
{Pilling}, S., {Carvalho}, G.~A., \& {Rocha}, W. R.~M. 2022, \apj, 925, 147

\bibitem[{{Pilling} {et~al.}(2023){Pilling}, {Rocha}, {Carvalho}, \& {de
  Abreu}}]{Pilling2023}
{Pilling}, S., {Rocha}, W. R.~M., {Carvalho}, G.~A., \& {de Abreu}, H.~A. 2023,
  ASR

\bibitem[{{Pontoppidan} {et~al.}(2004){Pontoppidan}, {van Dishoeck}, \&
  {Dartois}}]{Pontoppidan2004}
{Pontoppidan}, K.~M., {van Dishoeck}, E.~F., \& {Dartois}, E. 2004, \aap, 426,
  925

\bibitem[{{Prasad} \& {Tarafdar}(1983)}]{Prasad1983}
{Prasad}, S.~S. \& {Tarafdar}, S.~P. 1983, \apj, 267, 603

\bibitem[{{Qasim} {et~al.}(2018){Qasim}, {Chuang}, {Fedoseev}, {Ioppolo},
  {Boogert}, \& {Linnartz}}]{Qasim2018}
{Qasim}, D., {Chuang}, K.~J., {Fedoseev}, G., {et~al.} 2018, \aap, 612, A83

\bibitem[{{Raut} \& {Baragiola}(2011)}]{Raut2011}
{Raut}, U. \& {Baragiola}, R.~A. 2011, \apjl, 737, L14

\bibitem[{{Rimola} {et~al.}(2014){Rimola}, {Taquet}, {Ugliengo}, {Balucani}, \&
  {Ceccarelli}}]{Rimola2014}
{Rimola}, A., {Taquet}, V., {Ugliengo}, P., {Balucani}, N., \& {Ceccarelli}, C.
  2014, \aap, 572, A70

\bibitem[{{Roncero} {et~al.}(2018){Roncero}, {Zanchet}, \&
  {Aguado}}]{Roncero2018}
{Roncero}, O., {Zanchet}, A., \& {Aguado}, A. 2018, Physical Chemistry Chemical
  Physics (Incorporating Faraday Transactions), 20, 25951

\bibitem[{{Ruaud} {et~al.}(2015){Ruaud}, {Loison}, {Hickson}, {Gratier},
  {Hersant}, \& {Wakelam}}]{Ruaud2015}
{Ruaud}, M., {Loison}, J.~C., {Hickson}, K.~M., {et~al.} 2015, \mnras, 447,
  4004

\bibitem[{{Ruffle} \& {Herbst}(2000)}]{Ruffle2000}
{Ruffle}, D.~P. \& {Herbst}, E. 2000, \mnras, 319, 837

\bibitem[{{Santos} {et~al.}(2022){Santos}, {Chuang}, {Lamberts}, {Fedoseev},
  {Ioppolo}, \& {Linnartz}}]{Santos2022}
{Santos}, J.~C., {Chuang}, K.-J., {Lamberts}, T., {et~al.} 2022, \apjl, 931,
  L33

\bibitem[{{Schutte} {et~al.}(1991){Schutte}, {Tielens}, \&
  {Sandford}}]{Schutte1991}
{Schutte}, W.~A., {Tielens}, A.~G.~G., \& {Sandford}, S.~A. 1991, \apj, 382,
  523

\bibitem[{{Semenov} {et~al.}(2010){Semenov}, {Hersant}, {Wakelam}, {Dutrey},
  {Chapillon}, {Guilloteau}, {Henning}, {Launhardt}, {Pi{\'e}tu}, \&
  {Schreyer}}]{Semenov2010}
{Semenov}, D., {Hersant}, F., {Wakelam}, V., {et~al.} 2010, \aap, 522, A42

\bibitem[{{Shen} {et~al.}(2004){Shen}, {Greenberg}, {Schutte}, \& {van
  Dishoeck}}]{Shen2004}
{Shen}, C.~J., {Greenberg}, J.~M., {Schutte}, W.~A., \& {van Dishoeck}, E.~F.
  2004, \aap, 415, 203

\bibitem[{{Shimonishi} {et~al.}(2018){Shimonishi}, {Nakatani}, {Furuya}, \&
  {Hama}}]{Shimonishi2018}
{Shimonishi}, T., {Nakatani}, N., {Furuya}, K., \& {Hama}, T. 2018, \apj, 855,
  27

\bibitem[{{Shingledecker} {et~al.}(2020){Shingledecker}, {Lamberts}, {Laas},
  {Vasyunin}, {Herbst}, {K{\"a}stner}, \& {Caselli}}]{Shingledecker2020}
{Shingledecker}, C.~N., {Lamberts}, T., {Laas}, J.~C., {et~al.} 2020, \apj,
  888, 52

\bibitem[{{Shingledecker} {et~al.}(2019){Shingledecker}, {Vasyunin}, {Herbst},
  \& {Caselli}}]{Shingledecker2019}
{Shingledecker}, C.~N., {Vasyunin}, A., {Herbst}, E., \& {Caselli}, P. 2019,
  \apj, 876, 140

\bibitem[{{Skouteris} {et~al.}(2018){Skouteris}, {Balucani}, {Ceccarelli},
  {Vazart}, {Puzzarini}, {Barone}, {Codella}, \& {Lefloch}}]{Skouteris2018}
{Skouteris}, D., {Balucani}, N., {Ceccarelli}, C., {et~al.} 2018, \apj, 854,
  135

\bibitem[{{Thi} {et~al.}(2020{\natexlab{a}}){Thi}, {Hocuk}, {Kamp}, {Woitke},
  {Rab}, {Cazaux}, \& {Caselli}}]{Thi2020_feb}
{Thi}, W.~F., {Hocuk}, S., {Kamp}, I., {et~al.} 2020{\natexlab{a}}, \aap, 634,
  A42

\bibitem[{{Thi} {et~al.}(2020{\natexlab{b}}){Thi}, {Hocuk}, {Kamp}, {Woitke},
  {Rab}, {Cazaux}, {Caselli}, \& {D'Angelo}}]{Thi2020_mar}
{Thi}, W.~F., {Hocuk}, S., {Kamp}, I., {et~al.} 2020{\natexlab{b}}, \aap, 635,
  A16

\bibitem[{{Thi} {et~al.}(2011){Thi}, {Woitke}, \& {Kamp}}]{Thi2011}
{Thi}, W.~F., {Woitke}, P., \& {Kamp}, I. 2011, \mnras, 412, 711

\bibitem[{{Tielens} \& {Hagen}(1982)}]{Tielens1982}
{Tielens}, A.~G.~G.~M. \& {Hagen}, W. 1982, \aap, 114, 245

\bibitem[{Tramer \& Voltz(1979)}]{TRAMER1979281}
Tramer, A. \& Voltz, R. 1979, in Excited States, Vol.~4, Time-Resolved Studies
  of Excited Molecules, ed. E.~C. LIM (Elsevier), 281--394

\bibitem[{{van Dishoeck} \& {Visser}(2011)}]{vanDishoeck2011}
{van Dishoeck}, E.~F. \& {Visser}, R. 2011, arXiv e-prints, arXiv:1106.3917

\bibitem[{{Vasyunin} {et~al.}(2017){Vasyunin}, {Caselli}, {Dulieu}, \&
  {Jim{\'e}nez-Serra}}]{Vasyunin2017}
{Vasyunin}, A.~I., {Caselli}, P., {Dulieu}, F., \& {Jim{\'e}nez-Serra}, I.
  2017, \apj, 842, 33

\bibitem[{Vasyunin \& Herbst(2013)}]{Vasyunin2013}
Vasyunin, A.~I. \& Herbst, E. 2013, The Astrophysical Journal, 769, 34

\bibitem[{{Villadsen} {et~al.}(2022){Villadsen}, {Ligterink}, \&
  {Andersen}}]{Villadsen2022}
{Villadsen}, T., {Ligterink}, N.~F.~W., \& {Andersen}, M. 2022, \aap, 666, A45

\bibitem[{{Wakelam} {et~al.}(2017){Wakelam}, {Loison}, {Mereau}, \&
  {Ruaud}}]{Wakelam2017}
{Wakelam}, V., {Loison}, J.~C., {Mereau}, R., \& {Ruaud}, M. 2017, Molecular
  Astrophysics, 6, 22

\bibitem[{{Watanabe} \& {Kouchi}(2002)}]{Watanabe2002}
{Watanabe}, N. \& {Kouchi}, A. 2002, \apjl, 571, L173

\bibitem[{{Weaver} {et~al.}(2018){Weaver}, {Powers}, {McCabe}, \&
  {Zinga}}]{Weaver2018}
{Weaver}, S.~W., {Powers}, C.~R., {McCabe}, M.~N., \& {Zinga}, S. 2018, IAU
  Symposium, 332, 305

\bibitem[{{Westley} {et~al.}(1995){Westley}, {Baragiola}, {Johnson}, \&
  {Baratta}}]{Westley1995}
{Westley}, M.~S., {Baragiola}, R.~A., {Johnson}, R.~E., \& {Baratta}, G.~A.
  1995, \planss, 43, 1311

\bibitem[{{Woitke} {et~al.}(2009){Woitke}, {Kamp}, \& {Thi}}]{Woitke2009}
{Woitke}, P., {Kamp}, I., \& {Thi}, W.~F. 2009, \aap, 501, 383

\bibitem[{{Yang} {et~al.}(2022){Yang}, {Green}, {Pontoppidan}, {Bergner},
  {Cleeves}, {Evans}, {Garrod}, {Jin}, {Kim}, {Kim}, {Lee}, {Sakai},
  {Shingledecker}, {Shope}, {Tobin}, \& {van Dishoeck}}]{Yang2022}
{Yang}, Y.-L., {Green}, J.~D., {Pontoppidan}, K.~M., {et~al.} 2022, \apjl, 941,
  L13

\bibitem[{{Yocum} {et~al.}(2021){Yocum}, {Milam}, {Gerakines}, \& {Widicus
  Weaver}}]{Yocum2021}
{Yocum}, K.~M., {Milam}, S.~N., {Gerakines}, P.~A., \& {Widicus Weaver}, S.~L.
  2021, \apj, 913, 61

\end{thebibliography}

\begin{appendix}
\section{Photo-rate $\alpha$}
\label{ap_alpha}

The photodissociation rate is formally defined by:

\begin{equation}
    k_{phd} = \int_{\lambda_1}^{\lambda_2} \sigma_{\lambda} F_{\lambda} d\lambda
    \label{rate1}
\end{equation}
where $\sigma_{\lambda}$ is the photodissociation cross section and $F_{\lambda}$ is the flux in units of cm$^{-2}$ s$^{-2}$.

In a region shielded by dust, the photodissociation rate is given as:

\begin{equation}
    k_{phd} = \alpha \chi \mathrm{exp}\left(-\gamma A_V\right)
    \label{rate2}
\end{equation}
where $\alpha$ is the photo-rate of unshielded region, $\chi$ is the FUV strength in Draine-field, $\gamma$ is the dust attenuation factor, and $A_{\rm{V}}$ the visual extinction due to dust.

In a region without dust extinction, $A_{\rm{V}} = 0$, and from Equations~\ref{rate1} and \ref{rate2}, we have:
\begin{equation}
    \alpha \chi = \int_{\lambda_1}^{\lambda_2} \sigma_{\lambda} F_{\lambda} d\lambda.
    \label{general}
\end{equation}

In a context simulating the laboratory conditions, the flux can be given as a function of the flux lamp, namely,  $F_{\lambda} = F_{\lambda}^{lamp}$. To scale this flux with the radiation field from Draine \citep{Draine1978}, we consider the following equation: 
\begin{equation}
    \chi = \frac{\int_{91.2~nm}^{205~nm} F_{\lambda}^{\mathrm{lamp}} d\lambda}{F_{\mathrm{Draine}}}
    \label{Drfield}
\end{equation}
in which the integral boundaries cover the Lyman-$\alpha$ emission at 121~nm. In the denominator, the integrated flux from Draine is equal to 1.9921 $\times$ 10$^8$ cm$^{-2}$ s$^{-1}$ (see Equation~\ref{Draine_flux}).

Isolating the photo-rate term in Equation~\ref{general}, we have that:
\begin{equation}
    \alpha = \frac{\int_{\lambda_1}^{\lambda_2} F_{\lambda}^{lamp} \sigma_{\lambda}^{phd} d\lambda}{\int_{\lambda_1}^{\lambda_2} F_{\lambda}^{lamp} d\lambda} \cdot 1.9921 \times 10^{8} \; {\rm{s^{-1}}}
    \label{prate1}
\end{equation}

The ratio between the two intregrals in Equation~\ref{prate1} is equal to the average photodissociation cross-section ($\overline{\sigma}$), and can be rewritten as:

\begin{equation}
    \alpha = \overline{\sigma} \cdot 1.9921 \times 10^{8} \; {\rm{s^{-1}}}.
    \label{eq_photorate}
\end{equation}

\section{Chemical elements in the network}
The chemical elements, gas- and solid-phase chemical species adopted used in this study are shown in Table~\ref{all_elements}.

\label{all_elements}
\begin{table*}
\caption{Gas and solid species in the network}             
\centering                          
\begin{tabular}{lll}        
\hline\hline                 
chemical elements & H, He, C, N, O, Ne, Na, Mg, Si, S, Ar, Fe & 12  \\
pseudo elements   & *, !                                      & 2   \\
\hline
                  & \hspace{4cm} Gas species                \\
(H)               & H, H$^+$, H$^-$, H$_2$, H$_2^+$, H$_3^+$ , H$_2^{\rm{exc}}$    & 7   \\
(He)              & He, He$^+$, HeH$^+$                           & 3   \\
(C-H)             & C, C$^+$, C$^{++}$, CH, CH$^+$, CH$_2$, CH$_2^+$ CH$_3$, CH$_3^+$, CH$_4$, CH$_4^+$, CH$_5^+$  & 12  \\
(C-C)             & C$_2$, C$_2^+$, C$_2$H, C$_2$H$^+$, C$_2$H$_2$, C$_2$H$_2^+$, C$_2$H$_3$, C$_2$H$_3^+$, C$_2$H$_4$, C$_2$H$_4^+$, C$_2$H$_5$, C$_2$H$_5^+$, C$_2$H$_6$, & 15  \\
                  & C$_2$H$_6^+$ , C$_2$H$_7^+$ &     \\
(C-N)             & CN, CN$^+$, HCN, HCN$^+$, HNC, H$_2$CN, HCNH$^+$            & 7   \\
(C-O)             & CO, CO$^+$, HCO, HCO$^+$, H$_2$CO, H$_2$CO$^+$, CH$_3$O, H$_3$CO$^+$  & 12  \\
                  & CH$_3$OH$_2^+$                                         &     \\
(N-H)             & N, N$^+$, N$^{++}$, NH, NH$^+$, NH$_2$, NH$_2^+$, NH$_3$ , NH$_3^+$, NH$_4^+$ & 10  \\
(N-N)             & N$_2$ , N$_2^+$ , HN$_2^+$              & 3   \\
(N-O)             & NO, NO$^+$, NO$_2$ , NO$_2^+$ , HNO, HNO$^+$, H$_2$NO$^+$ & 7   \\
(O-H)             & O, O$^+$, O$^{++}$, OH, OH$^+$, H$_2$O, H$_2$O$^+$, H$_3$O$^+$  & 8   \\
(O-O)             & O$_2$, O$_2^+$, O$_2$H, O$_2$H$^+$, H$_2$O$_2$ & 5   \\
(S-H)             & S, S$^+$, S$^{++}$, HS, HS$^+$, H$_2$S, H$_2$S$^+$, H$_3$S$^+$ & 8   \\
(S-O)             & SO, SO$^+$, SO$_2$, SO$_2^+$ , HSO$_2^+$              & 5   \\
(S-C)             & CS, CS$^+$, HCS, HCS$^+$, H$_2$CS, H$_2$CS$^+$ , H$_3$CS$^+$  & 7   \\
(S-N)             & NS, NS$^+$, HNS$^+$       & 3   \\
(C-C-C)           & C$_3$, C$_3^+$, C$_3$H, C$_3$H$^+$, C$_3$H$_2$, C$_3$H$_2^+$ , C$_3$H$_3^+$ & 7   \\
(C-C-N)           & C$_2$N, C$_2$N$^+$, HC$_2$N, HC$_2$N$^+$, CH$_2$CN, CH$_3$CN, CH$_3$CNH$^+$ & 7   \\
(C-C-O)           & C$_2$O, C$_2$O$^+$, HC$_2$O$^+$ & 3   \\
(O-C-N)           & OCN, OCN$^+$   & 2   \\
(O-O-O)           & O$_3$                          & 1   \\
(O-C-O)           & CO$_2$, CO$_2^+$ , HCO$_2^+$, HCOOH, HCOOH$_2^+$ , HOCO   & 6   \\
(O-C-S)           & OCS, OCS$^+$, HOCS$^+$ & 3   \\
(C-C-C-C)         & C$_4$, C$_4^+$, C$_4$H$^+$ & 3   \\
(C-C-C-N)         & C$_3$N, HC$_3$N, HC$_3$N$^+$, C$_3$H$_2$N & 4   \\
(Si-H)            & Si, Si$^+$, Si$^{++}$, SiH, SiH$^+$, SiH$_2$ , SiH$_2^+$ , SiH$_3$, SiH$_3^+$ & 12  \\
(Si-O)            & SiO, SiO$^+$, SiOH$^+$ & 3   \\
(Si- )            & SiC, SiC$^+$, SiN, SiN$^+$, SiS, SiS$^+$, HCSi$^+$, HNSi$^+$, HSiS$^+$  & 9   \\
(Na)              & Na, Na$^+$, Na$^{++}$ & 3   \\
(Mg)              & Mg, Mg$^{+}$, Mg$^{++}$ & 3   \\
(Fe)              & Fe, Fe$^{+}$, Fe$^{++}$& 3   \\
(Ne)              & Ne, Ne$^{+}$, Ne$^{++}$ & 3   \\
(Ar)              & Ar, Ar$^{+}$, Ar$^{++}$ & 3   \\
\hline
                  & \hspace{4cm} Ice species                \\
(H)               & H\#, H$_2$\#  & 2   \\
(C-H)             & C\#, CH\#, CH$_2$\#, CH$_3$\#, CH$_4$\# & 5   \\
(C-C)             & C$_2$\#, C$_2$H\#, C$_2$H$_2$\#, C$_2$H$_3$\#, C$_2$H$_4$\#, C$_2$H$_5$\#, C$_2$H$_6$\#  & 7   \\
(C-N)             & CN\#, HCN\#, HNC\#, H$_2$CN\#& 4   \\
(C-C-C)           & C$_3$\#, C$_2$H\#, C$_3$H$_2$\# & 3   \\
(C-C-C-C)         & C$_4$\# & 1   \\
water             & O\#, O$_2$\#, O$_3$\#, OH\#, H$_2$O\#, O$_2$H\#, H$_2$O$_2$\#& 7   \\
methanol          & CO\#, HCO\#, H$_2$CO\#, CH$_3$O\#, CH$_2$OH\#, CH$_3$OH\# & 6   \\
(O-C-O)           & CO$_2$\#, HCOOH\#  & 2   \\
(C-C-O)           & C$_2$O\#  & 1   \\
(O-C-N)           & OCN\# & 1   \\
(N-H)             & N\#, NH\#, NH$_2$\#, NH$_3$\#, N$_2$\# & 5   \\
(N-O)             & NO\#, NO$_2$\#, HNO\# & 3   \\
(C-C-N)           & C$_2$N\#, HC$_2$N\#, CH$_2$CN\#, CH$_3$CN\# & 4   \\
(C-C-C-N)         & C$_3$N\#, HC$_3$N\#, C$_3$H$_2$N\#  & 3   \\
(S-H, S-O, S-N)   & S\#, HS\#, H$_2$S\#, SO\#, SO$_2$\#, NS\#  & 6   \\
(S-C)             & CS\#, HCS\#, H$_2$CS\#, OCS\#  & 4   \\
(Si- )            & Si\#, SiH\#, SiH$_2$\#, SiH$_3$\#, SiH$_4$\#, SiC\#, SiN\#, SiO\#, SiS\#  & 9   \\
                  & HOCO\# & 1   \\
metals            & Fe\#, Mg\#, Na\#  & 3   \\

\hline

special species   & *H\#, *H$_2$O\#, !H$_2$O\#, PAH, PAH$^-$, PAH$^+$, PAH$^{2+}$, PAH$^{3+}$, PAH\#, HPAH, HPAH$^{+}$,& 12  \\
                  & HPAH\#  &     \\
complexes         & H$_2$OH$_2$\#, H$_2$OC\#, H$_2$OCO\# & 3   \\
\hline
species           & total & 276 \\
\hline                                   
\end{tabular}
\end{table*}

\section{Ice photodissociation rates}
\label{ice_photo}
The photodissociation rates of the molecules in ices adopted in this work are shown in Table~\ref{ap_photdiss}.

\longtab[1]{
\renewcommand{\arraystretch}{1.2}
\begin{longtable}{lllllllll}
\caption{\label{ap_photdiss} Ice photodissociation rates ($\alpha$) and the dust attenuation factor ($\gamma$) for other photodissociation reactions used in Equation~\ref{rate2}.}\\
\hline
\hline
Reaction & & & & & & & $\alpha$  & $\gamma$\\
\hline
\endfirsthead
\caption{Continued.}\\
\hline
Reaction & & & & & & & $\alpha$  & $\gamma$\\
\hline
\endhead
\hline
\endfoot
\hline
\endlastfoot
\hline
CH$_2$OH\#   & + & h$\nu$ & $\rightarrow$ & CH$_2$\#   & + & OH\#          & 7.0 $\times$ 10$^{-10}$ & 2.30\\
CH$_3$O\#    & + & h$\nu$ & $\rightarrow$ & H$_2$CO\#  & + & H\#           & 1.0 $\times$ 10$^{-9}$  & 1.50\\
H$_2$CO\#    & + & h$\nu$ & $\rightarrow$ & CO\#       & + & H$_2$\#       & 7.0 $\times$ 10$^{-10}$ & 1.70\\
CO$_2$\#     & + & h$\nu$ & $\rightarrow$ & CO\#       & + & O\#           & 1.4 $\times$ 10$^{-9}$ & 2.50\\
CO\#         & + & h$\nu$ & $\rightarrow$ & C\#        & + & O\#           & 2.0 $\times$ 10$^{-10}$ & 2.50\\
CH$_4$\#     & + & h$\nu$ & $\rightarrow$ & CH\#       & + & H$_2$\# + H\# & 2.2 $\times$ 10$^{-10}$ & 2.20\\
CH$_3$\#     & + & h$\nu$ & $\rightarrow$ & CH\#       & + & H$_2$\#       & 2.5 $\times$ 10$^{-10}$ & 1.90\\
CH$_3$\#     & + & h$\nu$ & $\rightarrow$ & CH$_2$\#   & + & H\#           & 2.5 $\times$ 10$^{-10}$ & 1.90\\
CH$_2$\#     & + & h$\nu$ & $\rightarrow$ & CH\#       & + & H\#           & 8.6 $\times$ 10$^{-10}$ & 1.20\\
CH\#         & + & h$\nu$ & $\rightarrow$ & C\#        & + & H\#           & 1.0 $\times$ 10$^{-9}$ & 1.50\\
H$_2$O$_2$\# & + & h$\nu$ & $\rightarrow$ & OH\#       & + & OH\#          & 8.3 $\times$ 10$^{-10}$ & 1.80\\
H$_2$O\#     & + & h$\nu$ & $\rightarrow$ & OH\#       & + & H\#           & 5.9 $\times$ 10$^{-10}$ & 1.70\\
O$_2$H\#     & + & h$\nu$ & $\rightarrow$ & O$_2$\#    & + & H\#           & 8.0 $\times$ 10$^{-10}$ & 1.80\\
OH\#         & + & h$\nu$ & $\rightarrow$ & O\#        & + & H\#           & 3.9 $\times$ 10$^{-10}$ & 2.20\\
CN\#         & + & h$\nu$ & $\rightarrow$ & C\#        & + & N\#           & 1.1 $\times$ 10$^{-9}$ & 3.10\\
H$_2$S\#     & + & h$\nu$ & $\rightarrow$ & HS\#       & + & S\#           & 1.5 $\times$ 10$^{-9}$ & 1.90\\
H$_2$S\#     & + & h$\nu$ & $\rightarrow$ & S\#        & + & H$_2$\#       & 1.5 $\times$ 10$^{-10}$ & 1.90\\
HS\#         & + & h$\nu$ & $\rightarrow$ & H\#        & + & S\#           & 9.7 $\times$ 10$^{-10}$ & 1.40\\
SO\#         & + & h$\nu$ & $\rightarrow$ & S\#        & + & O\#           & 3.7 $\times$ 10$^{-9}$ & 2.00\\
SO$_2$\#     & + & h$\nu$ & $\rightarrow$ & SO\#       & + & O\#           & 1.9 $\times$ 10$^{-9}$ & 1.90\\
SiO\#        & + & h$\nu$ & $\rightarrow$ & Si\#       & + & O\#           & 1.0 $\times$ 10$^{-10}$ & 2.30\\
HCO\#        & + & h$\nu$ & $\rightarrow$ & CO\#       & + & H\#           & 1.1 $\times$ 10$^{-10}$ & 0.80\\
NO\#         & + & h$\nu$ & $\rightarrow$ & N\#        & + & O\#           & 4.3 $\times$ 10$^{-10}$ & 1.70\\
C$_2$H\#     & + & h$\nu$ & $\rightarrow$ & C$_2$\#    & + & H\#           & 5.1 $\times$ 10$^{-10}$ & 1.90\\
N$_2$\#      & + & h$\nu$ & $\rightarrow$ & N\#        & + & N\#           & 2.3 $\times$ 10$^{-10}$ & 3.80\\
HCN\#        & + & h$\nu$ & $\rightarrow$ & CN\#       & + & H\#           & 1.3 $\times$ 10$^{-9}$ & 2.10\\
C$_2$H$_2$\# & + & h$\nu$ & $\rightarrow$ & C$_2$H\#   & + & H\#           & 7.3 $\times$ 10$^{-10}$ & 1.80\\
O$_2$\#      & + & h$\nu$ & $\rightarrow$ & O\#        & + & O\#           & 6.9 $\times$ 10$^{-10}$ & 1.80\\
O$_3$\#      & + & h$\nu$ & $\rightarrow$ & O$_2$\#    & + & O\#           & 1.9 $\times$ 10$^{-10}$ & 1.85\\
CS\#         & + & h$\nu$ & $\rightarrow$ & C\#        & + & S\#           & 9.8 $\times$ 10$^{-10}$ & 2.40\\
H$_2$S\#     & + & h$\nu$ & $\rightarrow$ & CS\#       & + & H$_2$\#       & 1.9 $\times$ 10$^{-10}$ & 1.70\\
OCS\#        & + & h$\nu$ & $\rightarrow$ & S\#        & + & CO\#          & 3.7 $\times$ 10$^{-10}$ & 2.10\\
NS\#         & + & h$\nu$ & $\rightarrow$ & N\#        & + & S\#           & 2.0 $\times$ 10$^{-10}$ & 2.00\\
C$_3$N\#     & + & h$\nu$ & $\rightarrow$ & CN\#       & + & C$_2$\#       & 5.0 $\times$ 10$^{-10}$ & 1.80\\
HC$_3$N\#    & + & h$\nu$ & $\rightarrow$ & CN$_2$\#   & + & C$_2$H\#      & 5.6 $\times$ 10$^{-9}$  & 2.20\\
C$_2$N\#     & + & h$\nu$ & $\rightarrow$ & N\#        & + & C$_2$\#       & 5.0 $\times$ 10$^{-10}$ & 1.70\\
C$_2$N\#     & + & h$\nu$ & $\rightarrow$ & CN\#       & + & N\#           & 1.9 $\times$ 10$^{-10}$ & 1.70\\
CH$_2$CN\#   & + & h$\nu$ & $\rightarrow$ & CN\#       & + & CH$_2$\#      & 1.6 $\times$ 10$^{-9}$  & 1.90\\
CH$_3$CN\#   & + & h$\nu$ & $\rightarrow$ & CN\#       & + & CH$_3$\#      & 2.5 $\times$ 10$^{-9}$  & 2.60\\
NO$_2$\#     & + & h$\nu$ & $\rightarrow$ & O\#        & + & NO\#          & 1.4 $\times$ 10$^{-9}$ & 2.10\\
HNO\#        & + & h$\nu$ & $\rightarrow$ & NO\#       & + & H\#           & 1.7 $\times$ 10$^{-10}$ & 0.50\\
NH$_3$\#     & + & h$\nu$ & $\rightarrow$ & NH$_2$\#   & + & H\#           & 9.2 $\times$ 10$^{-10}$ & 2.10\\
NH$_2$\#     & + & h$\nu$ & $\rightarrow$ & NH\#       & + & H\#           & 7.5 $\times$ 10$^{-10}$ & 2.00\\
NH\#         & + & h$\nu$ & $\rightarrow$ & N\#        & + & H\#           & 5.0 $\times$ 10$^{-10}$ & 2.30\\
C$_2$O\#     & + & h$\nu$ & $\rightarrow$ & C$_2$\#    & + & O\#           & 5.0 $\times$ 10$^{-10}$ & 1.70\\
C$_2$O\#     & + & h$\nu$ & $\rightarrow$ & O$_2$\#    & + & C\#           & 5.0 $\times$ 10$^{-10}$ & 1.70\\
C$_4$\#      & + & h$\nu$ & $\rightarrow$ & C$_3$\#    & + & C\#           & 4.2 $\times$ 10$^{-9}$  & 1.90\\
C$_4$\#      & + & h$\nu$ & $\rightarrow$ & C$_3$H\#   & + & H\#           & 4.2 $\times$ 10$^{-9}$  & 1.90\\
C$_3$H$_2$\# & + & h$\nu$ & $\rightarrow$ & C$_3$\#    & + & C$_2$\#       & 1.4 $\times$ 10$^{-9}$  & 1.90\\
C$_3$H\#     & + & h$\nu$ & $\rightarrow$ & C$_3$\#    & + & H\#           & 1.1 $\times$ 10$^{-9}$  & 1.80\\
C$_3$\#      & + & h$\nu$ & $\rightarrow$ & C$_2$\#    & + & C\#           & 5.0 $\times$ 10$^{-9}$  & 2.10\\
HCN\#        & + & h$\nu$ & $\rightarrow$ & CN\#       & + & H$_2$\#       & 1.5 $\times$ 10$^{-9}$  & 2.10\\
H$_2$CN\#      & + & h$\nu$ & $\rightarrow$ & HCN\#         & + & H\#       & 5.5 $\times$ 10$^{-10}$ & 2.00\\
C$_2$\#        & + & h$\nu$ & $\rightarrow$ & C\#           & + & C\#       & 2.4 $\times$ 10$^{-10}$ & 2.60\\
C$_2$H$_3$\#   & + & h$\nu$ & $\rightarrow$ & C$_2$H$_2$\#  & + & H\#       & 1.9 $\times$ 10$^{-9}$  & 1.70\\
C$_2$H$_4$\#   & + & h$\nu$ & $\rightarrow$ & C$_2$H$_2$\#  & + & H$_2$\#   & 3.0 $\times$ 10$^{-9}$  & 2.10\\
C$_2$H$_5$\#   & + & h$\nu$ & $\rightarrow$ & C$_2$H$_3$\#  & + & H$_2$\#   & 1.0 $\times$ 10$^{-9}$  & 1.70\\
C$_2$H$_6$\#   & + & h$\nu$ & $\rightarrow$ & C$_2$H$_5$\#  & + & H\#       & 2.1 $\times$ 10$^{-9}$  & 2.94\\
H$_2$\#        & + & h$\nu$ & $\rightarrow$ & H\#           & + & H\#       & 5.7 $\times$ 10$^{-11}$ & 4.18\\
SiH$_2$\#      & + & h$\nu$ & $\rightarrow$ & SiH\#         & + & H\#       & 5.0 $\times$ 10$^{-11}$ & 1.70\\
SiH$_3$\#      & + & h$\nu$ & $\rightarrow$ & SiH\#         & + & H$_2$\#   & 3.0 $\times$ 10$^{-10}$ & 1.70\\
SiH$_3$\#      & + & h$\nu$ & $\rightarrow$ & SiH$_2$\#     & + & H\#       & 3.0 $\times$ 10$^{-10}$ & 1.70\\
SiH$_4$\#      & + & h$\nu$ & $\rightarrow$ & SiH\#         & + & H\# + H\# & 1.6 $\times$ 10$^{-10}$ & 2.20\\
SiH$_4$\#      & + & h$\nu$ & $\rightarrow$ & SiH$_2$\#     & + & H$_2$\#   & 4.8 $\times$ 10$^{-10}$ & 2.20\\
SiH$_4$\#      & + & h$\nu$ & $\rightarrow$ & SiH$_3$\#     & + & H\#       & 1.6 $\times$ 10$^{-10}$ & 2.20\\
\end{longtable}
}

\section{Bimolecular surface reactions rates and binding energies}
\label{bimol}
The bimolecular surface reaction rates used in this paper are shown in Table~\ref{ap_suf_reac} and the binding energies are shown in Table~\ref{ap_bindings}.

\longtab[1]{
\renewcommand{\arraystretch}{1.2}
\begin{longtable}{lllllllll}
\caption{\label{ap_suf_reac} List of bimolecular surface reactions used in this paper and their respective energies barrier ($\Delta E$). See Equation~\ref{eq_diff}.}\\
\hline
\hline
Reaction & & & & & & & $\Delta E$ (K) & Note/Reference\\
\hline
\endfirsthead
\caption{Continued.}\\
\hline
Reaction & & & & & & & $\Delta E$ (K) & Note/Reference\\
\hline
\endhead
\hline
\endfoot
\hline
\endlastfoot
\multicolumn{9}{c}{\bf{Methanol-related reactions}}\\
\hline
CH$_3$OH\#   & + & OH\# & $\rightarrow$ & CH$_2$OH\#   & + & H$_2$O\#          & 1000 & \citet{Esplugues2016}\\
H\#   & + & CH$_2$OH\# & $\rightarrow$ & CH$_3$OH\#   &  &    & 0 & \citet{Hasegawa1993}\\
H\#   & + & CH$_3$O\# & $\rightarrow$ & CH$_3$OH\#   &  &    & 0 & \citet{Esplugues2016}\\
OH\#   & + & CH$_3$\# & $\rightarrow$ & CH$_3$OH\#   &  &    & 0 & KIDA\\
H\#   & + & CH$_2$OH\# & $\rightarrow$ & CH$_3$OH   &  &    & 0 & \citet{Hasegawa1993}\\
H\#   & + & CH$_3$O\# & $\rightarrow$ & CH$_3$OH   &  &    & 0 & \citet{Esplugues2016}\\
OH\#   & + & CH$_3$\# & $\rightarrow$ & CH$_3$OH   &  &    & 0 & KIDA\\
\hline
\multicolumn{9}{c}{\bf{Other reactions}}\\
\hline
H\#   & + & H\# & $\rightarrow$ & H$_2$\#   &  &    & 0 & \citet{Hasegawa1993}\\
H\#&+&C\#&$\rightarrow$&CH\#& & &0&\citet{Hasegawa1993}\\
H\#&+&O\#&$\rightarrow$&OH\#& & &0&\citet{Hasegawa1993}\\
H\#&+&CH\#&$\rightarrow$&CH$_2$\#& & &0&\citet{Hasegawa1993}\\
H\#&+&OH\#&$\rightarrow$&H$_2$O\#& & &0&\citet{Hasegawa1993}, \citet{Ioppolo2008}\\
H\#&+&C$_2$\#&$\rightarrow$&C$_2$H\#& & &0&\citet{Hasegawa1993}\\
H\#&+&C$_2$H\#&$\rightarrow$&C$_2$H$_2$& & &0&\citet{Hasegawa1993}\\
H\#&+&O$_2$\#&$\rightarrow$&O$_2$H\#& & &1200&\citet{Hasegawa1993} \\
H\#&+&CO\#&$\rightarrow$&HCO\#& & & 520 & \citet{Fuchs2009}\\
H\#&+&CO\#&$\rightarrow$&HCO\#& & & 1106 & \citet{Rimola2014}\\
H\#&+&CO\#&$\rightarrow$&HCO\#& & & 2000 & \citet{Awad2005}\\
H\#&+&CO\#&$\rightarrow$&HOC\#& & &1106&\citet{Rimola2014}\\
H\#&+&CH$_2$\#&$\rightarrow$&CH$_3$\#& & &0&\citet{Hasegawa1993}\\
H\#&+&CO$_2$\#&$\rightarrow$&CO\#&+&OH\#&10000&\citet{Esplugues2016} \\
H\#&+&O$_2$H\#&$\rightarrow$&H$_2$O$_2$\#& & &0&\citet{Chaabouni2012}\\
H\#&+&O$_2$H\#&$\rightarrow$&OH\#&+&OH\#&0&\citet{Esplugues2016}\\
H\#&+&O$_3$\#&$\rightarrow$&O$_2$\#&+&OH\#&480&\citet{Esplugues2016}\\
H\#&+&C$_3$\#&$\rightarrow$&C$_3$H\#& & &0&\citet{Hasegawa1993}\\
H\#&+&HOC\#&$\rightarrow$&CHOH\#& & &0&\citet{Hasegawa1993}\\
H\#&+&HCO\#&$\rightarrow$&H$_2$CO\#& & &0&\citet{Hasegawa1993}\\
H\#&+&HCO\#&$\rightarrow$&H$_2$CO& & &0&\citet{Hasegawa1993}\\
H\#&+&HCO\#&$\rightarrow$&CO\#&+&H$_2$\#&0&\citet{Minissale2016}\\
H\#&+&H$_2$O$_2$\#&$\rightarrow$&H$_2$O\#&+&OH\#&1000&\citet{Esplugues2016}\\
H\#&+&H$_2$O$_2$\#&$\rightarrow$&H$_2$\#&+&O$_2$H\#&1900&\citet{Lamberts2016}\\
H\#&+&CH$_3$\#&$\rightarrow$&CH$_4$\#& & &0&\citet{Hasegawa1993}\\
H\#&+&CH$_3$\#&$\rightarrow$&CH$_4$& & &0&\citet{Hasegawa1993}\\
H\#&+&C$_3$H\#&$\rightarrow$&C$_3$H$_2$\#& & &0&\citet{Hasegawa1993}\\
H\#&+&C$_4$\#&$\rightarrow$&C$_4$H\#& & &0&\citet{Hasegawa1993}\\
H\#&+&C$_2$H$_2$\#&$\rightarrow$&C$_2$H$_3$\#& & &1200&\citet{Hasegawa1993}\\
H\#&+&CHOH\#&$\rightarrow$&CH$_2$OH\#& & &0&\citet{Hasegawa1993}\\
H\#&+&CHOH\#&$\rightarrow$&CH$_2$OH& & &0&\citet{Hasegawa1993}\\
H\#&+&H$_2$CO\#&$\rightarrow$&CH$_2$OH\#& & &5660& \citet{Skouteris2018}\\
H\#&+&H$_2$CO\#&$\rightarrow$&CH$_2$OH& & &5660& \citet{Skouteris2018}\\
H\#&+&H$_2$CO\#&$\rightarrow$&CH$_3$O\#& & &1890&\citet{Skouteris2018}\\
H\#&+&H$_2$CO\#&$\rightarrow$&CH$_3$O& & &1890&\citet{Skouteris2018}\\
H\#&+&H$_2$CO\#&$\rightarrow$&HCO\#&+&H$_2$\#&3030&\citet{Skouteris2018}\\
H\#&+&HOCO\#&$\rightarrow$&HCOOH\#& & &0&KIDA\\
H\#&+&HOCO\#&$\rightarrow$&HCOOH& & &0&KIDA\\
H\#&+&C$_4$H\#&$\rightarrow$&C$_4$H$_2$\#& & &0&\citet{Hasegawa1993}\\
H\#&+&C$_5$\#&$\rightarrow$&C$_5$H\#& & &0&\citet{Hasegawa1993}\\
H\#&+&C$_5$\#&$\rightarrow$&C$_5$H& & &0&\citet{Hasegawa1993}\\
H\#&+&C$_2$H$_3$\#&$\rightarrow$&C$_2$H$_4$\#& & &0&\citet{Hasegawa1993}\\
H\#&+&C$_2$H$_3$\#&$\rightarrow$&C$_2$H$_4$& & &0&\citet{Hasegawa1993}\\
H\#&+&C$_3$H$_2$\#&$\rightarrow$&C$_3$H$_3$\#& & &1210&\citet{Hasegawa1993}\\
H\#&+&C$_3$H$_2$\#&$\rightarrow$&C$_3$H$_3$& & &1210&\citet{Hasegawa1993}\\
H\#&+&CH$_3$O\#&$\rightarrow$&H$_2$CO\#&+&H$_2$\#&850&\citet{Li2004}\\
H\#&+&C$_2$H$_4$\#&$\rightarrow$&C$_2$H$_5$\#& & &1570&\citet{Hasegawa1993}\\
H\#&+&C$_2$H$_4$\#&$\rightarrow$&C$_2$H$_5$& & &1570&\citet{Hasegawa1993}\\
H\#&+&C$_3$H$_3$\#&$\rightarrow$&C$_3$H$_4$\#& & &0&\citet{Hasegawa1993}\\
H\#&+&C$_4$H$_2$\#&$\rightarrow$&C$_4$H$_3$\#& & &1200&\citet{Hasegawa1993}\\
H\#&+&C$_5$H\#&$\rightarrow$&C$_5$H$_2$\#& & &0&\citet{Hasegawa1993}\\
H\#&+&C$_6$\#&$\rightarrow$&C$_6$H\#& & &0&\citet{Hasegawa1993}\\
H\#&+&C$_2$H$_5$\#&$\rightarrow$&C$_2$H$_6$\#& & &0&\citet{Hasegawa1993}\\
H\#&+&C$_4$H$_3$\#&$\rightarrow$&C$_4$H$_4$\#& & &0&\citet{Hasegawa1993}\\
H\#&+&C$_5$H$_2$\#&$\rightarrow$&C$_5$H$_3$\#& & &1210&\citet{Hasegawa1993}\\
H\#&+&C$_6$H\#&$\rightarrow$&C$_6$H$_2$\#& & &0&\citet{Hasegawa1993}\\
H\#&+&C7\#&$\rightarrow$&C$_7$H\#& & &0&\citet{Hasegawa1993}\\
H\#&+&C$_5$H$_3$\#&$\rightarrow$&C$_5$H$_4$\#& & &0&\citet{Hasegawa1993}\\
H\#&+&C$_6$H$_2$\#&$\rightarrow$&C$_6$H$_3$\#& & &1210&\citet{Hasegawa1993}\\
H\#&+&C$_7$H\#&$\rightarrow$&C$_7$H$_2$\#& & &0&\citet{Hasegawa1993}\\
H\#&+&C$_5$H$_3$\#&$\rightarrow$&C$_5$H$_4$\#& & &0&\citet{Hasegawa1993}\\
H\#&+&C$_6$H$_2$\#&$\rightarrow$&C$_6$H$_3$\#& & &1210&\citet{Hasegawa1993}\\
H\#&+&C$_7$H$_2$\#&$\rightarrow$&C7H$_3$\#& & &1210&\citet{Hasegawa1993}\\
H\#&+&C$_8$\#&$\rightarrow$&C$_8$H\#& & &0&\citet{Hasegawa1993}\\
H\#&+&C$_6$H$_3$\#&$\rightarrow$&C$_6$H$_4$\#& & &0&\citet{Hasegawa1993}\\
H\#&+&C$_7$H$_2$\#&$\rightarrow$&C$_7$H$_3$\#& & &1210&\citet{Hasegawa1993}\\
H\#&+&C$_8$H\#&$\rightarrow$&C$_8$H$_2$\#& & &0&\citet{Hasegawa1993}\\
H\#&+&C$_9$\#&$\rightarrow$&C$_9$H\#& & &0&\citet{Hasegawa1993}\\
H\#&+&C$_7$H$_3$\#&$\rightarrow$&C$_7$H$_4$\#& & &0&\citet{Hasegawa1993}\\
H\#&+&C$_8$H$_2$\#&$\rightarrow$&C$_8$H$_3$\#& & &1210&\citet{Hasegawa1993}\\
H\#&+&C$_9$H\#&$\rightarrow$&C$_9$H$_2$\#& & &0&\citet{Hasegawa1993}\\
H\#&+&C$_8$H$_3$\#&$\rightarrow$&C$_8$H$_4$\#& & &0&\citet{Hasegawa1993}\\
H\#&+&C$_9$H$_2$\#&$\rightarrow$&C$_9$H$_3$\#& & &1210&\citet{Hasegawa1993}\\
H\#&+&C$_9$H$_3$\#&$\rightarrow$&C$_9$H$_4$\#& & &1210&\citet{Hasegawa1993}\\
H\#&+&H$_2$C$_2$O$_2$\#&$\rightarrow$&H$_3$C$_2$O$_2$\#& & &0& \citet{Paardekooper2016}\\
H\#&+&H$_3$C$_2$O$_2$\#&$\rightarrow$&H$_4$C$_2$O$_2$\#& & &0&\citet{Paardekooper2016}\\
H\#&+&H$_4$C$_2$O$_2$\#&$\rightarrow$&H$_5$C$_2$O$_2$\#& & &0&\citet{Paardekooper2016}\\
H\#&+&H$_5$C$_2$O$_2$\#&$\rightarrow$&H$_6$C$_2$O$_2$\#& & &0&\citet{Paardekooper2016}\\
H\#&+&H$_4$C$_2$O$_2$\#&$\rightarrow$&H$_5$C$_2$O$_2$\#& & &0&\citet{Paardekooper2016}\\
H\#&+&H$_5$C$_2$O$_2$\#&$\rightarrow$&H$_6$C$_2$O$_2$\#& & &0&\citet{Paardekooper2016}\\
C\#&+&C\#&$\rightarrow$&C$_2$\#& & &161&\citet{Hasegawa1993}\\
C\#&+&C\#&$\rightarrow$&C$_2$& & &161&\citet{Hasegawa1993}\\
C\#&+&O\#&$\rightarrow$&CO\#& & &0&\citet{Hasegawa1993}\\
C\#&+&O\#&$\rightarrow$&CO& & &0&\citet{Hasegawa1993}\\
C\#&+&C$_2$\#&$\rightarrow$&C$_3$\#& & &0&\citet{Hasegawa1993}\\
C\#&+&C$_2$\#&$\rightarrow$&C$_3$& & &0&\citet{Hasegawa1993}\\
C\#&+&OH\#&$\rightarrow$&HOC\#& & &0&\citet{Hasegawa1993}\\
C\#&+&OH\#&$\rightarrow$&HOC& & &0&\citet{Hasegawa1993}\\
C\#&+&O$_2$\#&$\rightarrow$&CO\#&+&O\#&0&\citet{Hasegawa1993}\\
C\#&+&CH$_2$\#&$\rightarrow$&C$_2$H$_2$\#& & &0&\citet{Hasegawa1993}\\
C\#&+&C$_2$H\#&$\rightarrow$&C$_3$H\#& & &0&\citet{Hasegawa1993}\\
C\#&+&CH$_3$\#&$\rightarrow$&C$_2$H$_3$\#& & &0&\citet{Hasegawa1993}\\
C\#&+&C$_3$\#&$\rightarrow$&C$_4$\#& & &0&\citet{Hasegawa1993}\\
C\#&+&C$_3$\#&$\rightarrow$&C$_4$& & &0&\citet{Hasegawa1993}\\
C\#&+&C$_3$H\#&$\rightarrow$&C$_4$H\#& & &0&\citet{Hasegawa1993}\\
C\#&+&C$_4$\#&$\rightarrow$&C$_5$\#& & &0&\citet{Hasegawa1993}\\
C\#&+&C$_5$\#&$\rightarrow$&C$_6$\#& & &0&\citet{Hasegawa1993}\\
C\#&+&C$_2$H$_3$\#&$\rightarrow$&CH$_2$C$_2$H\#& & &1210&KIDA\\
C\#&+&C$_2$H$_3$\#&$\rightarrow$&CH$_2$C$_2$H& & &1210&KIDA\\
C\#&+&C$_6$\#&$\rightarrow$&C$_7$\#& & &0&\citet{Hasegawa1993}\\
C\#&+&C$_7$\#&$\rightarrow$&C$_8$\#& & &0&\citet{Hasegawa1993}\\
C\#&+&C$_8$\#&$\rightarrow$&C$_9$\#& & &0&\citet{Hasegawa1993}\\
O\#&+&O\#&$\rightarrow$&O$_2$\#& & &0&\citet{Hasegawa1993}\\
O\#&+&O\#&$\rightarrow$&O$_2$& & &0&\citet{Hasegawa1993}\\
O\#&+&H$_2$\#&$\rightarrow$&OH\#&+&H\#&2040&\citet{Lamberts2014}\\
O\#&+&CH\#&$\rightarrow$&HCO\#& & &0&\citet{Hasegawa1993}\\
O\#&+&CH\#&$\rightarrow$&HCO& & &0&\citet{Hasegawa1993}\\
O\#&+&CH\#&$\rightarrow$&CO\#&+&H\#&0&\citet{Hasegawa1993}\\
O\#&+&OH\#&$\rightarrow$&O$_2$H\#& & &0&\citet{Hasegawa1993}\\
O\#&+&OH\#&$\rightarrow$&O$_2$H& & &0&\citet{Hasegawa1993}\\
O\#&+&OH\#&$\rightarrow$&O$_2$\#&+&H\#&0&\citet{Hasegawa1993}\\
O\#&+&CO\#&$\rightarrow$&CO$_2$\#& & &650&\citet{Raut2011}\\
O\#&+&CO\#&$\rightarrow$&CO$_2$& & &650&\citet{Goumans2008}\\
O\#&+&O$_2$\#&$\rightarrow$&O$_3$\#& & &0&\citet{Hasegawa1993}\\
O\#&+&O$_2$\#&$\rightarrow$&O$_3$& & &0&\citet{Hasegawa1993}\\
O\#&+&O$_2$H\#&$\rightarrow$&O$_2$\#&+&OH\#&0&\citet{Hasegawa1993}\\
O\#&+&CH$_2$\#&$\rightarrow$&H$_2$CO\#& & &0&\citet{Hasegawa1993}\\
O\#&+&CH$_2$\#&$\rightarrow$&H$_2$CO& & &0&\citet{Hasegawa1993}\\
O\#&+&C$_3$\#&$\rightarrow$&C$_3$O\#& & &0&\citet{Hasegawa1993}\\
O\#&+&HCO\#&$\rightarrow$&CO$_2$\#&+&H\#&0&\citet{Hasegawa1993, Esplugues2016}\\
O\#&+&HOCO\#&$\rightarrow$&CO$_2$\#&+&OH\#&0&KIDA\\
O\#&+&H$_2$CO\#&$\rightarrow$&CO$_2$\#&+&H$_2$\#&335&\citet{Minissale2015}\\
O\#&+&CH$_3$O\#&$\rightarrow$&CH$_3$\#&+&O$_2$\#&240& \citet{Hasegawa1993}\\
CH\#&+&CH\#&$\rightarrow$&C$_2$H$_2$\#& & &0&\citet{Hasegawa1993}\\
CH\#&+&OH\#&$\rightarrow$&CHOH\#& & &0&\citet{Hasegawa1993}\\
CH\#&+&C$_2$\#&$\rightarrow$&C$_3$H\#& & &0&\citet{Hasegawa1993}\\
CH\#&+&CH$_3$\#&$\rightarrow$&C$_2$H$_4$\#& & &0&\citet{Hasegawa1993}\\
CH\#&+&C$_3$\#&$\rightarrow$&C$_4$H\#& & &0&\citet{Hasegawa1993}\\
CH\#&+&C$_4$\#&$\rightarrow$&C$_5$H\#& & &0&\citet{Hasegawa1993}\\
CH\#&+&C$_5$\#&$\rightarrow$&C$_6$H\#& & &0&\citet{Hasegawa1993}\\
CH\#&+&C$_6$\#&$\rightarrow$&C$_7$H\#& & &0&\citet{Hasegawa1993}\\
CH\#&+&C7\#&$\rightarrow$&C$_8$H\#& & &0&\citet{Hasegawa1993}\\
CH\#&+&C8\#&$\rightarrow$&C$_9$H\#& & &0&\citet{Hasegawa1993}\\
C$_2$H\#&+&CH\#&$\rightarrow$&C$_3$H$_2$\#& & &0&\citet{Hasegawa1993}\\
C$_2$H\#&+&CH\#&$\rightarrow$&C$_3$H$_2$& & &0&\citet{Hasegawa1993}\\
CH$_3$\#&+&CH$_3$\#&$\rightarrow$&C$_2$H$_6$\#& & &0&KIDA\\
H$_2$\#&+&OH\#&$\rightarrow$&H$_2$O\#&+&H\#&2100&\citet{Oba2012}\\
H$_2$\#&+&CH\#&$\rightarrow$&CH$_3$\#& & &116&NIST\\
H$_2$\#&+&CH\#&$\rightarrow$&CH$_3$& & &116&NIST\\
H$_2$\#&+&H$_2$CO\#&$\rightarrow$&CH$_3$OH\#& & &35119&NIST\\
H$_2$\#&+&C$_2$H$_3$\#&$\rightarrow$&C$_2$H$_4$\#&+&H\#&2646&\citet{Fahr1995}, \citet{Lara2014}\\
H$_2$\#&+&C$_2$H\#&$\rightarrow$&C$_2$H$_2$\#&+&H\#&1443&\citet{Opansky1996}, \citet{Lara2014}\\
H$_2$\#&+&C$_2$\#&$\rightarrow$&C$_2$H\#&+&H\#&1469&\citet{Lara2014}\\
H$_2$\#&+&C$_4$H\#&$\rightarrow$&C$_4$H$_2$\#&+&H\#&998&\citet{Lavvas2008}, \citet{Lara2014}\\
OH\#&+&OH\#&$\rightarrow$&H$_2$O$_2$\#& & &0&\citet{Esplugues2016}\\
OH\#&+&OH\#&$\rightarrow$&H$_2$O$_2$& & &0&\citet{Esplugues2016}\\
OH\#&+&CH$_2$\#&$\rightarrow$&CH$_2$OH\#& & &0&\citet{Hasegawa1993}\\
OH\#&+&CH$_4$\#&$\rightarrow$&CH$_3$\#&+&H$_2$O\#&1160&\citet{Qasim2018}\\
OH\#&+&CO\#&$\rightarrow$&HOCO\#& & &150&\citet{Ruaud2015}\\
OH\#&+&CO\#&$\rightarrow$&HOCO& & &150&\citet{Ruaud2015}\\
OH\#&+&CO\#&$\rightarrow$&CO$_2$\#&+&H\#&150&\citet{Ruaud2015}\\
H$_2$CO\#&+&OH\#&$\rightarrow$&HCO\#&+&H$_2$O\#&0&KIDA\\
H$_2$CO\#&+&OH\#&$\rightarrow$&HCO&+&H$_2$O\#&0&KIDA\\
CH$_3$OH\#&+&OH\#&$\rightarrow$&CH$_2$OH&+&H$_2$O\#&1000&\citet{Esplugues2016}\\
H$_2$\#&+&H$_2$O\#&$\rightarrow$&H$_2$OH$_2$\#& & &0& Guess\\
H$_2$OCO\#&+&OH\#&$\rightarrow$&H$_2$O\#&+&CO\#$_2$ + H\#&150&Guess\\
H$_2$OCO\#&+&O\#&$\rightarrow$&H$_2$O\#&+&CO$_2$\#&650&Guess\\
H$_2$OCO\#&+&H\#&$\rightarrow$&H$_2$O\#&+&HCO\#&0&Guess\\
H$_2$OCO\#&+&CH\#&$\rightarrow$&CH$_2$OH\#&+&CO\#&0&Guess\\
H$_2$OCO\#&+&CH\#&$\rightarrow$&H$_2$CO\#&+&CO\# + H\#&0&Guess\\
H$_2$OCO\#&+&C\#&$\rightarrow$&CH$_2$O\#&+&CO\#&0&Guess\\
H$_2$OCO\#&+&OH\#&$\rightarrow$&H$_2$O\#&+&HOCO\#&0&Guess\\
H$_2$OCO\#&+&CH$_2$\#&$\rightarrow$&H$_2$O\#&+&H$_2$CO\# + C\#&0&Guess\\
CO\#&+&H$_2$O\#&$\rightarrow$&H$_2$OCO\#& & &0&Guess\\
C\#&+&H$_2$O\#&$\rightarrow$&H$_2$OC\#& & &0&Guess\\
CH$_2$OH\#&+&O\#&$\rightarrow$&HCOOH\#&+&H\#&0&Guess\\
HCO\#&+&OH\#&$\rightarrow$&HCOOH\#& & &0&Guess\\
HCO\#&+&OH\#&$\rightarrow$&HCOOH& & &0&Guess\\
HOCO\#&+&H\#&$\rightarrow$&HCOOH\#& & &0&Guess\\
HOCO\#&+&H\#&$\rightarrow$&HCOOH& & &0&Guess\\
\end{longtable}
}

\longtab[2]{
\renewcommand{\arraystretch}{1.2}
\begin{longtable}{lll}
\caption{\label{ap_bindings} List of binding energies.}\\
\hline
\hline
Species & $E_b$ (K) & Reference\\
\hline
\endfirsthead
\caption{Continued.}\\
\hline
Species & $E_b$ (K) & Reference\\
\hline
\endhead
\hline
\endfoot
\hline
\endlastfoot
\hline
H&600&\citet{Cazaux2002}\\
H$_2$&430&\citet{Garrod2006}\\
He&100&\citet{Tielens1982}\\
C&10000&\citet{Wakelam2017}\\
CH&925&\citet{Garrod2006}\\
CH$_2$&1050&\citet{Garrod2006}\\
CH$_3$&1175&\citet{Garrod2006}\\
CH$_4$&1090&\citet{Herrero2010}\\
O&1600&\citet{Wakelam2017}\\
OH&4600&\citet{Minissale2016}\\
H$_2$O&4800&\citet{Brown2007}\\
C$_2$&1600&\citet{Garrod2006}\\
C$_2$H&2137&\citet{Garrod2006}\\
C$_2$H$_2$&2587&\citet{Collings2004}\\
C$_2$H$_3$&3037&\citet{Garrod2006}\\
C$_2$H$_4$&3487&\citet{Garrod2006}\\
CO&1300&\citet{Minissale2016}\\
C$_2$H$_5$&3937&\citet{Garrod2006}\\
HCO&1600&\citet{Garrod2006}\\
HOC&3650&\citet{Garrod2006}\\
C$_2$H$_6$&2300&\citet{Oberg2009}\\
CHOH&4634&\citet{Garrod2006}\\
H$_2$CO&4500&\citet{Wakelam2017}\\
HOCH&1910&\citet{Hasegawa1993}\\
CH$_2$OH&4400&\citet{Wakelam2017}\\
CH$_3$O&4400&\citet{Wakelam2017}\\
CH$_3$OH&5000&\citet{Wakelam2017}\\
O$_2$&1000&\citet{Garrod2006}\\
O$_2$H&3650&\citet{Garrod2006}\\
H$_2$O$_2$&5700&\citet{Garrod2006}\\
C$_3$&2400&\citet{Garrod2006}\\
C$_3$H&2937&\citet{Garrod2006}\\
C$_3$H$_2$&3387&\citet{Garrod2006}\\
H$_2$CCC&2110&\citet{Hasegawa1993}\\
C$_3$H$_3$&3837&\citet{Garrod2006}\\
C$_2$O&1950&\citet{Garrod2006}\\
C$_3$H$_4$&4287&\citet{Garrod2006}\\
CH$_3$CCH&2470&\citet{Hasegawa1993}\\
C$_3$H$_5$&4737&\citet{Garrod2006}\\
HC$_2$O&2400&\citet{Garrod2006}\\
C$_3$H$_6$&5187&\citet{Garrod2006}\\
CH$_2$CO&2200&\citet{Garrod2006}\\
C$_3$H7&5637&\citet{Garrod2006}\\
C$_3$H8&6087&\citet{Garrod2006}\\
CH$_3$CHO&3800&\citet{Oberg2009}\\
CO$_2$&2990&\citet{Edridge2010}\\
C$_2$H$_5$OH&5200&\citet{Oberg2009}\\
CH$_3$OCH$_3$&3300&\citet{Oberg2009}\\
HCOOH&5000&\citet{Oberg2009}\\
C$_4$&3200&\citet{Garrod2006}\\
O$_3$&1800&\citet{Garrod2006}\\
C$_4$H&3737&\citet{Garrod2006}\\
C$_4$H$_2$&4187&\citet{Garrod2006}\\
C$_4$H$_3$&4637&\citet{Garrod2006}\\
C$_3$O&2750&\citet{Garrod2006}\\
C$_4$H$_4$&5087&\citet{Garrod2006}\\
C$_4$H$_5$&5537&\citet{Garrod2006}\\
HC$_3$O&3200&\citet{Garrod2006}\\
C$_4$H$_6$&5987&\citet{Garrod2006}\\
H$_2$C$_3$O&3650&\citet{Garrod2006}\\
CH$_3$OCH$_3$&3300&\citet{Oberg2009}\\
C$_5$&4000&\citet{Garrod2006}\\
HCOOCH$_3$&4000&\citet{Oberg2009}\\
C$_5$H&4537&\citet{Garrod2006}\\
C$_5$H$_2$&4987&\citet{Garrod2006}\\
C$_5$H$_3$&5437&\citet{Garrod2006}\\
CH$_3$C$_4$H&5887&\citet{Garrod2006}\\
C$_6$&4800&\citet{Garrod2006}\\
C$_6$H&5337&\citet{Garrod2006}\\
C$_6$H$_2$&5787&\citet{Garrod2006}\\
C$_6$H$_3$&6237&\citet{Garrod2006}\\
C$_6$H$_4$&6687&\citet{Garrod2006}\\
C$_6$H$_6$&7587&\citet{Garrod2006}\\
C$_7$&5600&\citet{Garrod2006}\\
C$_7$H&6137&\citet{Garrod2006}\\
C$_7$H$_2$&6587&\citet{Garrod2006}\\
C$_7$H$_3$&7037&\citet{Garrod2006}\\
C$_7$H$_4$&7487&\citet{Garrod2006}\\
CH$_3$C$_6$H&7487&\citet{Garrod2006}\\
C$_8$&6400&\citet{Garrod2006}\\
C$_8$H&6937&\citet{Garrod2006}\\
C$_8$H$_2$&7387&\citet{Garrod2006}\\
C$_8$H$_3$&7837&\citet{Garrod2006}\\
C$_8$H$_4$&8287&\citet{Garrod2006}\\
C$_9$&7200&\citet{Garrod2006}\\
C$_9$H&7737&\citet{Garrod2006}\\
C$_9$H$_2$&8187&\citet{Garrod2006}\\
C$_9$H$_3$&8637&\citet{Garrod2006}\\
C$_9$H$_4$&9087&\citet{Garrod2006}\\
C$_{10}$&8000&\citet{Garrod2006}\\
\end{longtable}
}

\section{Gallery of fits at 20~K, 30~K, 50~K and 70~K}
\label{Ap_all_models}

\begin{figure}
   \centering
   \includegraphics[width=9cm]{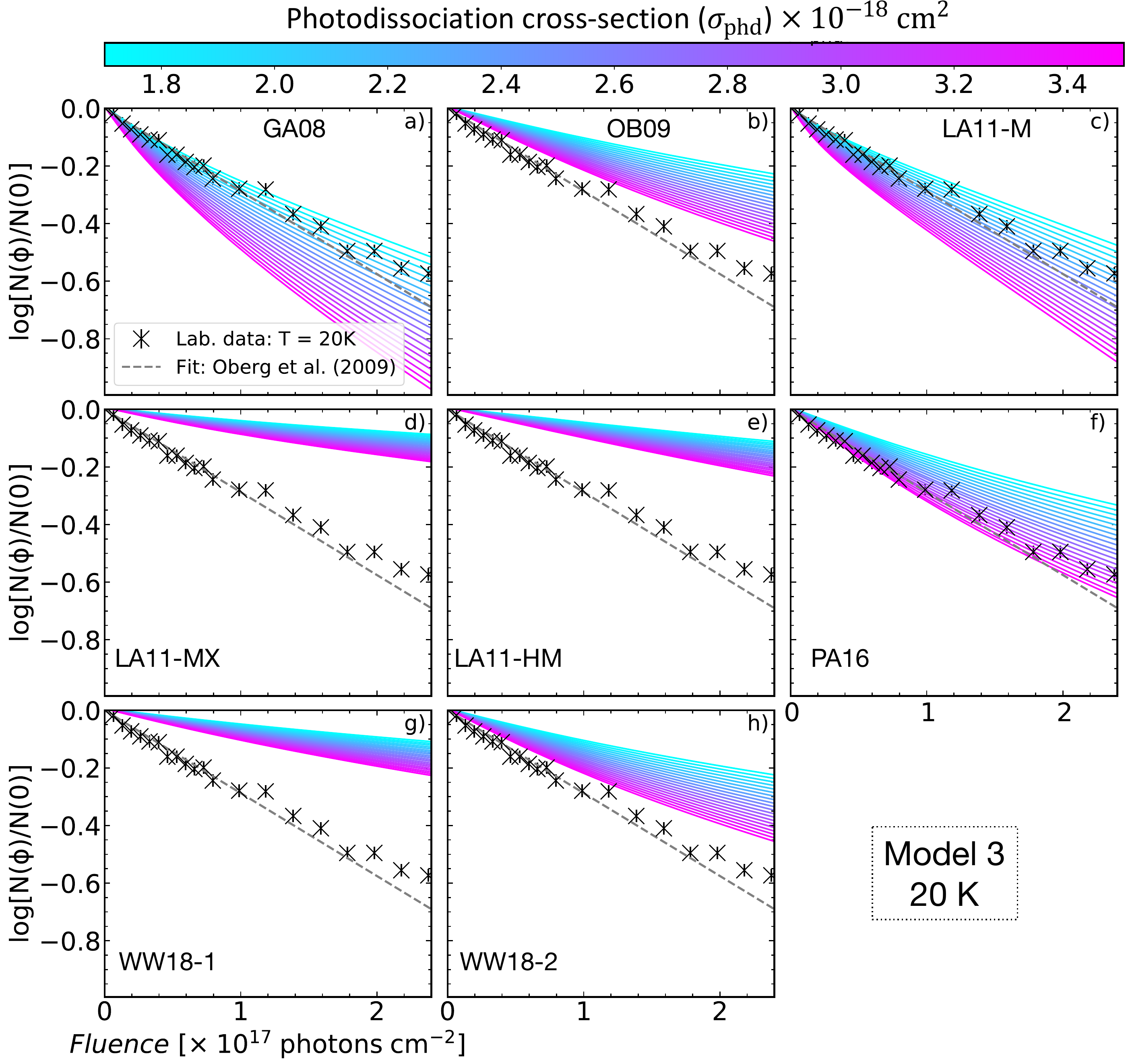}
      \caption{Grid of methanol ice photodissociation models at 20~K assuming different photodissociation cross-sections and BRs. The Model~3 in Table~\ref{t7} is adopted in these plots.}
         \label{M3}
   \end{figure}

\begin{figure}
   \centering
   \includegraphics[width=9cm]{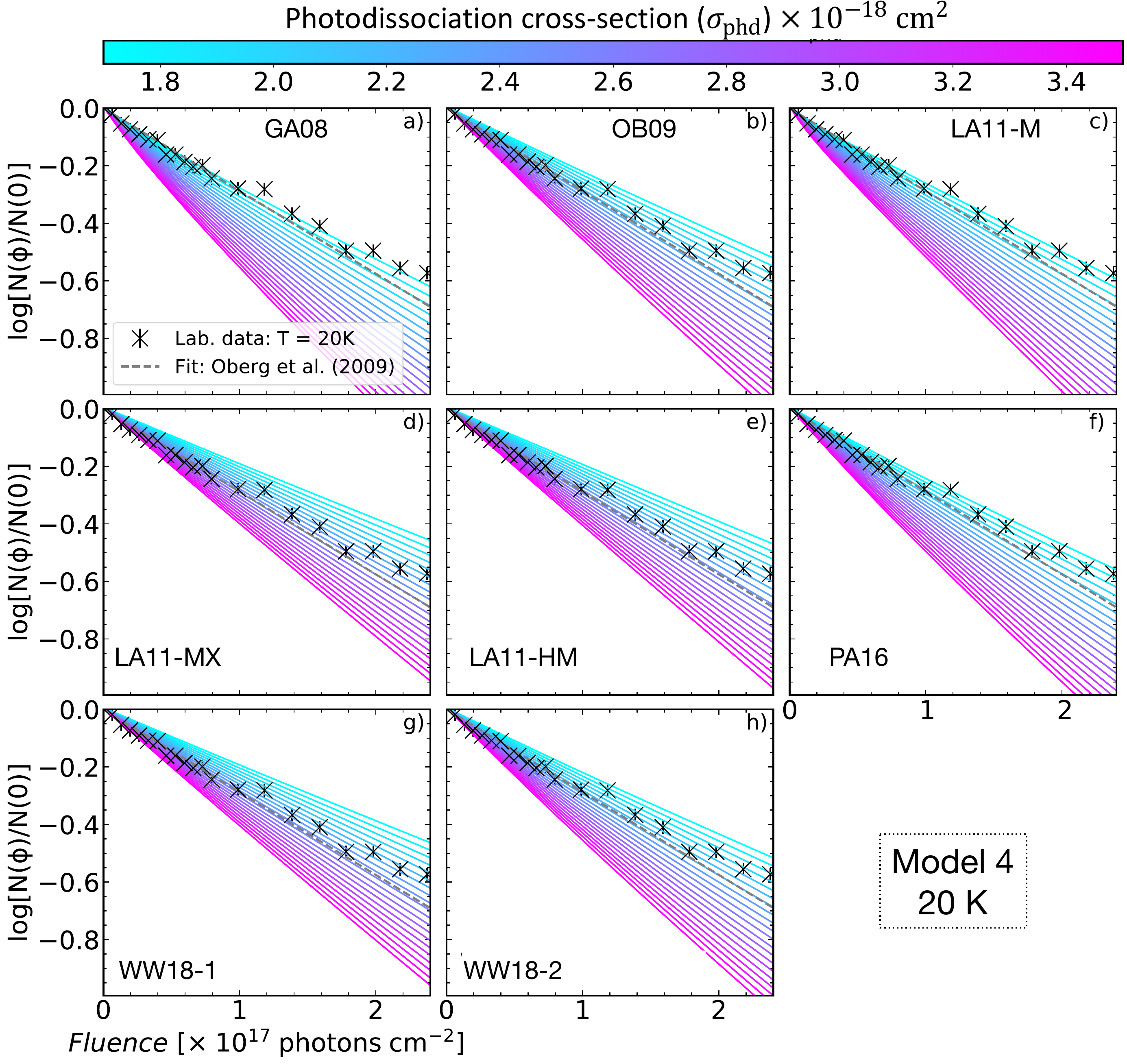}
      \caption{Same as Figure~\ref{M3}, but adopting Model~4.}
         \label{M4}
   \end{figure}

\begin{figure}
   \centering
   \includegraphics[width=9cm]{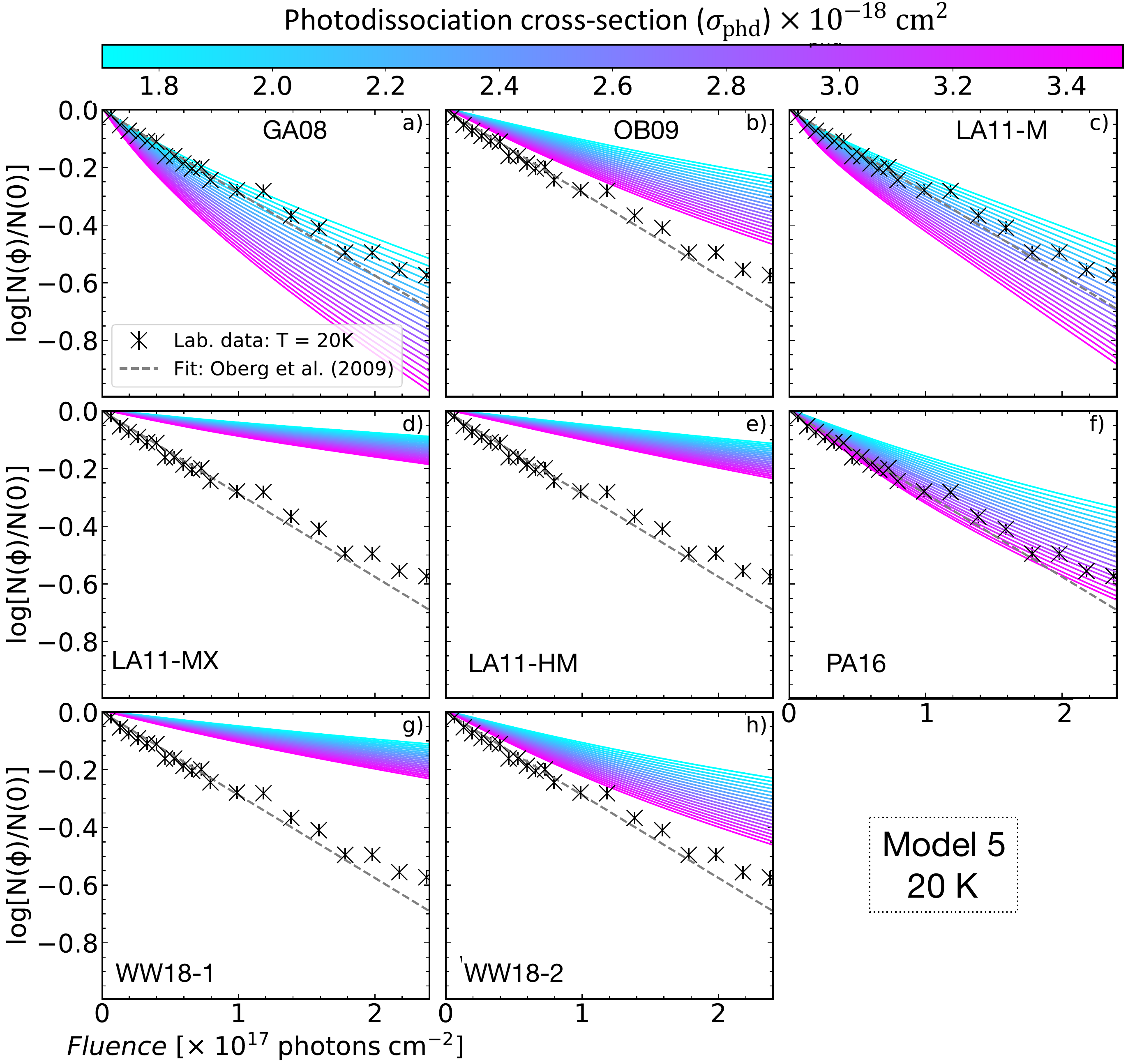}
      \caption{Same as Figure~\ref{M3}, but adopting Model~5.}
         \label{M5}
   \end{figure}

\begin{figure}
   \centering
   \includegraphics[width=9cm]{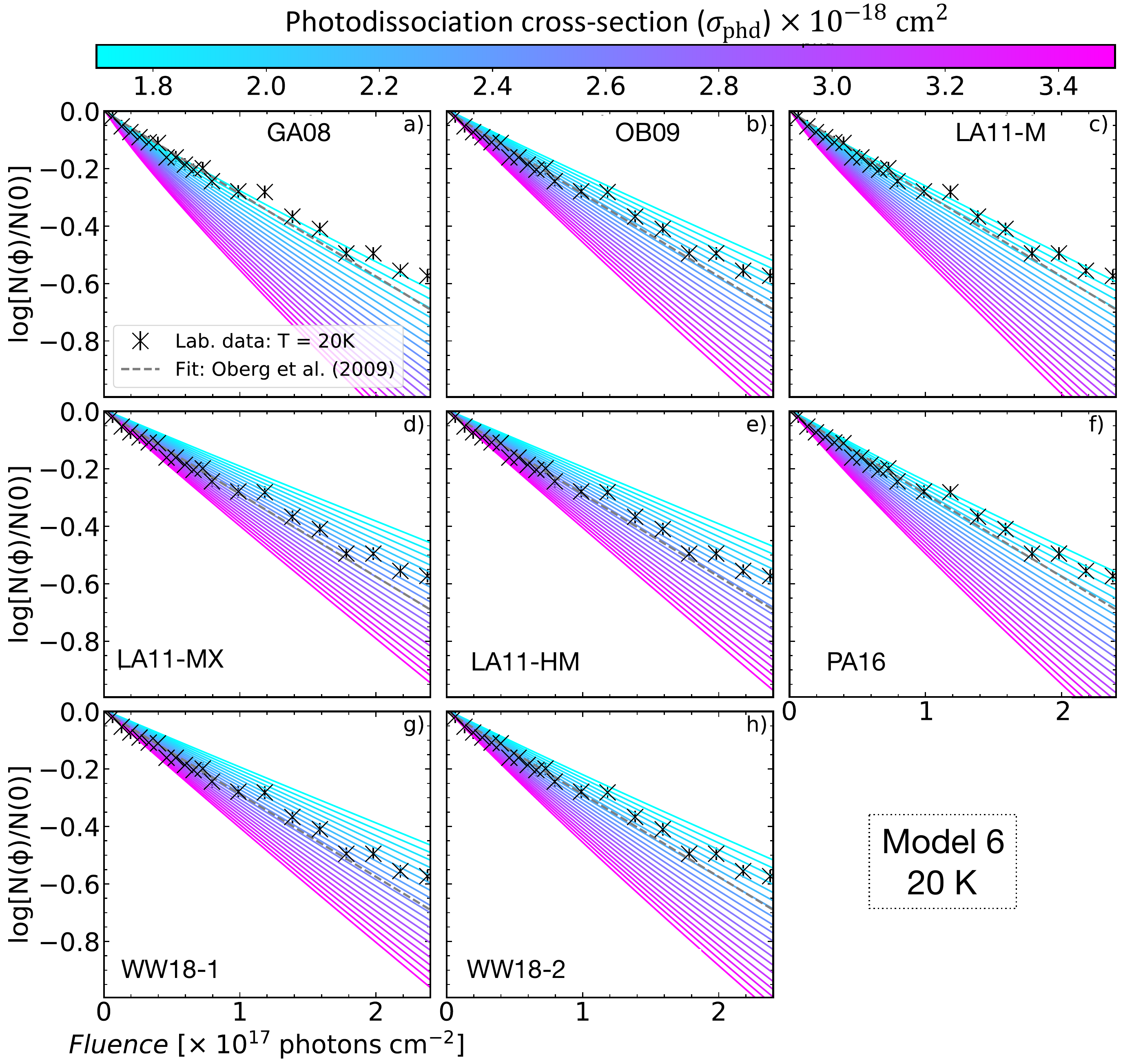}
      \caption{Same as Figure~\ref{M3}, but adopting Model~6.}
         \label{M6}
   \end{figure}

\begin{figure}
   \centering
   \includegraphics[width=9cm]{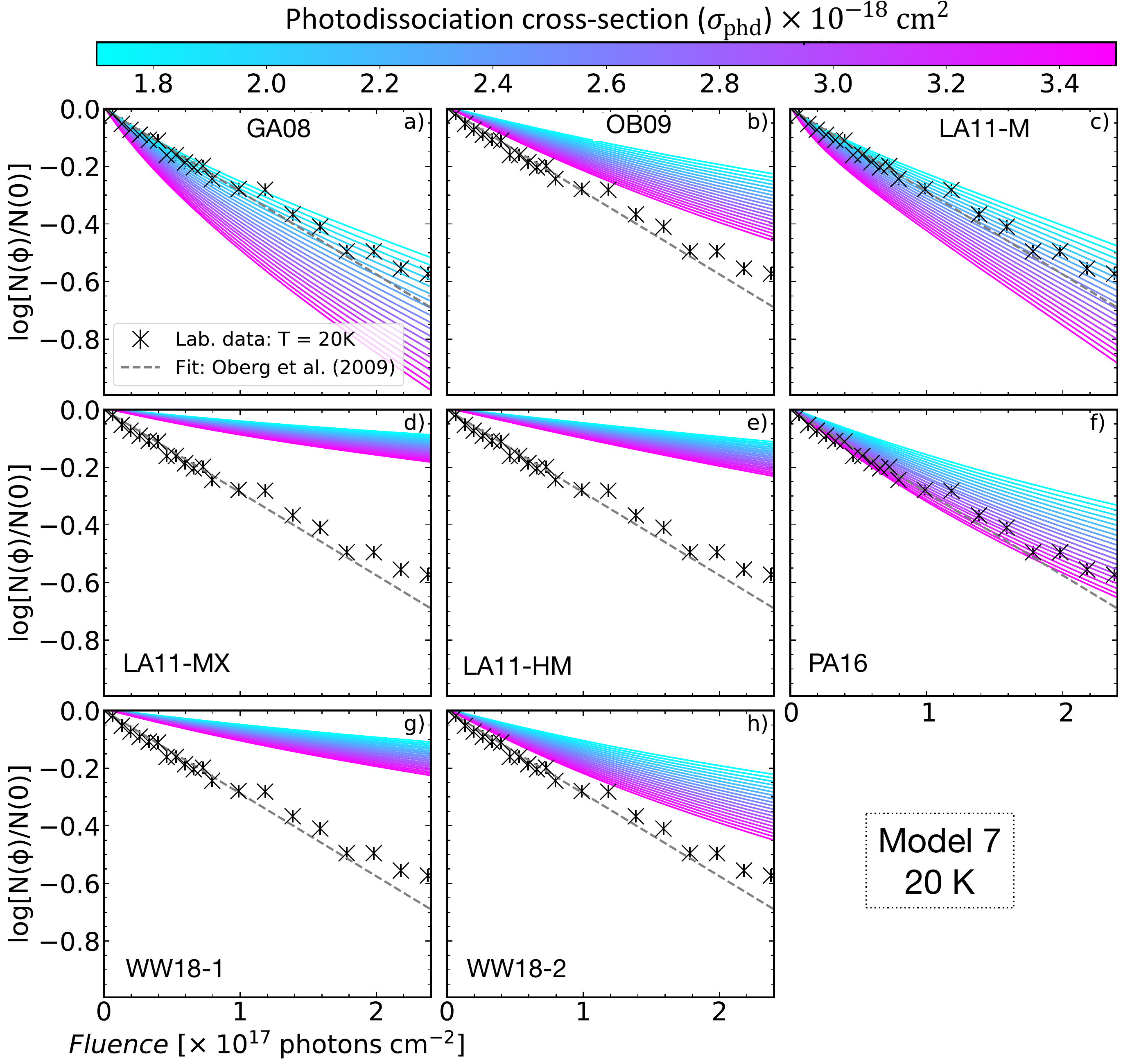}
      \caption{Same as Figure~\ref{M3}, but adopting Model~7.}
         \label{M7}
   \end{figure}

\begin{figure}
   \centering
   \includegraphics[width=9cm]{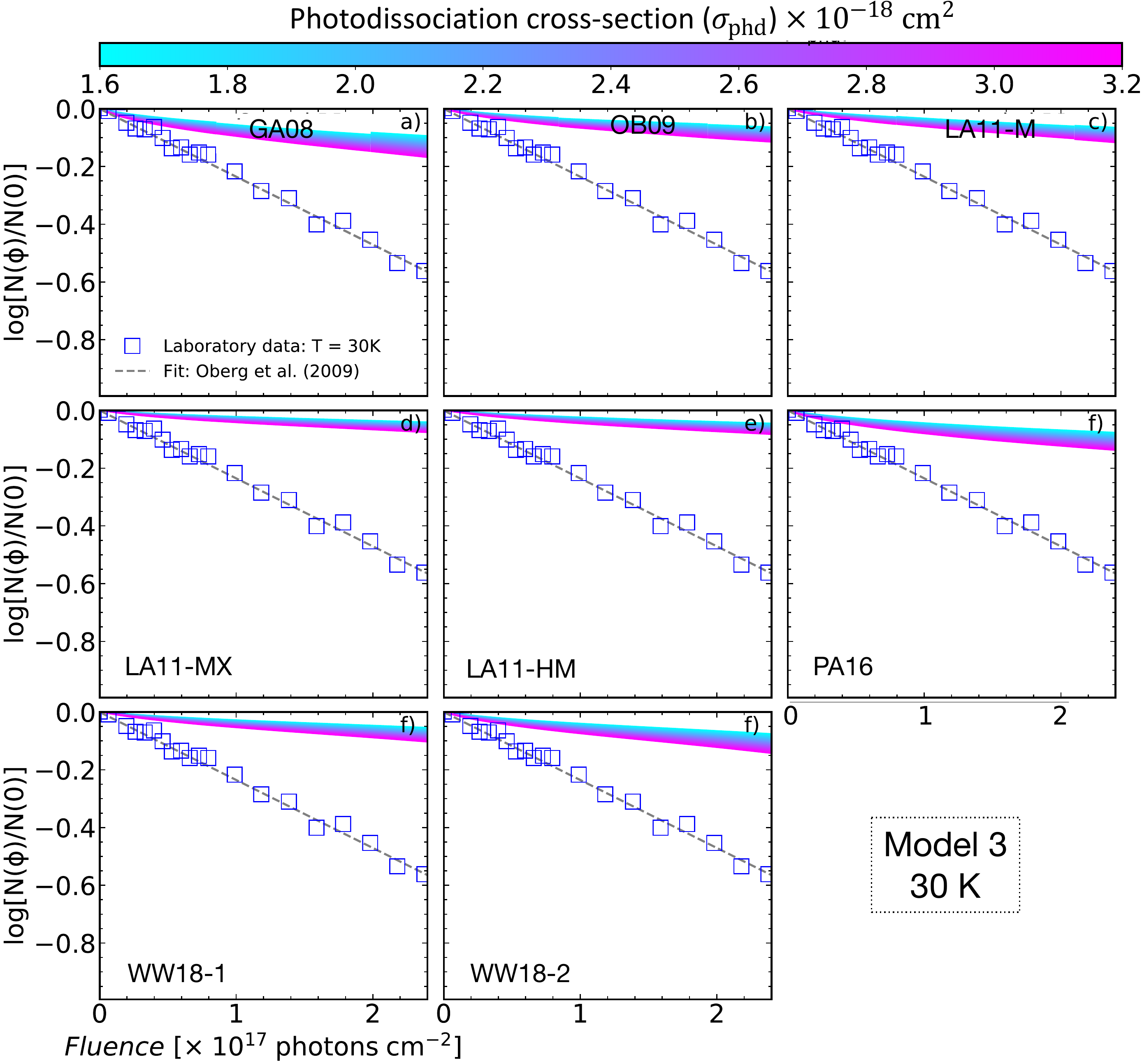}
      \caption{Grid of methanol ice photodissociation models at 30~K assuming different photodissociation cross-sections and BRs. The Model~3 in Table~\ref{t7} is adopted in these plots.}
         \label{M3_30}
   \end{figure}

\begin{figure}
   \centering
   \includegraphics[width=9cm]{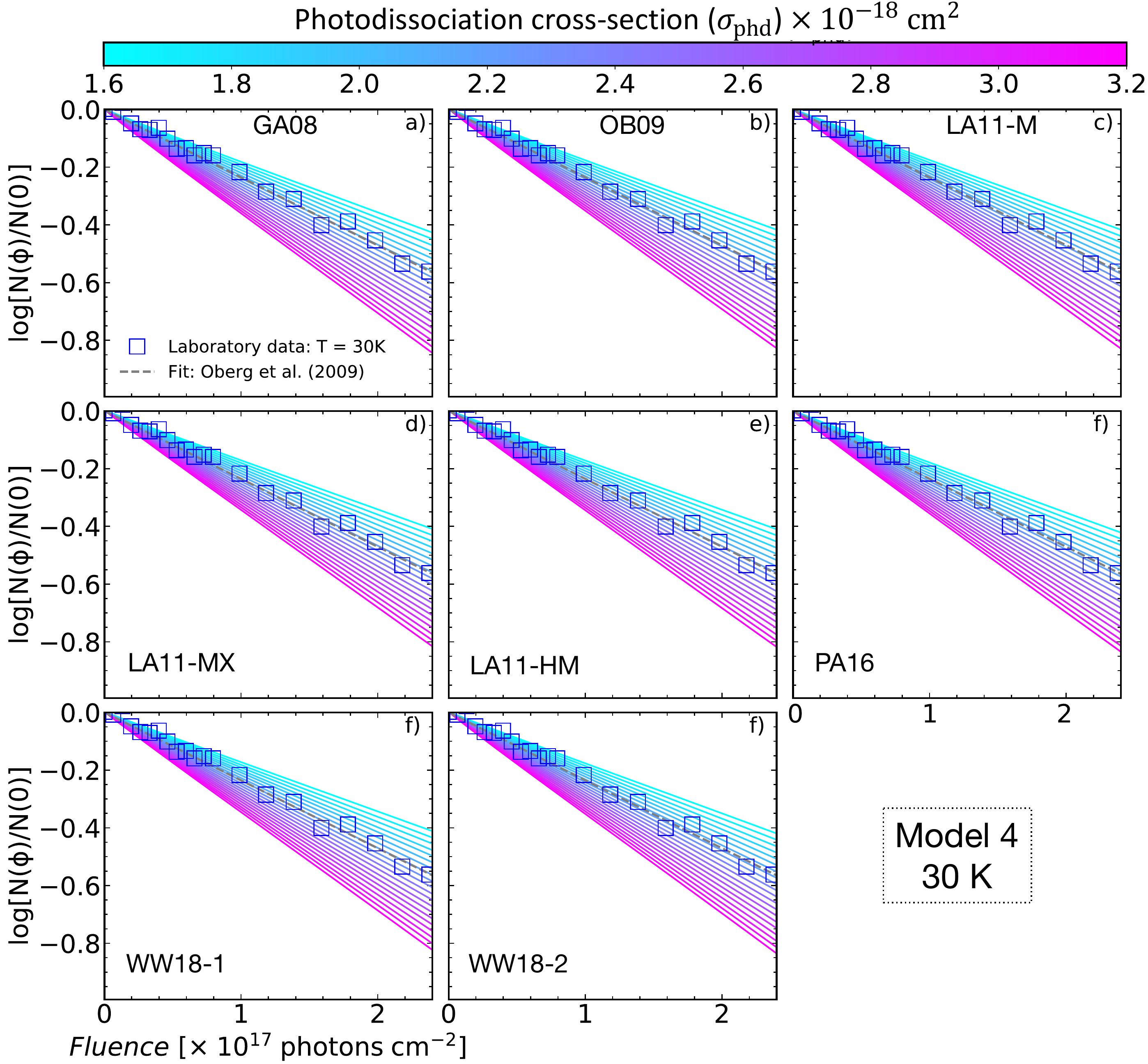}
      \caption{Same as Figure~\ref{M3_30}, but adopting Model~4.}
         \label{M4_30}
   \end{figure}

\begin{figure}
   \centering
   \includegraphics[width=9cm]{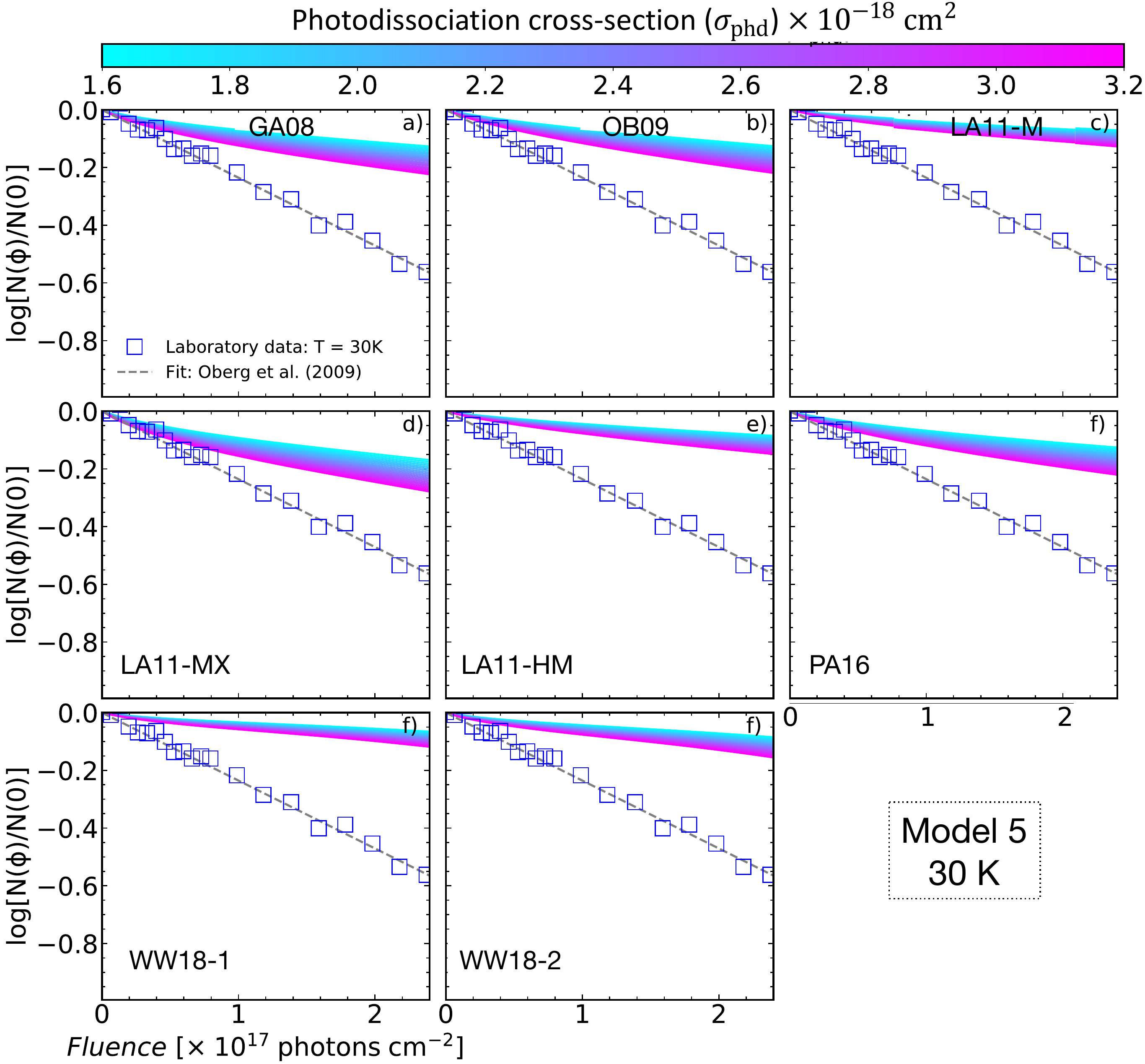}
      \caption{Same as Figure~\ref{M3_30}, but adopting Model~5.}
         \label{M5_30}
   \end{figure}

\begin{figure}
   \centering
   \includegraphics[width=9cm]{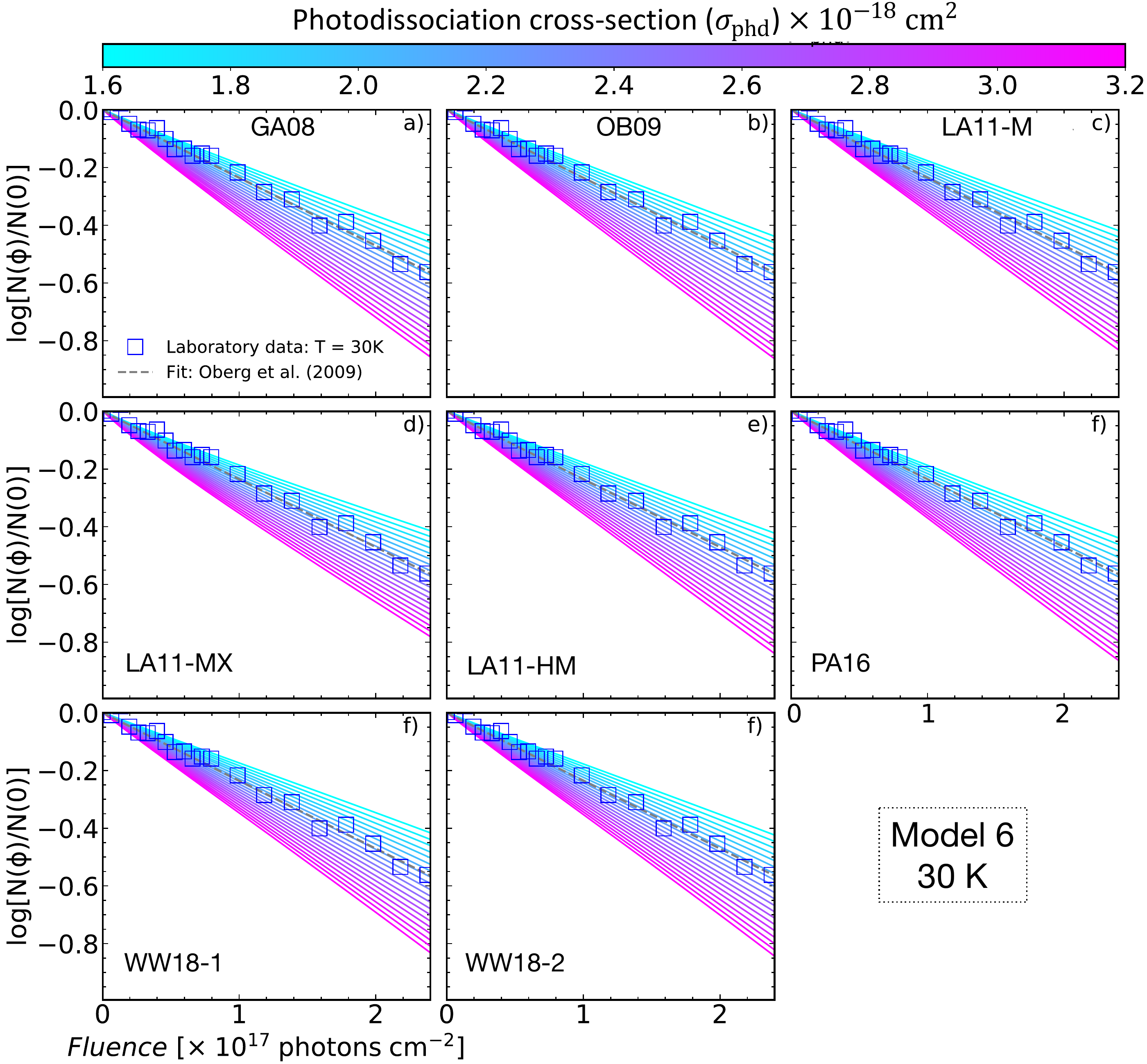}
      \caption{Same as Figure~\ref{M3_30}, but adopting Model~6.}
         \label{M6_30}
   \end{figure}

\begin{figure}
   \centering
   \includegraphics[width=9cm]{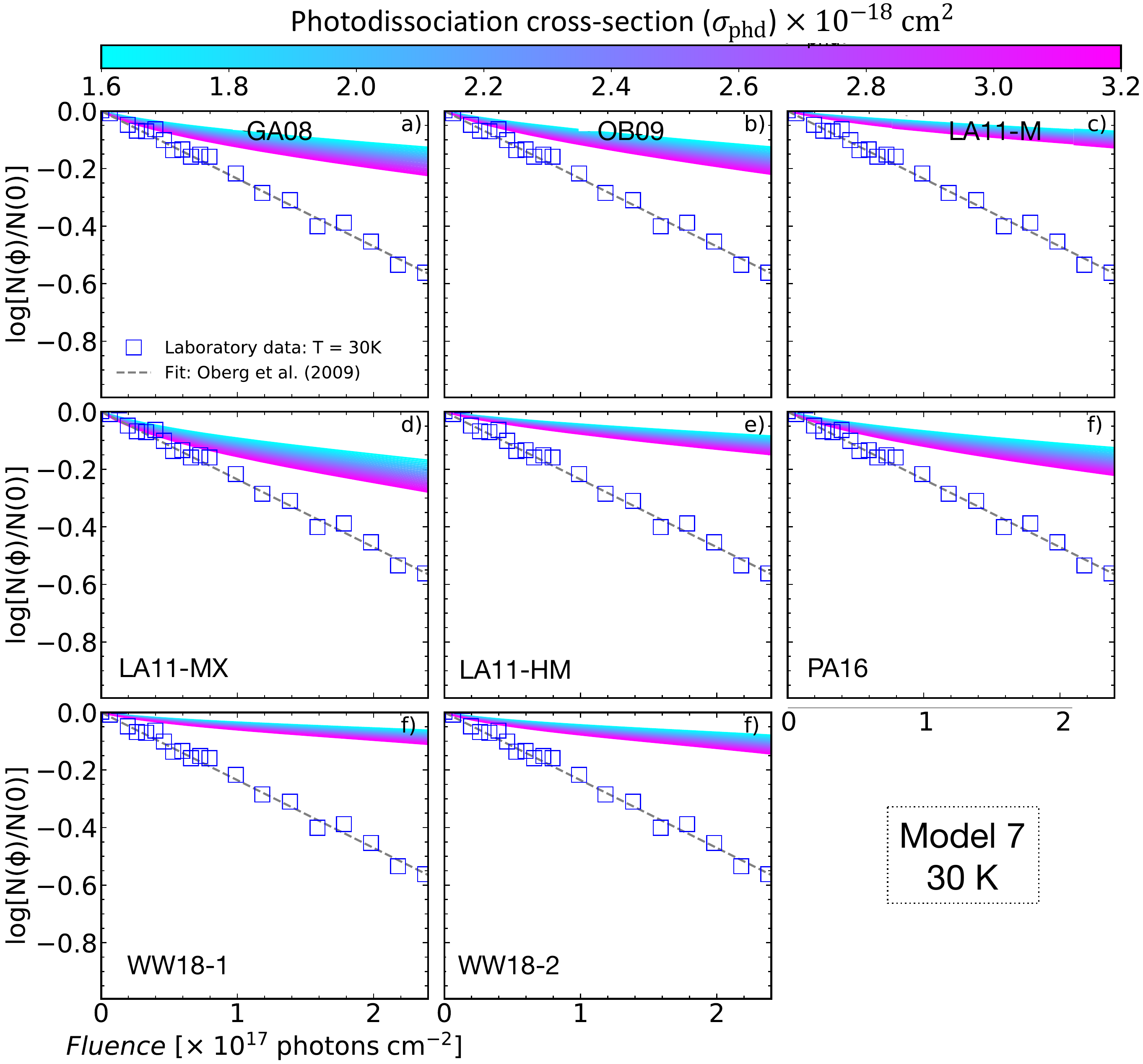}
      \caption{Same as Figure~\ref{M3_30}, but adopting Model~7.}
         \label{M7_30}
   \end{figure}

\begin{figure}
   \centering
   \includegraphics[width=9cm]{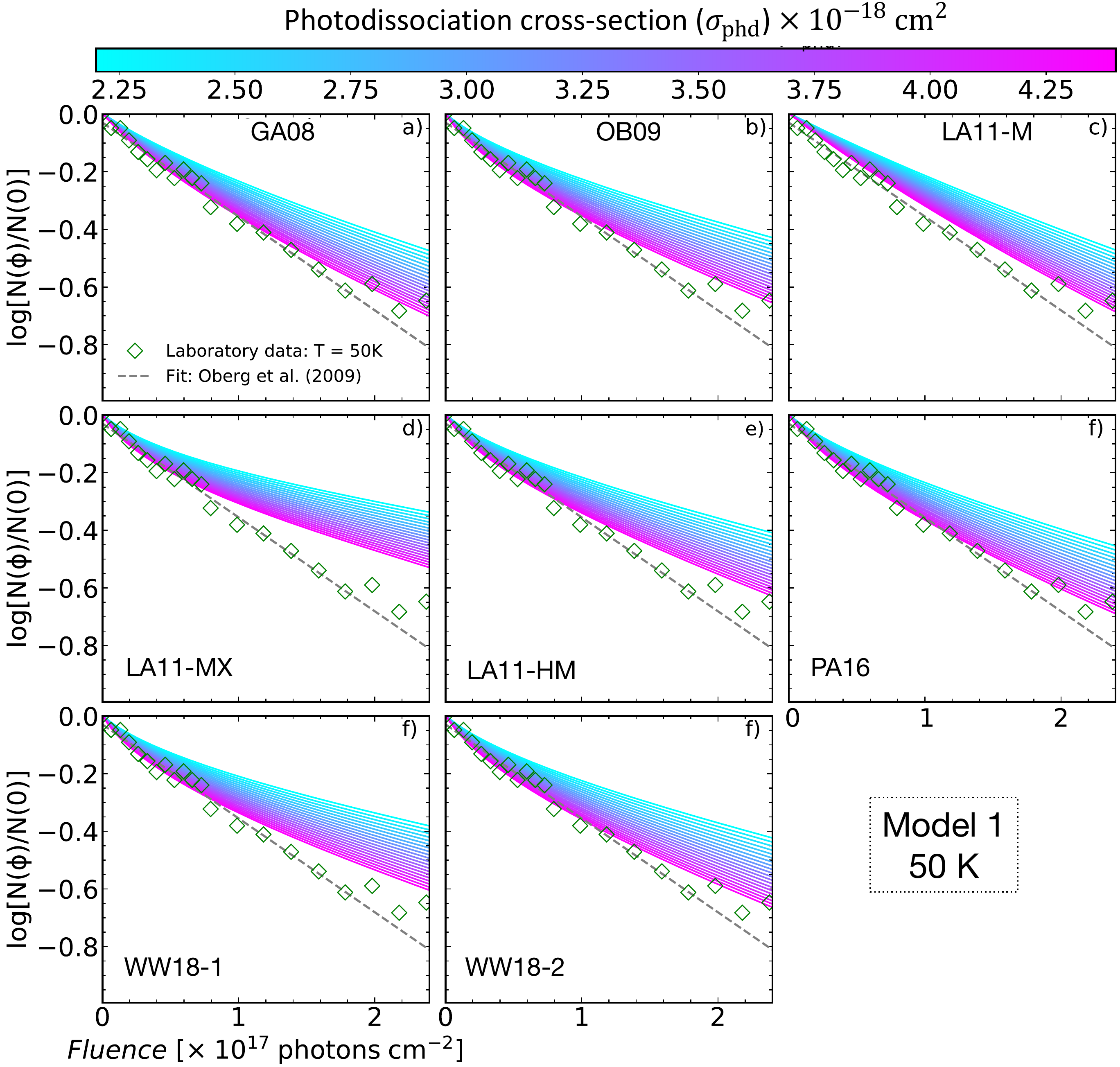}
      \caption{Grid of methanol ice photodissociation models at 50~K assuming different photodissociation cross-sections and BRs. The Model~1 in Table~\ref{t7} is adopted in these plots.}
         \label{M1_50}
   \end{figure}

\begin{figure}
   \centering
   \includegraphics[width=9cm]{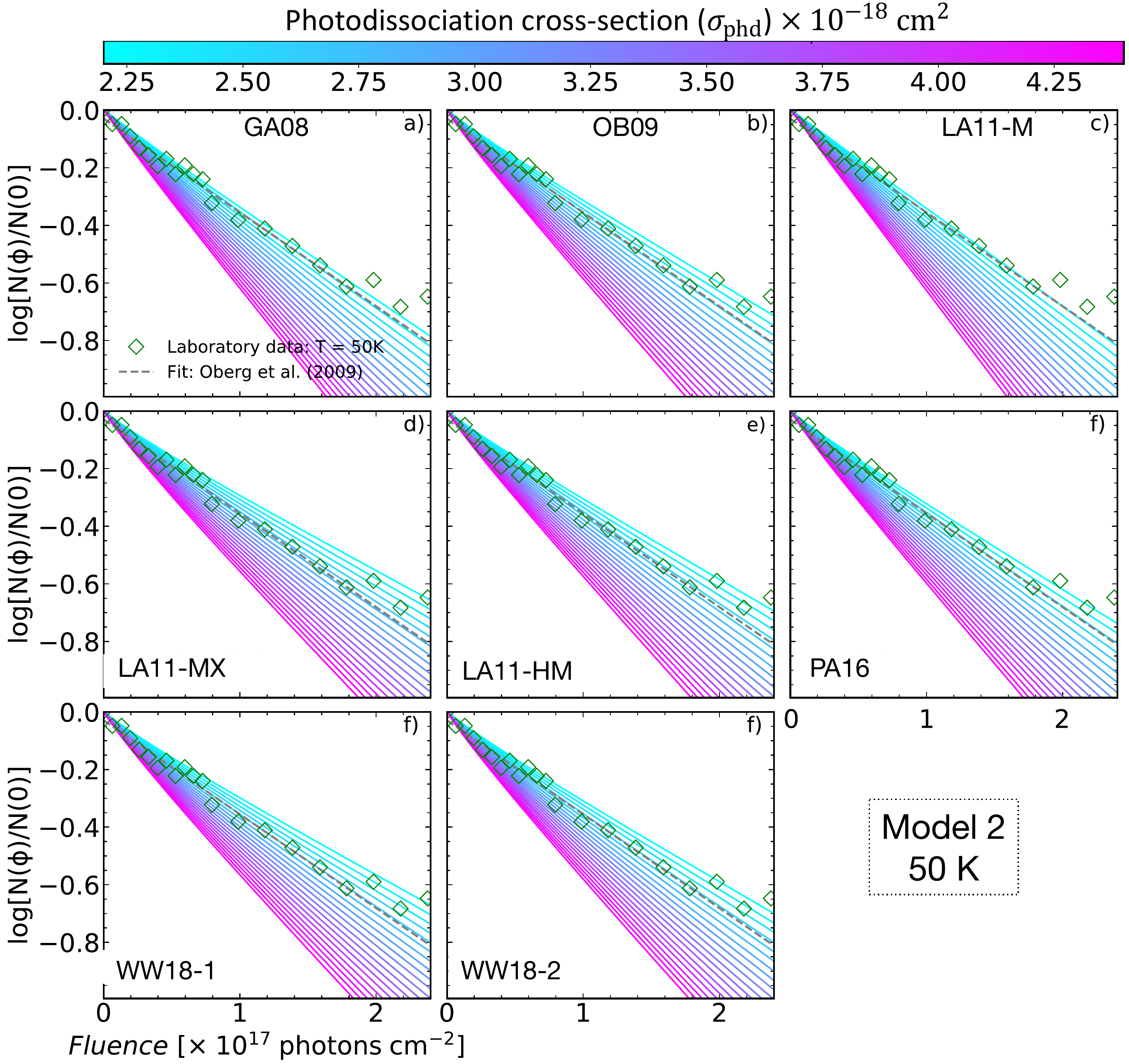}
      \caption{Same as Figure~\ref{M1_50}, but adopting Model~2.}
         \label{M2_50}
   \end{figure}

\begin{figure}
   \centering
   \includegraphics[width=9cm]{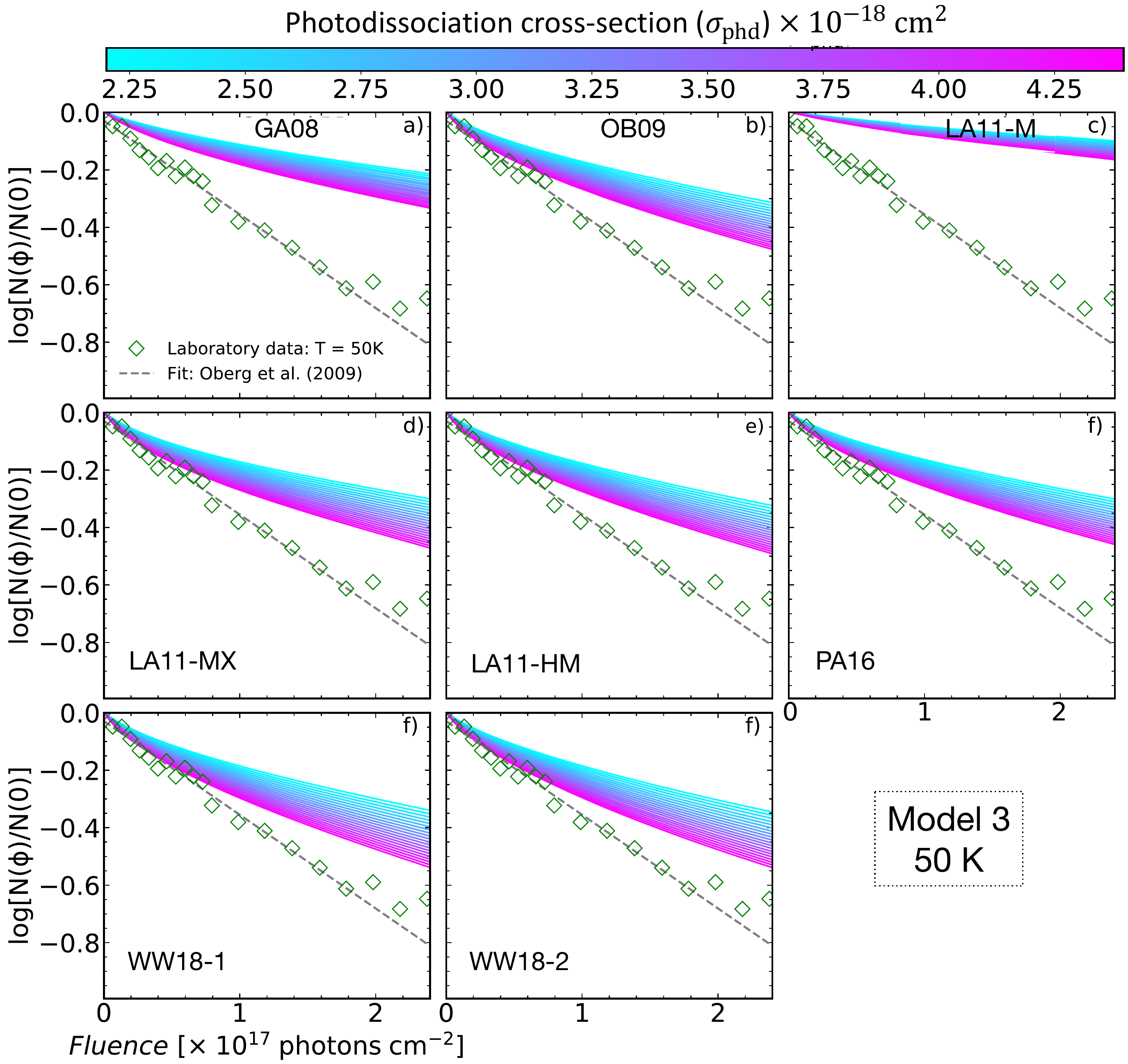}
      \caption{Same as Figure~\ref{M1_50}, but adopting Model~3.}
         \label{M3_50}
   \end{figure}

\begin{figure}
   \centering
   \includegraphics[width=9cm]{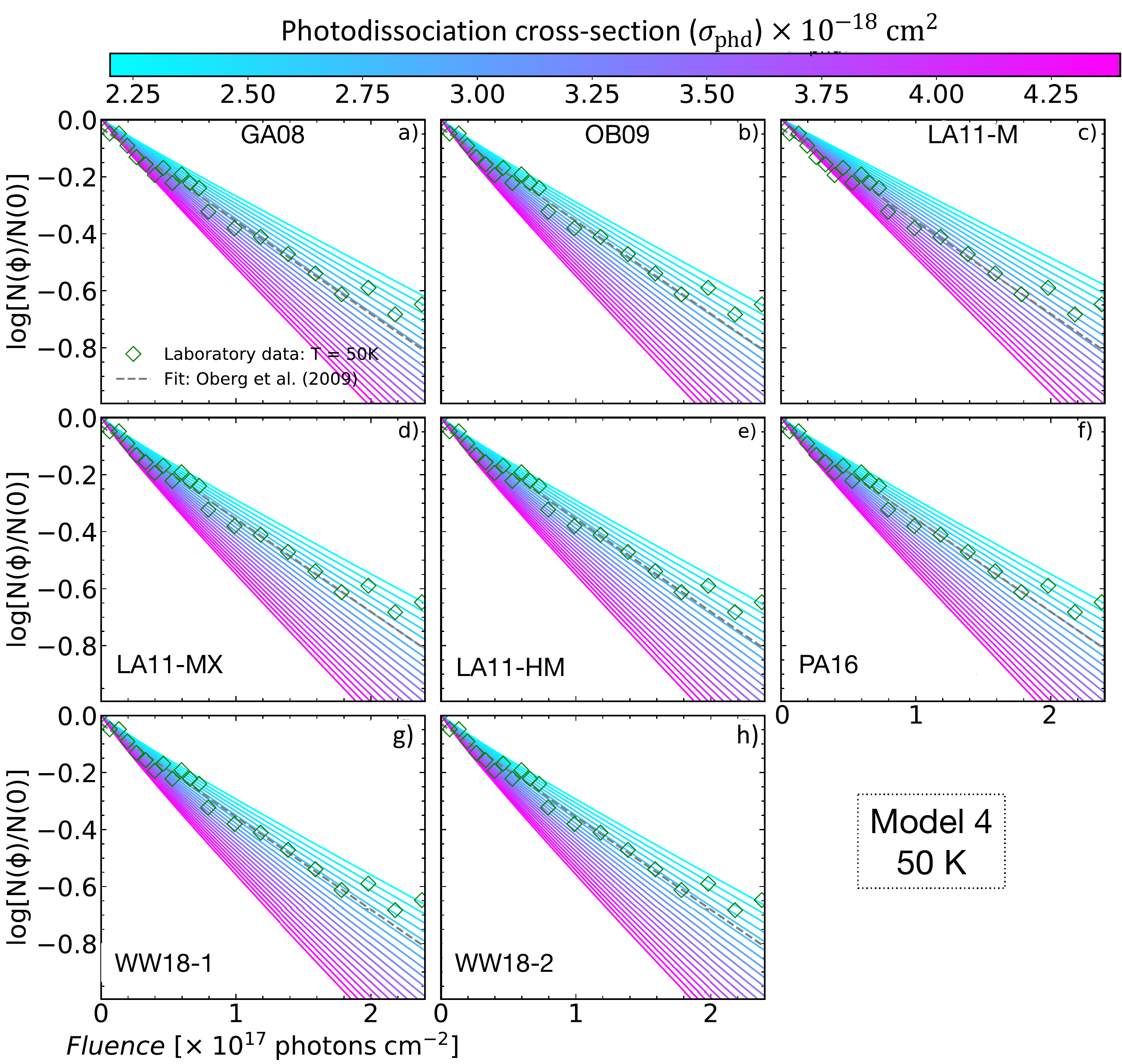}
      \caption{Same as Figure~\ref{M1_50}, but adopting Model~4.}
         \label{M4_50}
   \end{figure}

\begin{figure}
   \centering
   \includegraphics[width=9cm]{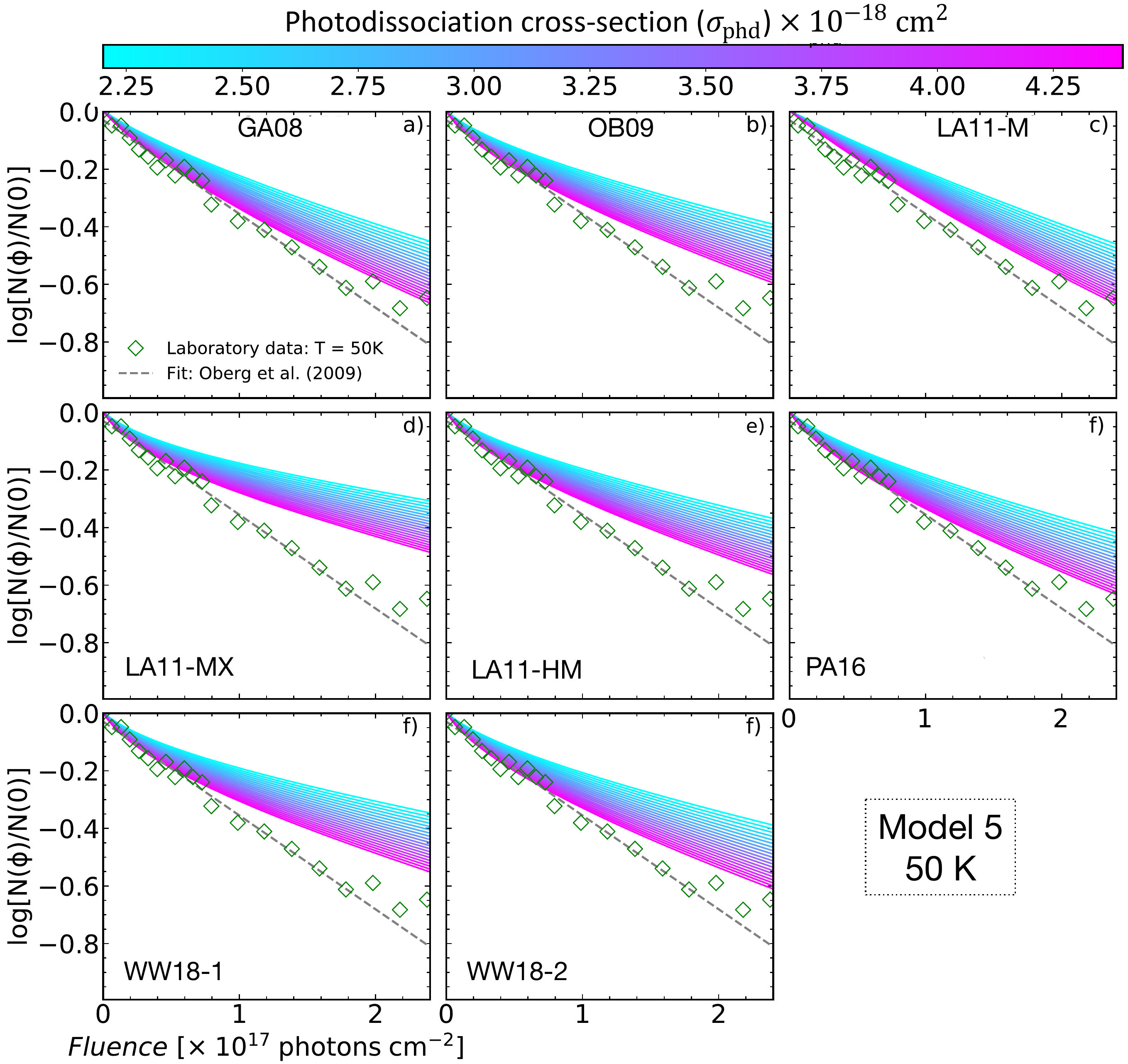}
      \caption{Same as Figure~\ref{M1_50}, but adopting Model~5.}
         \label{M5_50}
   \end{figure}

\begin{figure}
   \centering
   \includegraphics[width=9cm]{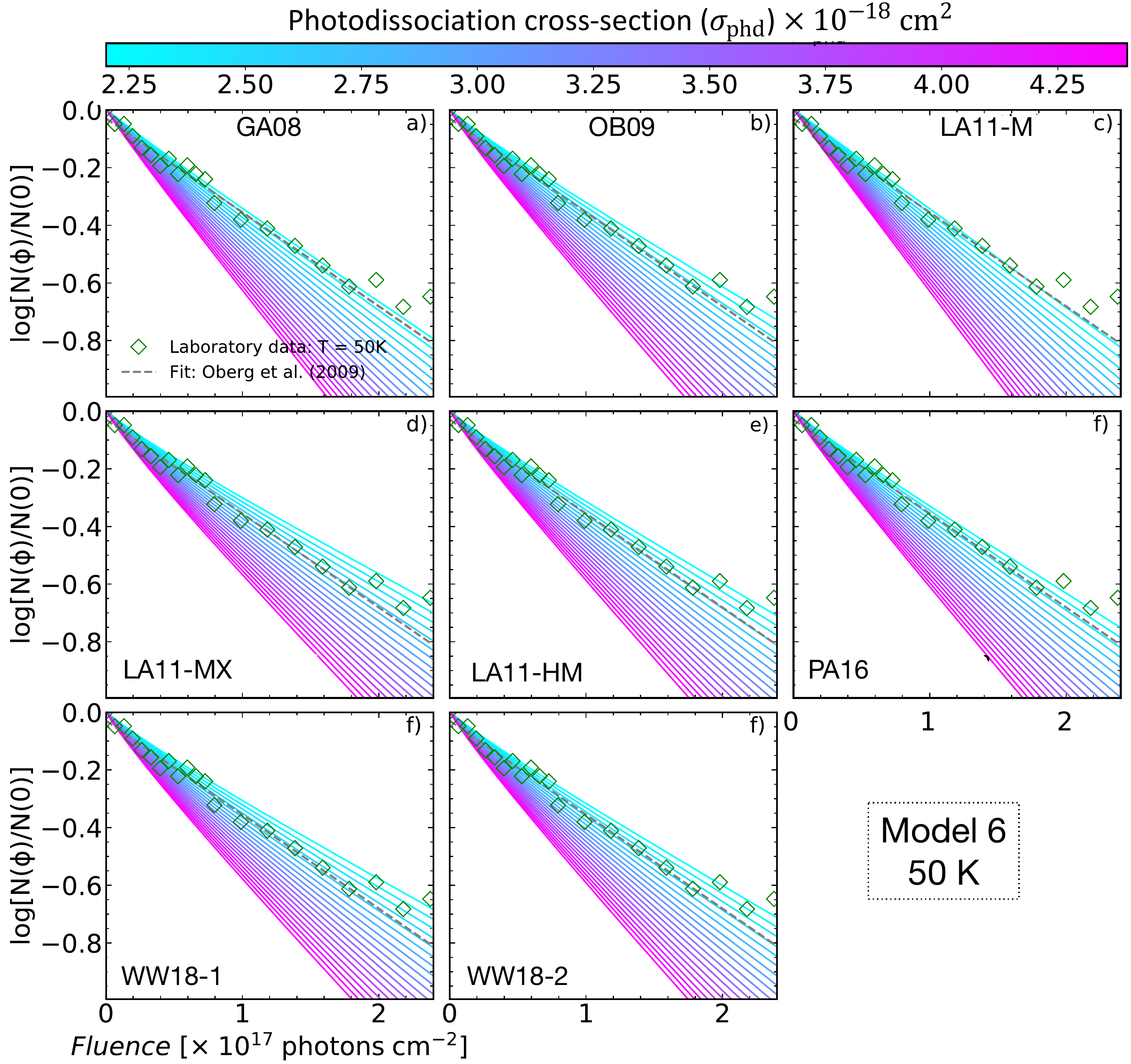}
      \caption{Same as Figure~\ref{M1_50}, but adopting Model~6.}
         \label{M6_50}
   \end{figure}

\begin{figure}
   \centering
   \includegraphics[width=9cm]{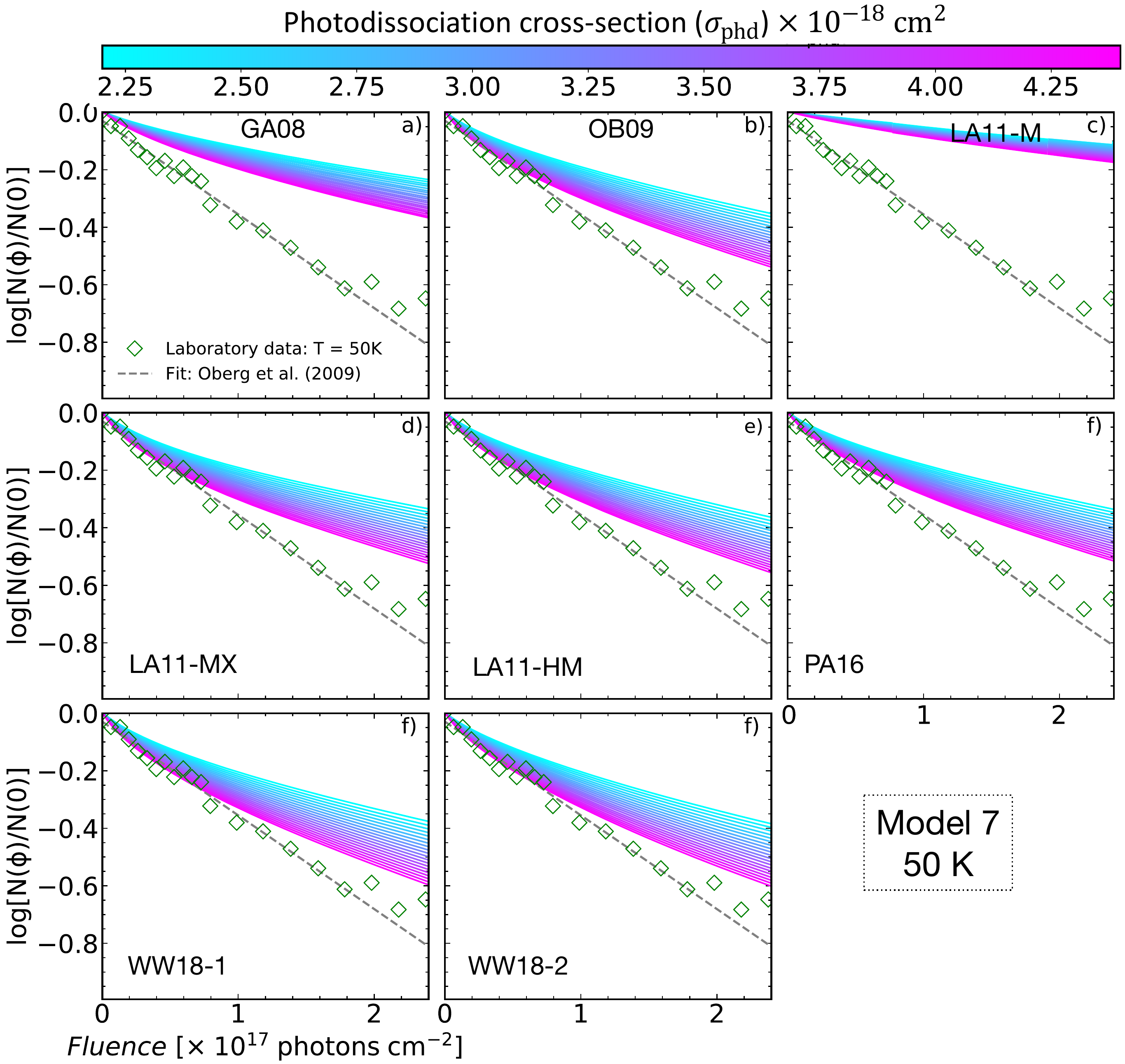}
      \caption{Same as Figure~\ref{M1_50}, but adopting Model~7.}
         \label{M7_50}
   \end{figure}

\begin{figure}
   \centering
   \includegraphics[width=9cm]{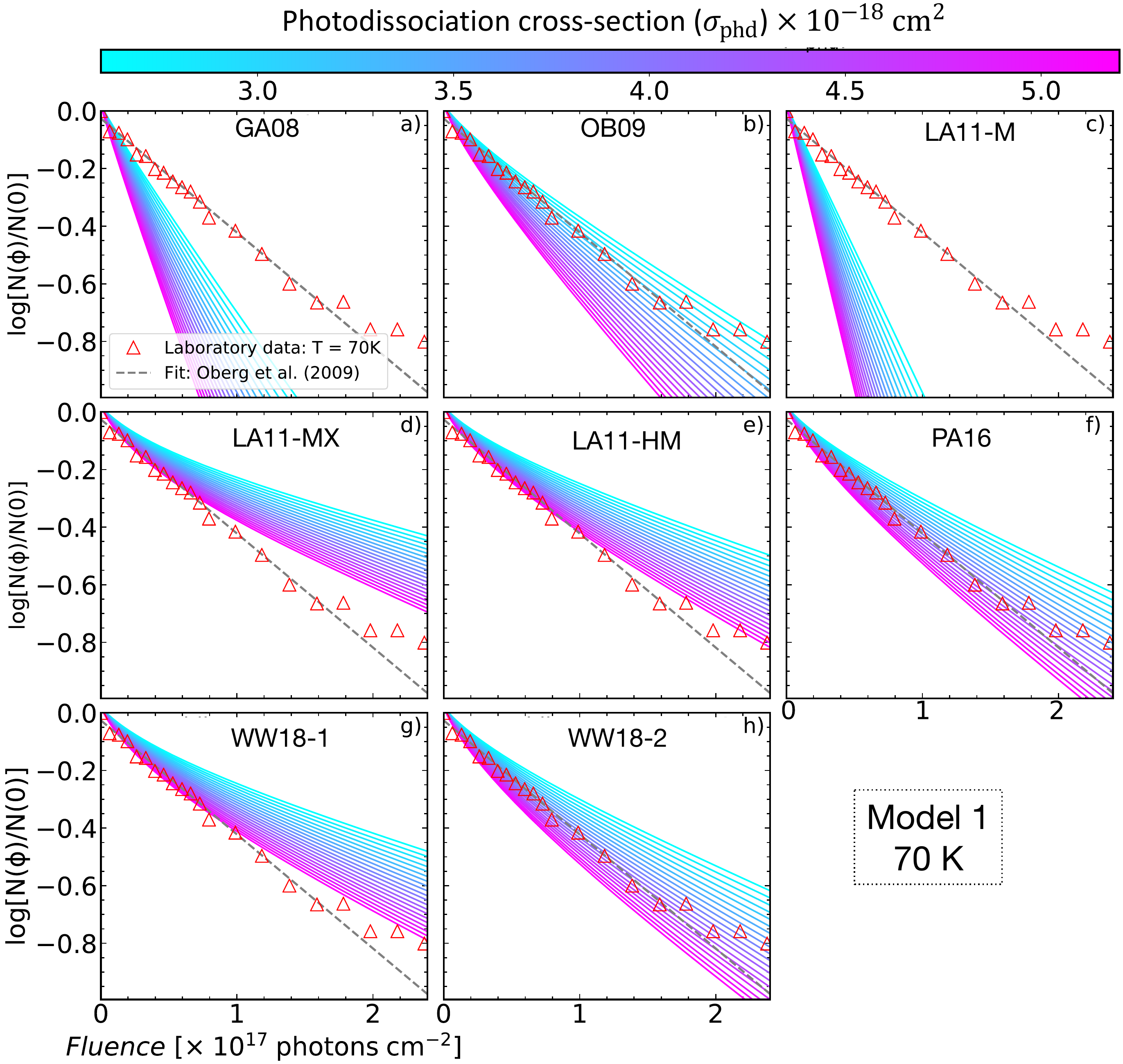}
      \caption{Grid of methanol ice photodissociation models at 70~K assuming different photodissociation cross-sections and BRs. The Model~1 in Table~\ref{t7} is adopted in these plots.}
         \label{M1_70}
   \end{figure}

\begin{figure}
   \centering
   \includegraphics[width=9cm]{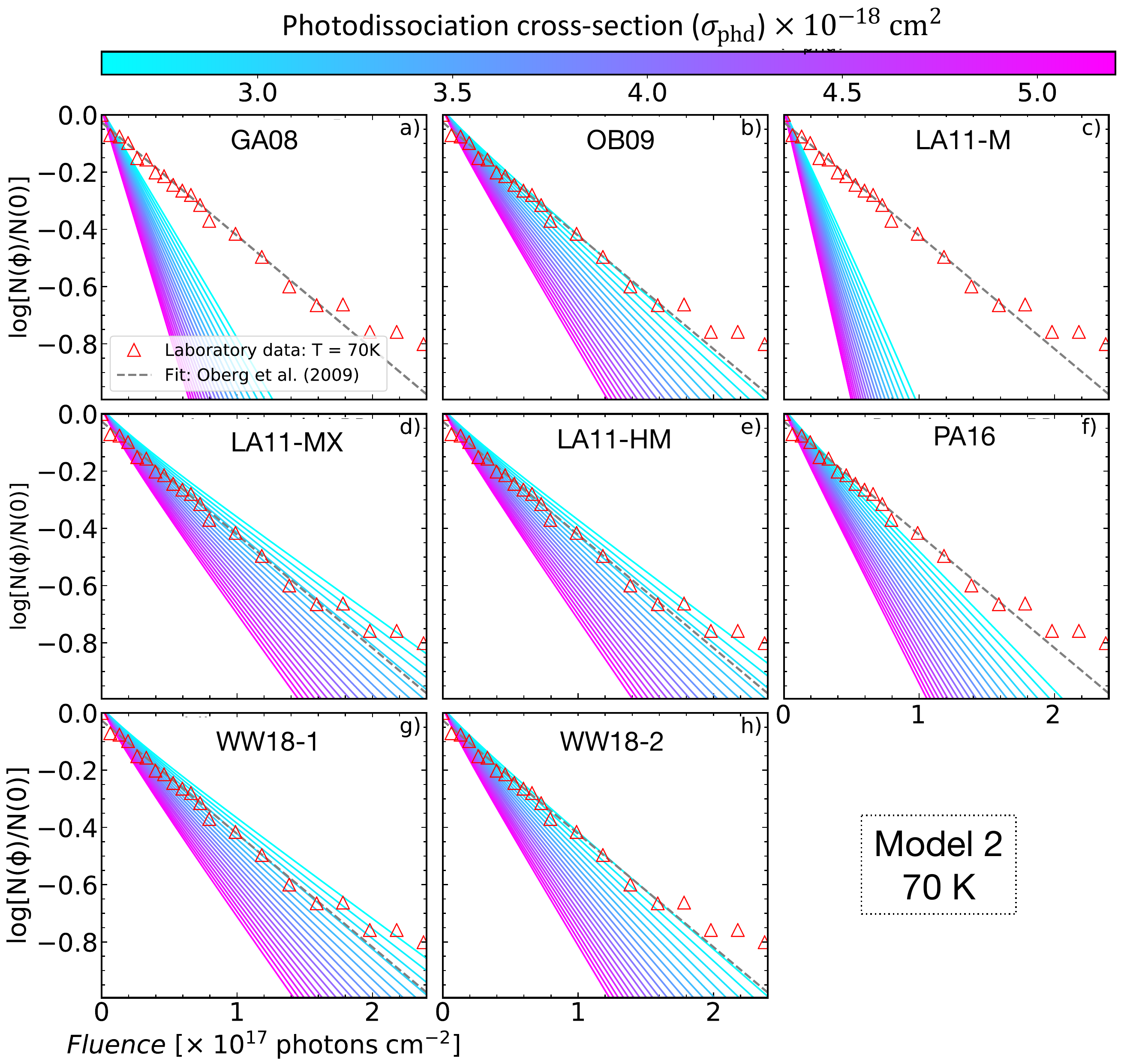}
      \caption{Same as Figure~\ref{M1_70}, but adopting Model~2.}
         \label{M2_70}
   \end{figure}

\begin{figure}
   \centering
   \includegraphics[width=9cm]{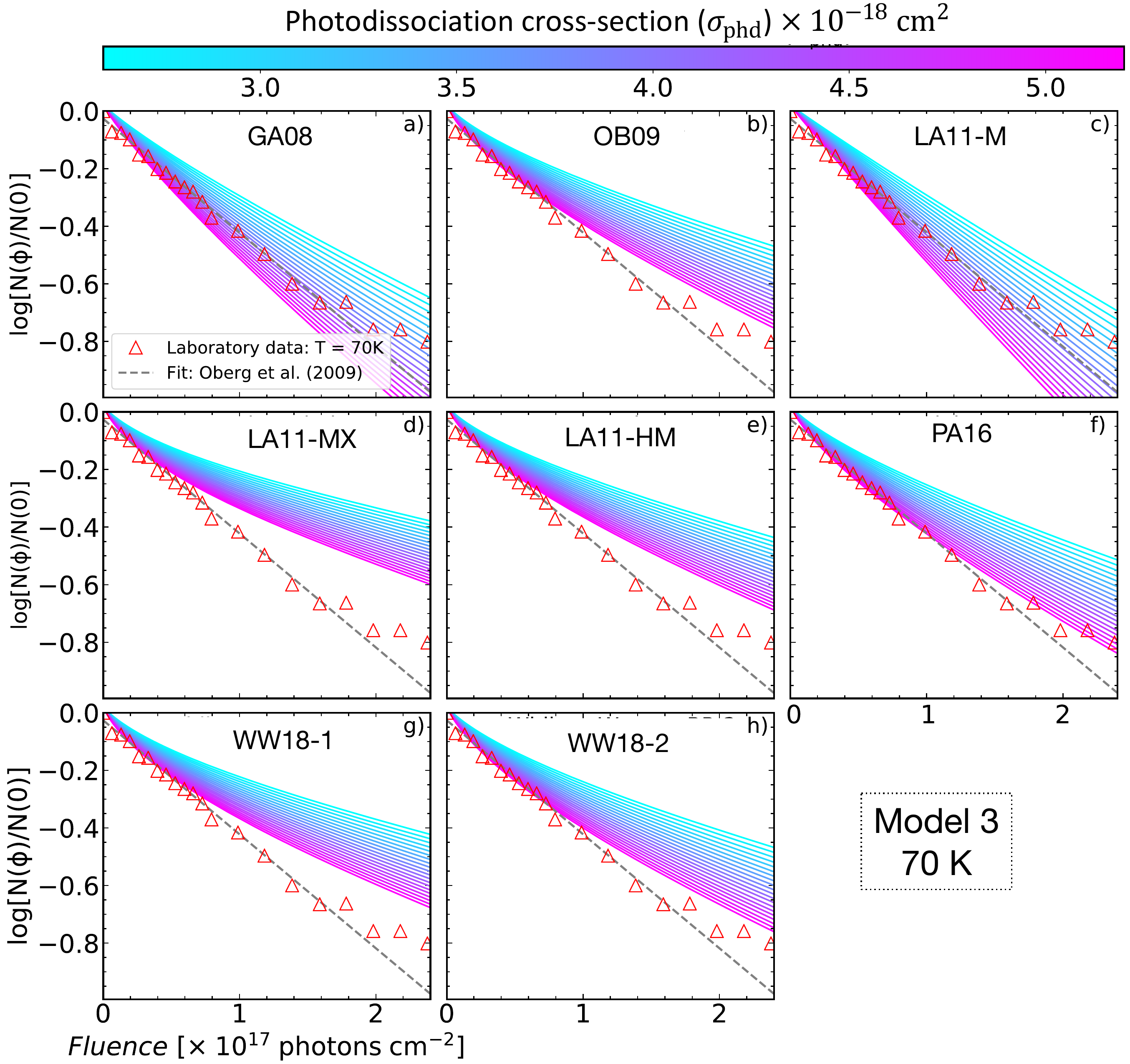}
      \caption{Same as Figure~\ref{M1_50}, but adopting Model~3.}
         \label{M3_70}
   \end{figure}

\begin{figure}
   \centering
   \includegraphics[width=9cm]{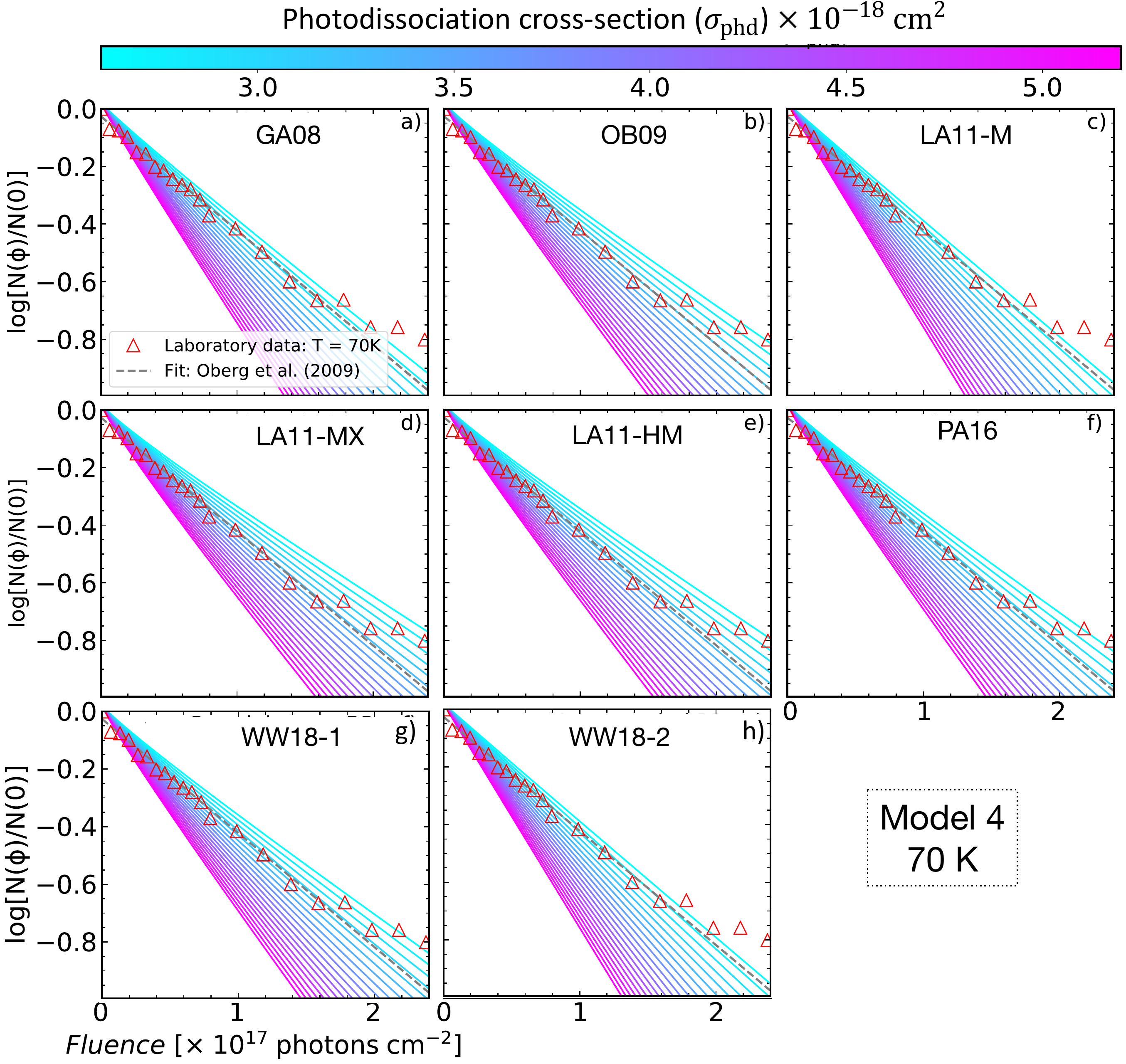}
      \caption{Same as Figure~\ref{M1_50}, but adopting Model~4.}
         \label{M4_70}
   \end{figure}

\begin{figure}
   \centering
   \includegraphics[width=9cm]{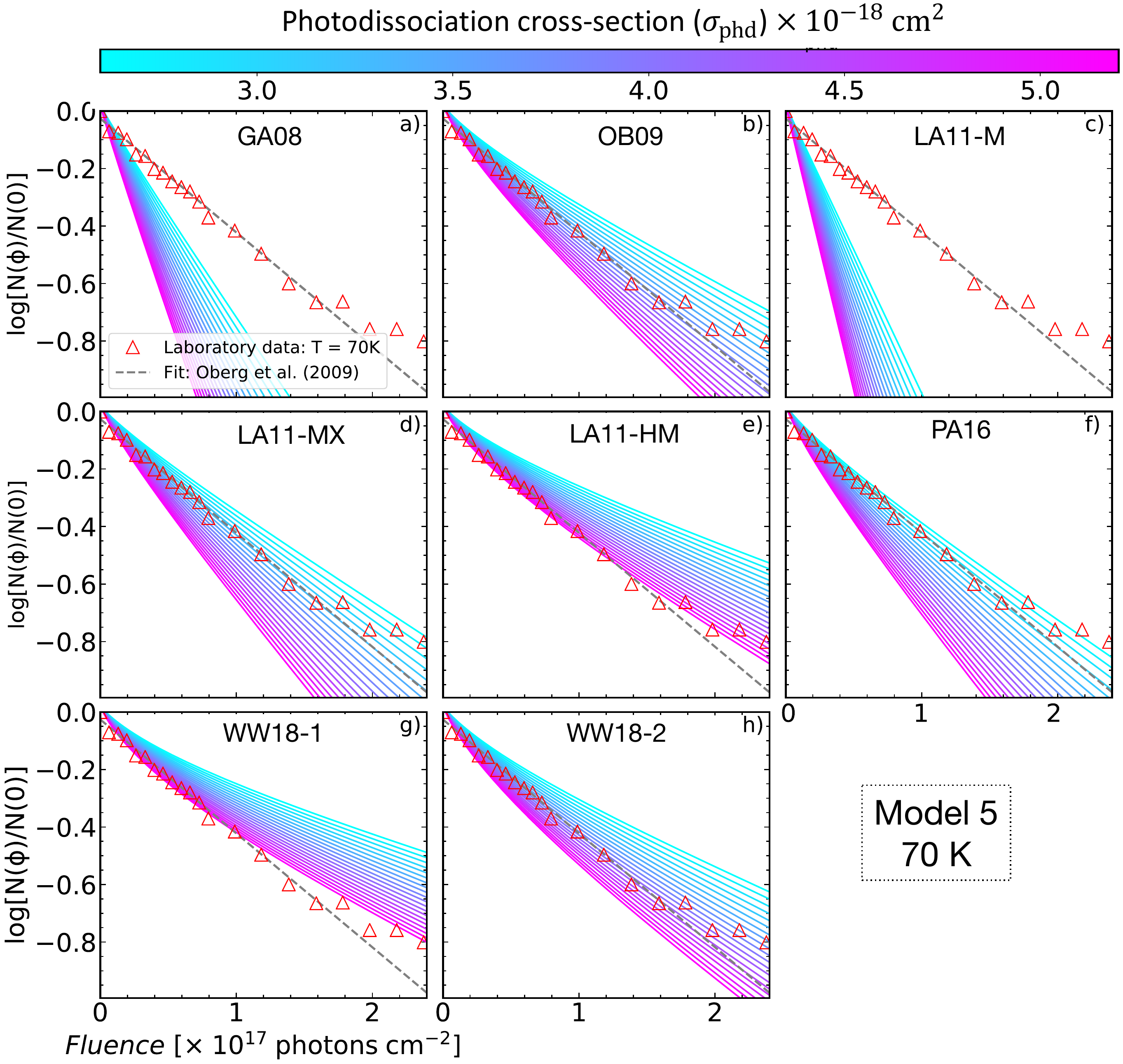}
      \caption{Same as Figure~\ref{M1_50}, but adopting Model~5.}
         \label{M5_70}
   \end{figure}

\begin{figure}
   \centering
   \includegraphics[width=9cm]{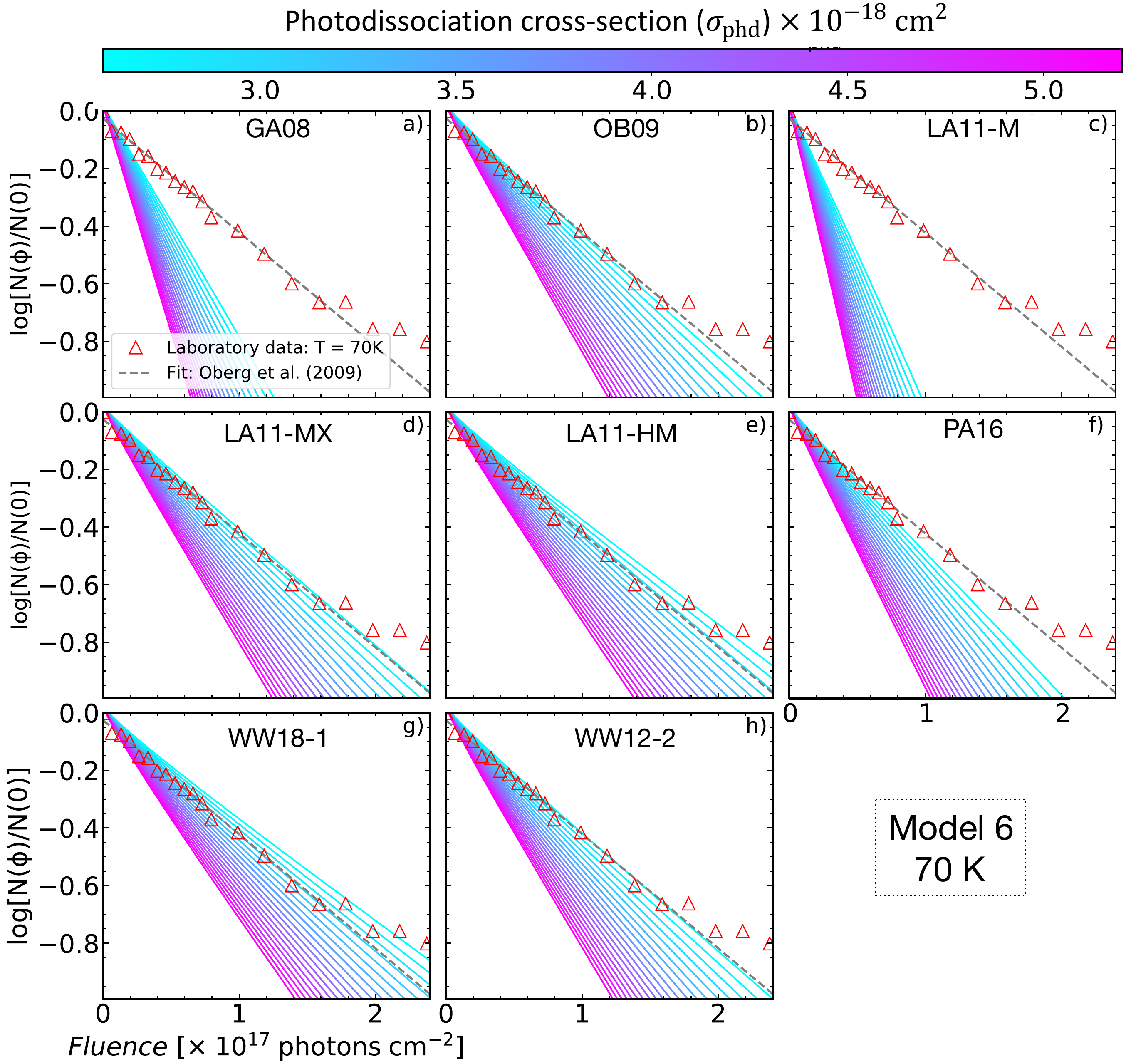}
      \caption{Same as Figure~\ref{M1_50}, but adopting Model~6.}
         \label{M6_70}
   \end{figure}

\begin{figure}
   \centering
   \includegraphics[width=9cm]{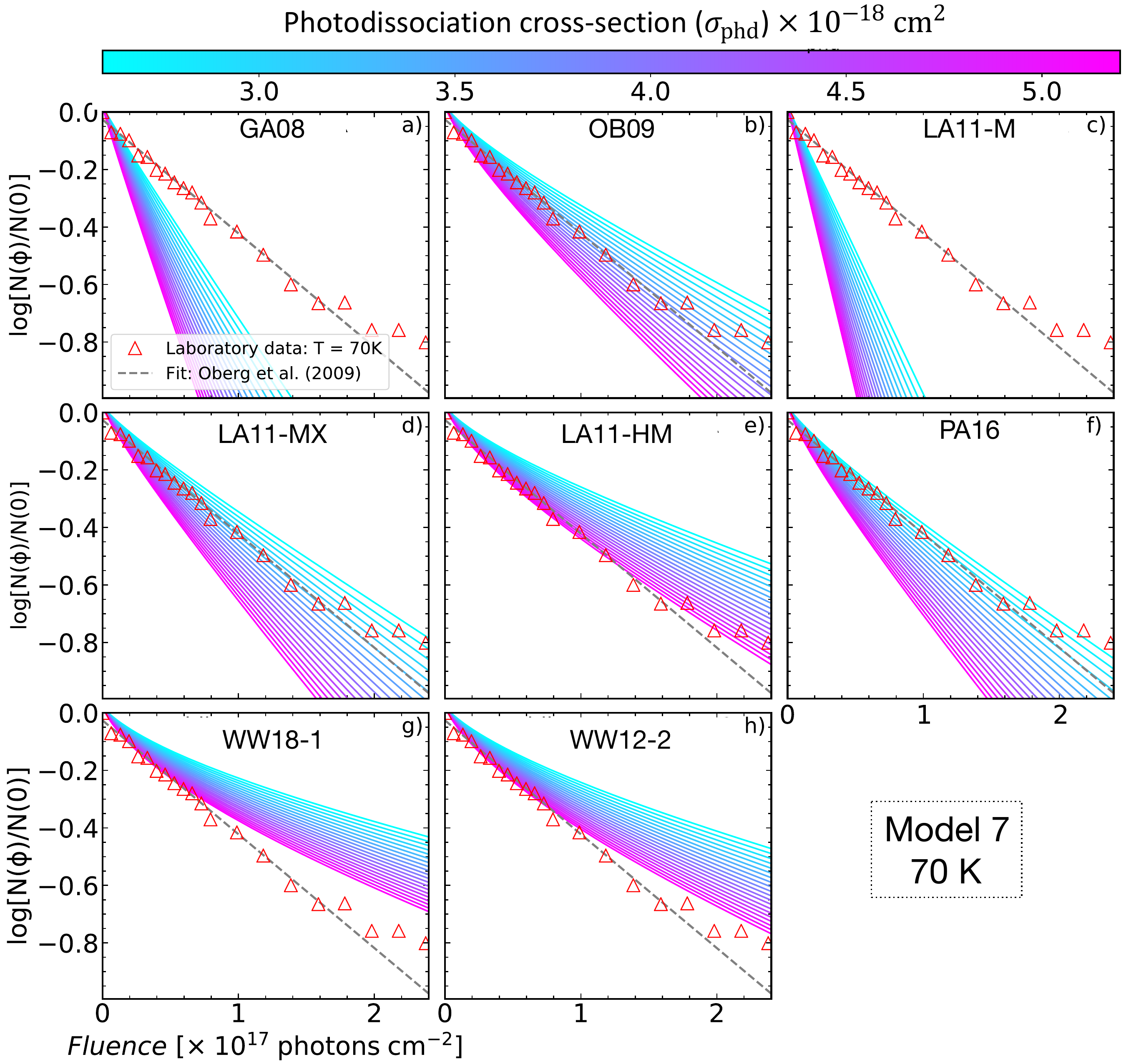}
      \caption{Same as Figure~\ref{M1_50}, but adopting Model~7.}
         \label{M7_70}
   \end{figure}

The grid of photodissociation models at four temperatures and adopting different surface chemistry parameters are shown in this appendix section. Figures~\ref{M3}-\ref{M7} show the fits at 20~K for Models~3$-$7. Figures~\ref{M3_30}-\ref{M7_30} show the fits at 30~K for Models~3$-$7. In these fits, only the models adopting reactive desorption from \citet{Minissale2016} fit well the experimental data. Figures~\ref{M1_50}-\ref{M7_50} show the fits at 50~K for Models~1$-$7. Figures~\ref{M1_70}-\ref{M7_70} show the fits at 70~K for Models~1$-$7. 

\section{CH$_3$OH gas-phase abundances at 30~K, 50~K, and 70~K in the molecular cloud model}
\label{Ap_mc_highT}
The methanol gas-phase abundances in the molecular cloud model at three temperatures and adopting different numerical densities and H$_2$ ionization rates. In these cases, that difference between models including or excluding methanol ice photolysis is negligible. Additionally, the abundances mostly deviate from the observations toward the Orion Bar.

\begin{figure}
   \centering
   \includegraphics[width=\hsize]{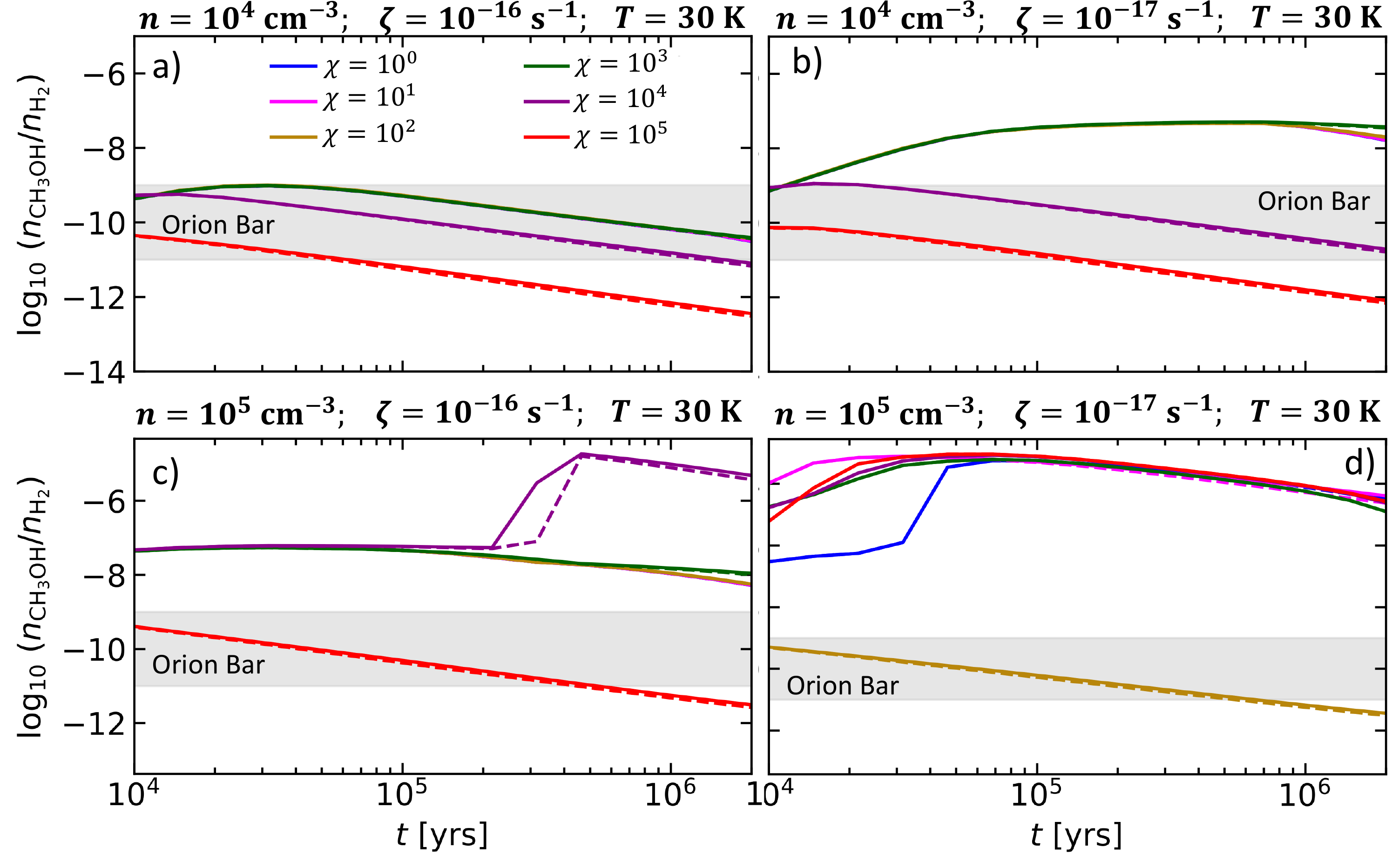}
      \caption{Abundances of gas-phase methanol at different physical conditions and $T =$ 30~K. The solid and dashed lines show the abundances in the models with and without methanol ice photolysis, respectively. The line colours indicate the strength of the UV radiation field. The grey shaded area indicate methanol abundance in the Orion Bar PDR region \citep[][]{Cuadrado2017} only for comparison purposes with an astrophysical environment.}
         \label{abund30}
   \end{figure}

\begin{figure}
   \centering
   \includegraphics[width=\hsize]{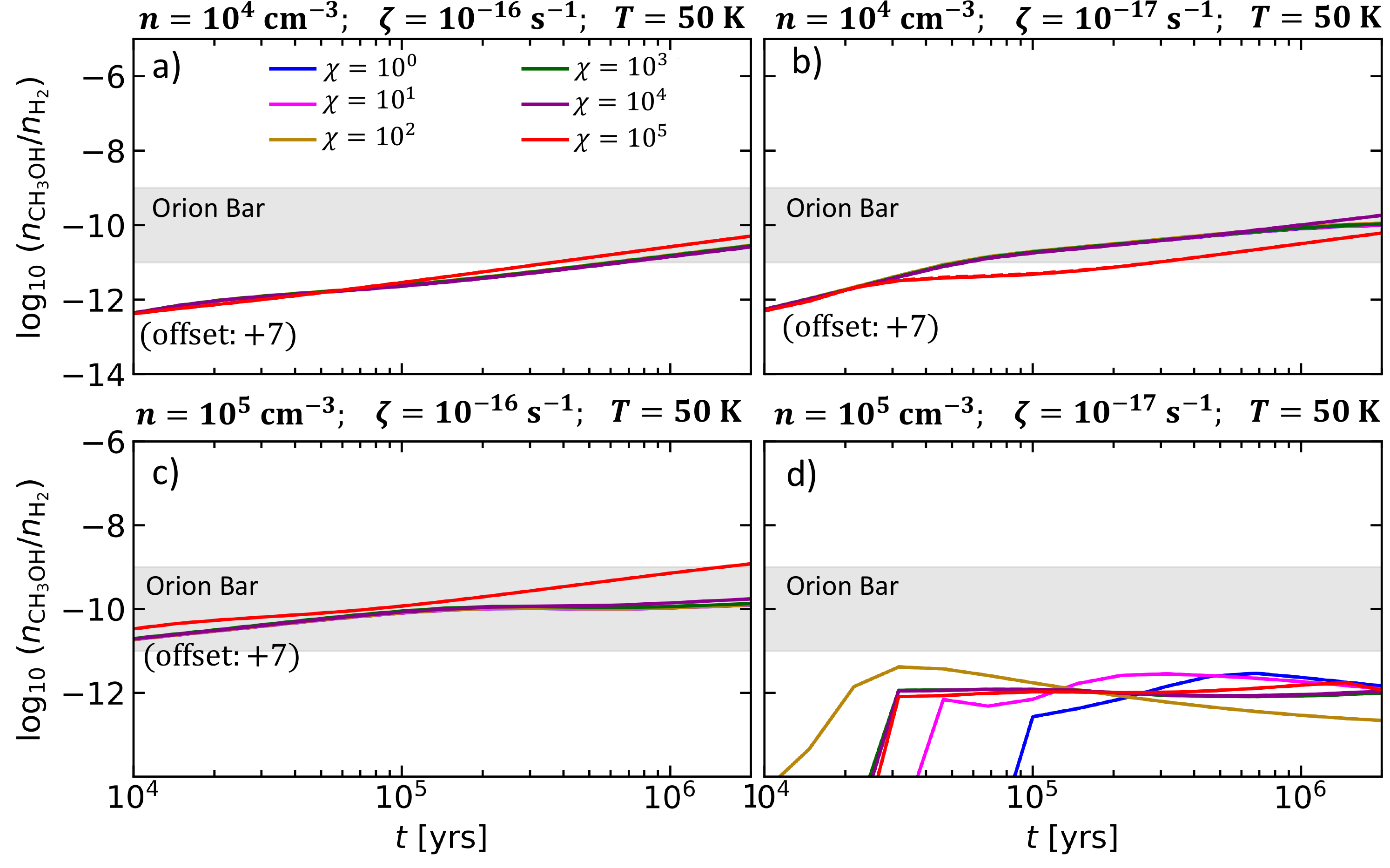}
      \caption{Same as Figure~\ref{abund30}, but for $T =$ 50~K.}
         \label{abund50}
   \end{figure}

\begin{figure}
   \centering
   \includegraphics[width=\hsize]{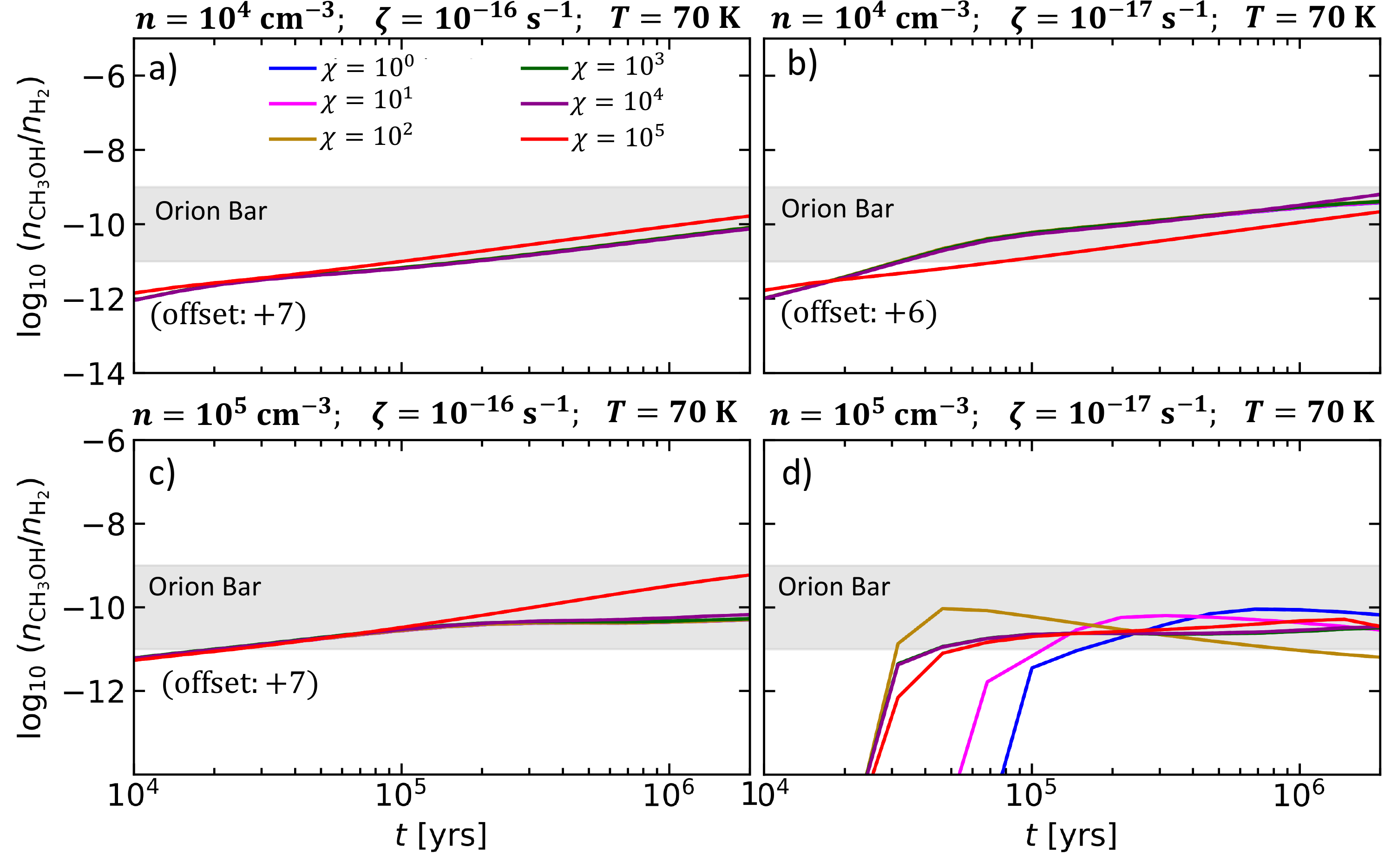}
      \caption{Same as Figure~\ref{abund30}, but for $T =$ 70~K.}
         \label{abund70}
   \end{figure}

\end{appendix}

\end{document}